\pdfoutput=1
\documentclass[acmsmall,nonacm]{acmart}

\usepackage{macros}

\usepackage{equational-macros}

\knowledge{notion}
 | rational functions
 | rational function
 | Rational functions

\knowledge{notion}
 | rational relation
 | rational relations

\knowledge{notion}
 | Types
 | Type
 | type
 | types

\knowledge{notion}
 | rational terms
 | rational term
 | Rational terms

\knowledge{notion}
 | atomic terms
 | atomic term
 | Atomic terms
 | atomic rational terms
 | atomic rational term

 \knowledge{notion}
 | ATOMIC LIFT TERMS
 | atomic lift
 | atomic lifts
 | atomic lift term
 | atomic lift terms

\knowledge{notion}
 | ATOMIC DROP TERMS
 | atomic drop
 | atomic drops
 | atomic drop term
 | atomic drop terms

\knowledge{notion}
 | lift
 | Lift
 | lifts
 | Lifts
 | lift term
 | lift terms

\knowledge{notion}
 | drop
 | Drops
 | drops
 | Drop
 | drop term
 | drop terms

\knowledge{notion}
  | simple rational term
  | simple rational terms
  | Simple rational terms
  | simple terms
  | simple term

\knowledge{notion}
  | aperiodic term
  | aperiodic terms
  | aperiodic rational term
  | aperiodic rational terms

\knowledge{notion}
 | value
 | Values
 | values
 | Value

\knowledge{notion}
 | concatenation map
 | Concatenation maps
 | concatenation maps
 | Concatenation map

\knowledge{notion}
 | multivariate concatenation maps
 | multivariate concatenation map
 | Multivariate concatenation maps

\knowledge{notion}
 | Rebracketing
 | Rebracketings
 | rebracketing
 | rebracketings

\knowledge{notion}
  | finite alphabet
  | finite alphabets
  | finite output alphabet

\knowledge{notion}
  | binary alphabet

\knowledge{ref=def:isomorphism, style=notion}
 | isomorphism
 | isomorphisms

\knowledge{ref=def:injective, style=notion}
 | injective
 | Injective
 | injectivity

\knowledge{ref=def:surjective, style=notion}
 | surjective
 | Surjective
 | surjectivity

\knowledge{notion}
 | denotation
 | denotations
 | Denotation
 | Denotations

\knowledge{notion}
 | string representation
 | string representations
 | String representation
 | String representations

\knowledge{notion}
 | copairing
 | copairings
 | Copairing

\knowledge{notion}
 | category of rational functions

\knowledge{notion}
 | category of rational terms

\knowledge{notion}
 | first-order definable rational functions
 | first-order definable

\knowledge{notion}
 | prefix multiplication term
 | group prefix multiplication
 | Group prefix multiplication

\knowledge{notion}
 | unprefixing term
 | unprefixing terms

\knowledge{notion}
 | bundle term
 | bundle terms

\knowledge{notion}
 | projections coherence laws
 | projections coherence law

\knowledge{notion}
 | multivariate list-polynomial functor
 | multivariate list-polynomial functors
 | list-polynomial functor
 | list-polynomial functors

\knowledge{notion}
 | injective type
 | injective types

\knowledge{notion}
 | normalised drop term
 | normalised drop terms
 | normalised drop
 | normalised drops

\knowledge{notion}
 | composition rank
 | rank

\knowledge{ref=lem:simon-lift, style=notion}
 | Simon lift
 | Simon lifts

\knowledge{ref=lem:group-injection, style=notion}
 | group injection
 | group injections

\knowledge{notion}
 | group-lifts
 | group-lift
 | Group-lifts

\knowledge{ref=lem:lift-then-drop, style=notion}
 | lift-drop factorisation
 | lift-drop factorisations

\knowledge{ref=lem:grouplift-then-drop, style=notion}
 | group-lift--drop factorisation

\acmJournal{PACMPL}

\begin{document}

\title{The Equational Theory of Rational String-to-String Functions}
\thanks{This is the full version, with all proofs, of the paper published in \emph{Proc.\ ACM Program.\ Lang.}\ 11, POPL, Article~101 (January 2027), \url{https://doi.org/10.1145/3871803}.}

\author{Miko{\l}aj Boja\'nczyk}
\orcid{0000-0002-7758-1072}
\email{bojan@mimuw.edu.pl}
\affiliation{%
  \institution{Institute of Informatics, University of Warsaw}
  \city{Warsaw}
  \country{Poland}
}
\author{Kacper Lewandowski}
\orcid{0009-0004-8066-5287}
\email{kacper.lewandowski@mimuw.edu.pl}
\affiliation{%
  \institution{Institute of Informatics, %
  University of Warsaw}
  \city{Warsaw}
  \country{Poland}
}
\author{Katzper Michno}
\orcid{0009-0003-9400-698X}
\email{katzper.michno@mimuw.edu.pl}
\affiliation{%
  \institution{Institute of Informatics, %
  University of Warsaw}
  \city{Warsaw}
  \country{Poland}
}

\begin{abstract}
   We give a sound and complete equational axiomatisation of the rational string-to-string functions, which can be seen as a transducer counterpart of the classical axiomatisations of regular languages by Salomaa [JACM'66], Krob [TCS'91], Bloom and {\'E}sik [MSCS'93], and Kozen [IC'94]. The axioms are stated over a compositional, algebraic syntax for rational functions, namely the rational fragment of the syntax for regular functions introduced by Boja\'nczyk, Daviaud and Krishna [LICS'18]. The axiomatisation is finitary, as it consists of finitely many uniform families of axioms, each parameterised by the types involved. It is also purely equational: it consists only of equations, with no implication rules. 
\end{abstract}

\begin{CCSXML}
<ccs2012>
   <concept>
       <concept_id>10003752.10003766.10003773.10003774</concept_id>
       <concept_desc>Theory of computation~Transducers</concept_desc>
       <concept_significance>500</concept_significance>
       </concept>
   <concept>
       <concept_id>10003752.10003790.10003798</concept_id>
       <concept_desc>Theory of computation~Equational logic and rewriting</concept_desc>
       <concept_significance>500</concept_significance>
       </concept>
 </ccs2012>
\end{CCSXML}

\ccsdesc[500]{Theory of computation~Transducers}
\ccsdesc[500]{Theory of computation~Equational logic and rewriting}

\keywords{equational axiomatisation, rational functions, transducers, completeness, string-to-string functions}

\maketitle 

\section{Introduction}\label{sec:intro}
\paragraph*{Equational Theories of Regular Languages.}
One of the landmark achievements in the theory of automata is the axiomatisation of Kleene algebra. In his original paper on regular expressions~\cite[p.~32]{Kleene1956}, Kleene asked: ``Is there an effective procedure for deciding whether two regular expressions are equal, or equivalent?'' A positive answer to this question is now  known to be a simple consequence of decidable equivalence for finite automata, however the question was quickly picked up in the following meaning: is there an equational axiomatisation, in which every two equivalent regular expressions can be transformed into each other by using a sequence of valid equations such as  $(L +K)^* = L^* \cdot (K  \cdot L^*)^*$? 

In 1964, Redko proved that there cannot be any  axiomatisation that consists of a finite set of equations~\cite{Redko1964}. However, the problem continued to attract a lot of attention, including an entire book by Conway~\cite{Conway1971}. One line of work keeps the axiomatisation purely equational, at the cost of using infinitely many equations. Conway proposed such an infinite family of equational axioms, containing certain \emph{monoid identities}~\cite[pp.~116--119]{Conway1971}, and its completeness was later proved by Krob~\cite[Theorem 0.1]{Krob1991}. Around the same time, Bloom and \'Esik proved completeness for another infinite family of equations, in which the monoid identities are replaced by more general \emph{commutative identities}~\cite[Corollary 2.14]{Bloom_Ésik_1993}. Another line of work strives for axiomatisations that are finite, but at the cost of using certain rules instead of only equations. An important example is Salomaa's axiomatisation~\cite[Theorems 1 and 2]{Salomaa1966}, which uses certain induction rules. Another example -- which is the state of the art in the field -- is the axiomatisation by Kozen~\cite[Theorem 5.5]{Kozen1994}, which uses a finite set of equations and two implications: 
\begin{align*}
L \cdot K \subseteq K \quad \text{implies} \quad L^* \cdot K \subseteq K
\qquad \text{and} \qquad 
K \cdot L \subseteq K \quad \text{implies} \quad K \cdot L^* \subseteq K.
\end{align*}
Subsequent work gave complete axiomatisations of many related systems, including: Kleene algebra without the identity element~\cite[Theorems 10 and 11]{DoumanePous2018}, Kleene algebra with tests~\cite[Theorem 8]{KozenSmith1996}, a network version of Kleene algebra with tests~\cite[Section 4]{Anderson2014NetKAT}, a version of Kleene  
algebra for tree languages~\cite[Theorem 8]{Esik1998}, and a version of Kleene algebra for regular expressions quotiented by bisimulation~\cite[Theorem 5.1]{Grabmayer2022}. Other relevant work includes complete axiomatisations of: iteration theories~\cite[Chapter 14]{BloomEsik1993}, algebras that describe graphs of treewidth at most three~\cite[Theorem 5.5]{DoumaneHumeauPous2024}, and the propositional $\mu$-calculus~\cite[Theorem 37]{Walukiewicz2000}.

\paragraph*{Transducers.}
The work on Kleene algebra concerns regular languages, which can be seen as string-to-Boolean functions. The goal of this paper is to study the equational theories of transducers,  which define string-to-string functions. To the best of our knowledge, this is the first work of this kind.
The particular transducer model that is studied in this paper is the \emph{rational functions}, which we describe below, although in future work we intend to consider more powerful models, such as the regular functions~\cite[Part C]{Bojanczyk2026}, or even the polyregular functions~\cite{Bojanczyk2022}.

A good way to describe the \kl{rational functions}, which highlights the connections with the classical work on regular expressions, is to think of them as a special case of the more general \emph{rational relations}. Rational relations are defined to be binary relations between strings that can be defined by a regular expression in the product monoid $A^* \times B^*$, and rational functions are the \kl{rational relations} that are also functions. They are one of the central models in the transducer literature, with numerous equivalent characterisations, including:  compositions of sequential machines~\cite[Proposition 7.4]{Elgot1965}, Eilenberg  bimachines~\cite[Chapter XI]{Eilenberg1974}, relabellings definable in \mso~\cite[Section C.4.2]{Bojanczyk2026}, as well as origin transductions with monotone origins~\cite[Theorem 3]{Bojanczyk2014}.

A natural idea would be to find a complete axiomatisation of the rational relations. However, this is not possible, since such an axiomatisation (at least with a recursively enumerable set of axioms) would imply that equivalence for rational relations is recursively enumerable, which is known to be false (see~\cite[p.~412]{Griffiths1968} and~\cite[Theorem 7]{FischerRosenberg1968}). 
This problem does not occur for rational functions as their equivalence is decidable, which can be proved using a canonical form shown in~\cite[Theorem 2]{ReutenauerSchutzenberger1991}.\footnote{Alternatively, the problem can be reduced to the equivalence of weighted automata over a field~\cite[Section B.3]{Bojanczyk2026}.}

\paragraph*{Our Contribution.} The contribution of this paper is \cref{thm:main}, which presents a complete, finitary equational axiomatisation of the rational functions. In order to have such an axiomatisation, we need to develop an algebraic syntax for rational functions, the \emph{rational terms}, in which new \kl{rational terms} are constructed from simpler ones by applying certain operations (similarly to regular expressions, which are constructed using sum, concatenation and Kleene star). We therefore begin by defining such an algebraic syntax, although this can hardly be called an original contribution, since our syntax is obtained by  isolating a fragment  of a more expressive system from~\cite{BojanczykDaviaudKrishna2018}. The idea behind our system is that it describes rational functions as morphisms in a certain category, in which the objects are constructed from a singleton \kl{type} $\onetype$ by applying constructors for products, coproducts and lists. A typical morphism in this category would have a type like
\begin{align*}
(\onetype + \onetype)^* \to \onetype^*,
\end{align*}
which describes rational functions that input strings over a \kl{binary alphabet} and output strings over a unary alphabet. The system is defined by identifying certain \kl{atomic terms}, such as the operations of the list monad
\begin{align*}
A^{**} \xrightarrow \concatterm A^* \qquad \text{and} \qquad A \xrightarrow \pureterm A^*,
\end{align*}
and then closing them under certain combinators, such as the composition $f\comp{}g$. The atomic terms and combinators are defined so that they represent exactly the rational functions. The equational axiomatisation consists of axioms such as
\begin{equation}
    \label{eq:intro-examples-of-equations}
\myunderbrace{(f\comp{}g)\comp{}h \equiv f\comp(g\comp{}h)}{associativity of composition}
\qquad \text{or} \qquad 
\myunderbrace{\pureterm\comp\concatterm \equiv \idterm}{one of the axioms  of the list monad}.
\end{equation}

\Cref{thm:main} states that this axiomatisation is finitary and complete. This means that if two terms describe the same rational function, then their equality can be derived from a finite set of \emph{uniform families of axioms}. The axiomatisation is not finite in the literal sense, since the algebra has infinitely many sorts, which are parameterised by types: e.g.~the two axioms in~\eqref{eq:intro-examples-of-equations} are infinite families parameterised by the types of the terms involved. Moreover, unlike Kozen's axiomatisation of Kleene algebra \cite{Kozen1994}, ours uses no implications, only equalities.

\paragraph*{Canonical Models.} Our completeness result requires a substantial proof. One of the central difficulties is related to canonical models. In Kozen's completeness proof for Kleene algebra, the strategy is to prove that every  regular expression is equivalent to one  that describes a minimal automaton~\cite[Lemma 5.4]{Kozen1994}. Since the minimal automaton is unique up to isomorphism, every two regular expressions that describe the same language can be proved equivalent. Unfortunately, this strategy does not work for rational functions, due to the lack of a truly satisfactory canonical model (see~\cite[Section 4]{Filiot2018}). To overcome this difficulty, we use a strategy which is described in \cref{sec:rational-functions}, and which may be of interest outside the context of equational axiomatisations: although rational functions might not have canonical models, such models can be recovered if the input domain is appropriately annotated. Furthermore, the annotations are coherent: two rational functions of the same type can use the same annotation, and thus their canonical models (after annotation) can be meaningfully compared. This is, however, only one of many difficulties, and the others will be described as they arise.

\paragraph*{Structure of the Paper.} In \cref{sec:rational-functions}, we define the rational functions and present a simplified version of our equational axiomatisation. In \cref{sec:syntax}, we define rational terms, an algebraic syntax for rational functions based on~\cite{BojanczykDaviaudKrishna2018}. In \cref{sec:equational-theory} we state the axioms of our equational axiomatisation, and in Sections~\ref{sec:proof-overview}--\ref{sec:completeness-for-all-rational-terms} we prove its completeness. Technical proofs are given in \appendixplace\iffullversion\else{} of this paper, which is available on arXiv~\cite{fullversion}\fi.

\section{Rational Functions and a Baby Version of Our Equational Axiomatisation}
\label{sec:rational-functions}
In this section, we define the \kl{rational functions} and extend them to more structured \kl{types} than strings. We also present a baby version of our equational axiomatisation, which contains some of the core ideas of the main result, such as canonical forms.

\subsection{Definition of Rational Functions}
\label{sec:definition-of-rational-functions}
Given two \kl{finite alphabets} $A$ and $B$, we can consider a regular expression over their product, in which the atomic symbols are pairs in $A^* \times B^*$. Such an expression defines a binary relation $L \subseteq A^* \times B^*$, called a \intro{rational relation}. Here are some examples: 
\begin{gather*}
\myunderbrace{((a,a)+(b,b))^*}{identity function}
\qquad
\myunderbrace{((a,\varepsilon)+(b,b))^*}{delete every $a$}
\qquad
\myunderbrace{((a,aa)+(b,b))^*}{duplicate every $a$}
\\[2pt]
\myunderbrace{((a,a)+(b,b)+(\varepsilon,a)+(\varepsilon,b))^*}{scattered subsequence relation}
\qquad \quad
\myunderbrace{(aa,a)^*}{halve inputs of even length}.
\end{gather*}
We are interested in the case where the rational relation is a function, i.e.~for every input string there is exactly one output string, as in the identity, deletion and duplication examples above. (The halving example is not a function, since it is undefined on inputs of odd length.)

\begin{definition}
    [Rational function]
    \label{def:rational-function}
     A \intro{rational function} $f : A^* \to B^*$ is any function whose graph is a \kl{rational relation}.
\end{definition}

Equivalently, rational relations can be defined by nondeterministic finite automata whose transitions are labelled by pairs of (possibly empty) input and output strings. This definition is more convenient for proving closure properties, such as closure under composition, which follows from a standard product construction~\cite[Theorem 4.1]{Eilenberg1974}.

\subsection{Rational Functions on Types}
    As mentioned in our introduction, we think of \kl{rational functions} as operating on types that are built from a unit type by applying constructors for products, coproducts, and lists. 
A formal definition of these types  is given below.
\begin{definition}
    [Types]\label{def:types} \intro{Types} are expressions built using the following constructors:
\begin{align*}
    \myunderbrace{\onetype}{unit  type \\ which contains \\ a unique element $1 \in \onetype$} 
\qquad
\myunderbrace{A \times B}{product \\ type}
\qquad
\myunderbrace{A + B}{coproduct \\ type}
\qquad
\myunderbrace{A^*}{list \\ type}.
\end{align*}
\end{definition}

Instead of considering only string-to-string rational functions, we will consider type-to-type rational functions, i.e.~rational functions of type $A \to B$ where the input type  $A$ and the output type  $B$ are constructed as in the above definition. To extend the notion of rationality to such functions, we  use string representations. As the output alphabet for these representations,  we use  an eight-letter alphabet that we denote by $8$:
\begin{align*}
( \quad ) \quad [ \quad ] \quad , \quad \onelement \quad \leftterm \quad \rightterm.
\end{align*}
 The \intro{string representation} for elements of types is defined by induction:  the unique element of $\onetype$ is represented by  $\onelement$; elements of $A + B$ are represented by $\leftterm\  a$ or $\rightterm\  b$; pairs are represented by $(a,b)$, and lists are represented by $[a_1,\ldots,a_n]$.

\begin{definition}[Rational type-to-type functions]\label{def:rational-functions-on-types}
    Let $A$ and $B$ be \kl{types}. A function $A \xrightarrow f  B$ is called rational if  the following diagram commutes:
    \[
    \begin{tikzcd}
    A 
    \ar[r,"f"]
    \ar[d,"\text{string representation}"']
    &
    \ar[d,"\text{string representation}"]
    B
    \\
    8^* 
    \ar[r,"f'"']
    &
    8^*
    \end{tikzcd}
    \qquad \text{for some rational string-to-string function $f'$.}
    \]
\end{definition}
Although \cref{def:rational-functions-on-types} is simple, it is also robust. In particular, no information is lost by passing to string representations, since the string representation can be inverted by a rational function (see the proof of \cref{lem:provably-injective-string-representation}). Moreover, in \cref{sec:syntax} we give a syntax of \kl{rational terms}, which represent exactly the rational functions on types.

As we mentioned in \cref{sec:definition-of-rational-functions}, rational functions are closed under composition. Since the identity is also a rational function, it follows that the rational functions on types  form a category, which we call the \intro{category of rational functions}. The objects of this category are types, and the morphisms are rational functions. As we will see in \cref{sec:syntax}, this is a somewhat peculiar category: $+$ is a proper coproduct, but $\times$ is not a categorical product, since there is no pairing. In fact, $\times$ is not even symmetric, since swapping the components of a pair is not a rational function in general.

\subsection{Simple Rational Terms} Before we define general \kl{rational terms} in \cref{sec:syntax}, in this section, we study a restricted class of them, which we call the \emph{simple rational terms}. The rough idea behind simple terms is that these represent type-to-string functions that are executed recursively over the structure of the input \kl{type}, with no states, and in each step they can only add some string constants. This class of terms is very limited, but we will show in \cref{thm:simplification-of-rational-terms} that every \kl{rational function} can be reduced in a certain way to a function represented by a simple term. 

From now on, the letters $f, g, h,\ldots$ refer to terms rather than functions. The rational function represented by a term $f$ is called its \intro{denotation}, and is written $\sem{f}$. Simple rational terms output strings over a finite alphabet, which is represented as a type as follows.

\begin{definition}[Finite alphabet]\label{def:finite-alphabet}
    A \kl{type} is called a \intro{finite alphabet} if it has the form $\onetype + \cdots + \onetype$.
\end{definition}
In other words, a finite alphabet is a coproduct of several copies of $\onetype$, with one copy for each letter. For example, $\twotype$ is the \intro{binary alphabet}.

\begin{definition}[Simple rational terms]\label{def:simple-rational-terms}  Let $B$ be a \kl{finite output alphabet}. 
    \emph{\intro{Simple rational terms}} with outputs in $B^*$ are constructed using the following rules:
    \begin{itemize}
        \item \textbf{Empty.} The  term $[] :\onetype \to B^*$ with empty output is a \kl{simple rational term}.
        \item \textbf{Combinators.} If $w \in B^*$ is a string constant (by $\varepsilon$ we denote the empty string), and 
        \begin{align*}
        A_1 \xrightarrow {f_1} B^* \quad \text{and} \quad  A_2 \xrightarrow {f_2} B^*
        \end{align*}
        are  \kl{simple rational terms}, then the following are also \kl{simple rational terms}:
        \begin{align*}
        \myunderbrace{A_1 + A_2 \xrightarrow{\either {f_1} {f_2}}  B^*}{\textbf{copairing} \\ apply $\sem{f_1}$ to inputs from $A_1$\\ and $\sem{f_2}$ to inputs from $A_2$} 
        \quad 
        \myunderbrace{A_1 \times A_2 \xrightarrow { f_1 \, \fdot \, f_2 } B^*}{\textbf{concatenation} \\ concatenate the \\  outputs of \\ $\sem{f_1}$ and $\sem{f_2}$}
        \quad 
        \myunderbrace{A_1^* \xrightarrow { \hom f_1 } B^*}{\textbf{homomorphism} \\ apply $\sem{f_1}$ to each \\ list element and then \\ concatenate the outputs}
        \quad 
        \myunderbrace{A_1 \xrightarrow { w \, \wdot \, f_1 } B^*}{\textbf{prepend} \\ apply $\sem{f_1}$ \\ and prepend $w$}
        \quad
        \myunderbrace{A_1 \xrightarrow { f_1 \, \wdot \, w} B^*}{\textbf{append} \\ apply $\sem{f_1}$ \\ and append $w$}.
        \end{align*}
    \end{itemize}
\end{definition} 

Note that $\fdot$ combines the domains of the two terms using product, while $\wdot$ does not change the domain. One could also imagine a combinator that concatenates pointwise two terms of the same type $A \to B^*$, but as we will see in \cref{sec:atomic-for-products}, it is not admissible for rational terms, since it would require a form of input duplication that cannot be implemented by a rational function.

\begin{myexample}(Simple rational terms).
Let us discuss the possible simple rational terms of type $N^* \to M^*$, where $N$ and $M$ are finite alphabets. Such a term is necessarily of the form $w \wdot \hom f \wdot v$, where $f : N \to M^*$ is an arbitrary letter-to-string simple term. In other words, the simple rational terms of type $N^* \to M^*$ represent exactly the string-to-string homomorphisms, with some fixed constants at the beginning and at the end of the output string.
\end{myexample}

The main results of this section are: in Section~\ref{sec:canonical-form-for-simple-rational-terms} we show that functions represented by simple rational terms have a canonical form, which is unique up to a certain form of conjugacy, and in Section~\ref{sec:simplification-of-rational-terms} we show that every rational function is represented by a simple term after lifting the domain. Putting these two results together, we will obtain a certain canonical form for rational functions.
\subsection{Equivalence for Simple Rational Terms}
\label{sec:canonical-form-for-simple-rational-terms}

The definition of \kl{simple rational terms} is inductive: each simple rational term is constructed from the empty term $[]$ by using the five combinators in the definition. It is possible to define the same function with two different terms. One reason is monoidality of string concatenation, e.g.~$f$ and $\varepsilon \wdot f$  both describe the same function. Another related reason is an iterated form of associativity, which we call conjugacy, as illustrated in the following example.

\begin{myexample}(Conjugacy).
Consider the following two simple rational terms of type $\onetype^* \to \set{a,b}^*$, where $\set{a,b}$ denotes a two-letter alphabet:
\begin{align*}
    \hom (1 \mapsto ab) \wdot a \quad \text{and} \quad a \wdot \hom (1 \mapsto ba),
\end{align*}
where $1 \mapsto ab$ denotes the simple term $ab \wdot []$.
We adopt the notational convention that $\hom$ binds more strongly than $\wdot$, and hence in the first term we first apply the homomorphism and then we append $a$.
Both terms describe the same function, namely $1^n \mapsto (ab)^na = a(ba)^n$.
\end{myexample}

As the following theorem shows, conjugacy is the only nontrivial reason why two simple rational terms could describe the same function.
\begin{theorem}[Conjugacy completeness]\label{thm:conjugacy-completeness} For every two \kl{simple rational terms} $f$ and $g$, we have that $\sem{f} = \sem{g}$ if and only if $f \equiv g$, where $\equiv$ is defined using the following rules:
    \begin{enumerate}
        \item \label{item:congruence-rule} \textbf{Congruence:} $\equiv$ is stable under applying the constructors of \kl{simple rational terms}.
        \item \label{item:concatenation-monoidality-rule} \textbf{Concatenation monoidality:} for all \kl{simple rational terms} $f,g,h$ and strings $w,v$ we have
        \begin{gather*}
            f \wdot \varepsilon \equiv f \equiv \varepsilon \wdot f \qquad [] \wdot w \equiv w \wdot [] \\ 
            (f \wdot w) \wdot v \equiv f \wdot (w \cdot v) \qquad
            w \wdot (v \wdot f) \equiv (w \cdot v) \wdot f \qquad
            (w \wdot f) \wdot v \equiv w \wdot (f \wdot v) \\
            (f \wdot w) \fdot g \equiv f \fdot (w \wdot g) \qquad
            w \wdot (f \fdot g) \equiv (w \wdot f) \fdot g \qquad
            (f \fdot g) \wdot w \equiv f \fdot (g \wdot w).
        \end{gather*}
        \item \label{item:constants-copairing} \textbf{Constants and copairing:} for all \kl{simple rational terms} $f_1, f_2$ and strings $w$ we have
              \begin{align*}
                  w \wdot (\either {f_1} {f_2}) \equiv \either {(w \wdot f_1)} {(w \wdot f_2)} \qquad
                  (\either {f_1} {f_2}) \wdot w \equiv \either {(f_1 \wdot w)} {(f_2 \wdot w)}.
              \end{align*}
        \item \label{item:distributivity-rule} \textbf{Case splitting:} for all \kl{simple rational terms} $f_1,f_2,g_1,g_2,h$ we have
              \begin{align*}
                  \either {(f_1 \fdot h)} {(f_2 \fdot h)} \equiv \either {(g_1 \fdot h)} {(g_2 \fdot h)} \enspace \text{implies} & \enspace (\either {f_1} {f_2}) \fdot h \equiv (\either {g_1} {g_2}) \fdot h \\
                  \either {(h \fdot f_1)} {(h \fdot f_2)} \equiv \either {(h \fdot g_1)} {(h \fdot g_2)} \enspace \text{implies} & \enspace h \fdot (\either {f_1} {f_2}) \equiv h \fdot (\either {g_1} {g_2}).
              \end{align*}
        \item \label{item:conjugacy-rule} \textbf{Conjugacy:} for all \kl{simple rational terms} $f,g$ and string $w$ we have
              \begin{align*}
                  f \wdot w \equiv w \wdot g \quad \text{implies} \quad  \hom f \wdot w \equiv w \wdot \hom g.
              \end{align*}
    \end{enumerate}
\end{theorem}

Before proving the theorem, we underline how it falls short of a full completeness result. The first issue is that it concerns only the simple rational terms, which are too restricted to capture all \kl{rational functions}. The second issue is that it is based on implications, while we would like our system to be based only on equalities. Extension to all rational functions will be addressed informally in \cref{sec:simplification-of-rational-terms}, and formally in \cref{sec:completeness-for-all-rational-terms}, while the issue with implications will be resolved in \cref{sec:simple-rational-equationally}.

\begin{proof}[Proof of \cref{thm:conjugacy-completeness}]
    The soundness part of the theorem, which says that $f \equiv g$ implies that $f$ and $g$ define the same function, is straightforward. The soundness of the conjugacy rule is proved by induction on the length of the input list for the terms $\hom f \wdot w$ and $w \wdot \hom g$.
    
    We now turn to the completeness part, which says that if the terms $f$ and $g$ define the same function, then one can prove $f \equiv g$. This is shown by induction on the structure of the input \kl{type}. In the induction basis, when the input type is $A = \onetype$, both terms are necessarily constructed from the empty term $[]$ by prepending and appending constants. Hence, $f \equiv g$ can be proved using the rules for concatenation monoidality, where the rule $[] \wdot w \equiv w \wdot []$ is used to bring both terms to the form $w \wdot []$, with $w$ being their common output. In the induction step for coproducts, we first use the rules for constants and \kl{copairing} to push all string constants inside the copairing, so that $f = \either {f_1} {f_2}$ and $g = \either {g_1} {g_2}$. Then $f_1 \equiv g_1$ and $f_2 \equiv g_2$ follow from the induction hypothesis, and we finish by congruence.
    
    The interesting case is the induction step for products, which also treats the case of a list. By the monoidality and case-splitting rules, we can assume that the input type is a product of lists, possibly with some factors of the unit type:
    \begin{align*}
    A = (A_1)^* \times \cdots \times (A_n)^*.
    \end{align*}
    By monoidality, we can assume without loss of generality that $f$ and $g$ are of the form
    \begin{align*}
    f & = w_0 \wdot (\hom f_1 \wdot w_1) \fdot \cdots \fdot (\hom f_n \wdot w_n) \\
    g & = v_0 \wdot (\hom g_1 \wdot v_1) \fdot \cdots \fdot (\hom g_n \wdot v_n).
    \end{align*}
    The proof uses  a new nested induction, which is on the sum of lengths of the constants $w_i$ and $v_i$. If both constants $w_0$ and $v_0$ are nonempty, then they must begin with the same letter. Hence, we can strip away this letter, and then apply the induction hypothesis in the nested induction. The interesting case is when one of the constants $w_0$ or $v_0$ is empty; by symmetry we assume that $v_0$ is empty. We now  observe that  the lengths of the outputs need to be synchronised across each of the components $A_1,\ldots,A_n$. 
\begin{claim}
    For every $i \in \set{1,\ldots,n}$ and  $a \in A_i$, the strings $\sem{f_i}(a)$ and $\sem{g_i}(a)$ have the same length.
\end{claim}
\begin{proof}
    Toward a contradiction, suppose that there is some counterexample $a \in A_i$. Consider some input from $A$, and modify it by appending one copy of $a$ to the list on the $i$-th component. After this modification, the output lengths change in a different way, and hence the two functions cannot be equal.
\end{proof}
Apply the above claim to $i=1$, concluding that $\sem{f_1}$ and $\sem{g_1}$ have the same output lengths.
Consider an input from $A$ where all lists are empty. Then the two outputs will be  $w_0 \cdots w_n$ and $v_1 \cdots v_n$, since we have assumed that $v_0$ is empty.  Therefore, $w_0$  is a prefix of  $v_1 \cdots v_n$.
Now consider an input from $A$ where the first list has one item and all remaining lists are empty: 
\begin{align*}
([a],[],\ldots,[])  \qquad \text{for $a \in A_1$.}
\end{align*}
For this input, the two outputs will be 
\begin{align*}
w_0 \cdot \sem{f_1}(a) \cdot w_1 \cdots w_n \quad \text{and} \quad
\sem{g_1}(a) \cdot v_1 \cdots v_n.
\end{align*}
Since the outputs $\sem{f_1}(a)$ and $\sem{g_1}(a)$ have the same length by the  claim, and  $v_1 \cdots v_n$ begins with $w_0$, it follows that
\begin{align*}
w_0 \cdot \sem{f_1}(a) = \sem{g_1}(a) \cdot w_0.
\end{align*}
This holds for every $a \in A_1$, and hence $\sem{w_0 \cdot f_1} = \sem{g_1 \cdot w_0}$. These are simple rational terms with domain $A_1$, which is strictly simpler than $A$. Therefore, by the induction hypothesis, denotational equality implies provable equality. Thus, we have $w_0 \wdot f_1 \equiv g_1 \wdot w_0$. By the conjugacy rule, we further have $w_0 \wdot \hom f_1 \equiv \hom g_1 \wdot w_0$. After this replacement, we can strip away $\hom{g_1}$ which appears as a prefix on both sides, and then apply the induction hypothesis in the main induction.
\end{proof}

One way of presenting the above theorem is that a simple rational term for a given function is unique up to conjugacy (with rules~\ref{item:congruence-rule}--\ref{item:distributivity-rule} being implicit).
A simple corollary of the above theorem is a polynomial time equivalence test for  simple rational terms.
\begin{corollary}
    Equivalence for \kl{simple rational terms} is decidable in polynomial time.
\end{corollary}
\begin{proof}
We can check equivalence under $\equiv$ in polynomial time. As witnessed by the proof of \cref{thm:conjugacy-completeness}, the only conjugacies which are needed are those that use an infix of the constant string $w_0 \cdots w_n$, and there is a quadratic number of such choices.  
\end{proof}

\subsection{Reduction to Simple Rational Terms}
\label{sec:simplification-of-rational-terms}
In \cref{thm:conjugacy-completeness}, we showed that \kl{simple rational terms} are unique up to conjugacy, and their equivalence can be decided in polynomial time. One could think that this is an irrelevant result, because there are very few functions represented by simple rational terms. However, as the following theorem shows, every \kl{rational term} can be made simple by suitably lifting the domain.

\begin{theorem}
    [Simplification of rational terms]
    \label{thm:simplification-of-rational-terms}
    Consider \kl{rational terms}
    $f, g: N^* \to M^*$
    where $N$ and $M$ are \kl{finite alphabets}. There is a \kl{type} $C$ and a \kl{rational term}
    $\pi: C \to N^*$
    with surjective \kl{denotation}, such that  $\sem{\pi\comp{}f}$ and $\sem{\pi\comp{}g}$ are represented by \kl{simple rational terms}.
\end{theorem}

We do not give  a proof of \cref{thm:simplification-of-rational-terms} here for three reasons: (a) we have not yet defined rational terms (this is done in \cref{sec:syntax}), (b) we will prove a strictly stronger statement in the following sections (\cref{thm:provable-simplification}), and (c) an alternative proof can be obtained by modifying the proof of~\cite[Theorem 3.6]{BojanczykDaviaudKrishna2018}, which itself is based on a result from semigroup theory that is called   the Factorisation Forest Theorem~\cite[Theorem 6.1]{Simon1990}. 
The theorem is illustrated below, with its conclusion in blue:
\[
    \forall \; \blue \exists \qquad
    \begin{tikzcd}
        N^*
        \ar[r,"{f,\,g}"]
        & M^* \\
        \blue{C}
        \ar[two heads, blue, u,"\pi"{xshift=-0.5ex, yshift=-0.5ex}]
        \ar[ur,blue, "{\pi \comp f, \, \pi  \comp  g}"']
    \end{tikzcd}
\]
Since $\sem{\pi}$ is surjective, $f$ and $g$ describe the same function if and only if $\pi\comp{}f$ and $\pi\comp{}g$ do. Therefore, the theorem reduces the equivalence problem for rational terms to that for simple rational terms. To use this reduction in a completeness proof, however, it needs to be carried out within the equational system, i.e.~the conclusion of \cref{thm:simplification-of-rational-terms} needs to be derivable from the axioms. This is exactly what we do in \cref{sec:completeness-for-all-rational-terms}.

\section{Compositional Syntax of Rational Functions}
\label{sec:syntax}
As explained in \cref{sec:intro}, in order to have an equational axiomatisation, we need a compositional (algebraic) syntax for the \kl{rational functions} -- the \kl{rational terms} which are constructed from certain \kl{atomic terms} by applying certain combinators. We describe such a system in this section.

\subsection{Atomic Terms and Combinators}
\label{sec:atomic-terms-and-combinators}
\phantomintro{atomic terms}%
Our system of  atomic terms and combinators  is not entirely new, since it can be seen as a  part of a system that was described in~\cite{BojanczykDaviaudKrishna2018}. However, the system from~\cite{BojanczykDaviaudKrishna2018} was designed for a larger class of functions -- the regular functions -- and thus it contained terms representing functions that are not rational (such as string reversal) or combinators that do not preserve rationality (such as pairing of terms). For this reason, we will now carefully isolate the rational fragment of the system from~\cite{BojanczykDaviaudKrishna2018}. At each step of the discussion, we briefly argue about soundness of the system, understood in the sense that all atomic terms represent \kl{rational functions}, and that all combinators preserve rationality. We do this informally here; we refer to the lecture notes~\cite[Section C.5.2]{Bojanczyk2026} for a formal treatment.

\subsubsection{Composition}
Arguably the most important combinator is  composition of terms: 
\begin{align*}
    \combinator{
        A \xrightarrow  f B \quad B \xrightarrow g C
    } {
        A \xrightarrow {f \comp g} C
    }{\ (composition)}
\end{align*}
As we have argued in \cref{sec:definition-of-rational-functions}, \kl{rational functions} are closed under composition, and hence the combinator $\comp$ preserves rationality. Also, for every \kl{type} $A$ we have an \kl{atomic term} for the identity:
\begin{atomicterm}
    \label{atomic:identity}
    \aliaslabel{atomic:first}{atomic:identity}
\describeterm{\idterm}{A}{A}.
\end{atomicterm}
\subsubsection{Coproducts}
\label{sec:atomic-for-co-products}
To deal with coproducts, we introduce coprojections:
\begin{atomicterm}
    \label{atomic:coprojections}
\describeterm{\leftterm} {A}{A + B}
\qquad
\describeterm{\rightterm} {B}{A + B},
\end{atomicterm}
whose \kl{denotations} are easily seen to be rational.
Under \kl{string representation}, the denotation of a coprojection simply prepends one letter to the input, which is a \kl{rational function}.  We also have two combinators for coproducts:
\begin{align*}
\combinator{A_1 \xrightarrow {f_1} B \quad A_2 \xrightarrow {f_2} B }{A_1 + A_2 \xrightarrow {\either{f_1}{f_2}} B}{\ (copairing)}
\qquad 
\combinator{A_1 \xrightarrow {f_1} B_1 \quad A_2 \xrightarrow {f_2} B_2}{A_1 + A_2 \xrightarrow {{f_1 + f_2}} B_1 + B_2}{\ (functoriality of $+$)}.
\end{align*}
Note that copairing was already used in the definition of \kl{simple terms} (\cref{def:simple-rational-terms}).
Both combinators preserve rationality. For copairing, we modify the regular expressions for $\sem{f_1}$ and $\sem{f_2}$ so that they require the $\leftterm$ and $\rightterm$ letters at the beginning of the input, respectively, and then we take their union. For $+$, the construction is the same, except that we need to keep the $\leftterm$ and $\rightterm$ letters in the output as well. 

One can derive $+$ from copairing, since $f_1 + f_2$ is obtained by copairing terms that are obtained by composing $f_1$ and $f_2$ with the corresponding coprojections. To obtain copairing from $+$, we need an extra \kl{atomic term} for the codiagonal:
\begin{atomicterm}
    \label{atomic:co-diagonal}   
    \codiagterm : A + A \to A.
\end{atomicterm}
In terms of string representations, the denotation of the codiagonal simply deletes the first letter of the input string. As a result, in our system we can have either copairing, or $+$ and the codiagonal. Since we will use the functorial approach for the remaining \kl{type} constructors (pairs and lists), we choose the second option, and hence \intro{copairing} becomes a derived combinator:
\begin{align*}
\either {f_1} {f_2} \quad \eqdef \quad (f_1 + f_2)\comp \codiagterm.
\end{align*}
More generally, we have the \emph{$n$-fold copairing}:
\begin{align*}
A_1 + \dots + A_n \xrightarrow{\eithermany{n}{f_1 \cdots f_n}} B \qquad \text{ where } A_i \xrightarrow{f_i} B \text{ for every } i \in [n].
\end{align*}
We can assume that $n$-fold coproducts (and products) are right-associated, although it is not important -- all bracketings are equivalent. We justify this in \appref{sec:distributivity}.

\subsubsection{Products}
\label{sec:atomic-for-products}
We now discuss  \kl{rational terms} that operate on products. Our system for rational terms deviates from the system for regular terms in~\cite{BojanczykDaviaudKrishna2018} where products are concerned. In particular, products in the rational system will be noncommutative, unlike in the regular system. The first \kl{atomic terms} for the product \kl{type} are the projections:
\begin{atomicterm}
    \label{atomic:projections}
\describeterm{\fstterm} {A \times B}{A}
\qquad 
\describeterm{\sndterm} {A \times B}{B}.
\end{atomicterm}
These represent \kl{rational functions}, although the proof is a bit more difficult than for coprojections, since the function must identify the separating comma in a pair $(a,b)$. This is possible because, by a simple induction on the type, for every type $A$ the language $\setbuild{ \text{string representation of $a$}}{$a \in A$}$ is regular. Thanks to this observation, the unique decomposition of the \kl{string representation} $(a,b)$ into the string representations of $a$ and $b$ can be guessed by the function, and thus $\sem{\fstterm}$ and $\sem{\sndterm}$ are rational.

Things get more interesting with the combinators for products:
\begin{align*}
\red{\myunderbrace{\combinator{A \xrightarrow {f_1} B_1 \quad A \xrightarrow {f_2} B_2}{A \xrightarrow{(f_1,f_2)} B_1 \times B_2}{\ (pairing)}}{not a part of our system}} \qquad  
\combinator{A_1 \xrightarrow {f_1} B_1 \quad A_2 \xrightarrow {f_2} B_2}{A_1 \times A_2 \xrightarrow {{f_1 \times f_2}} B_1 \times B_2}{\ (functoriality of $\times$)}.
\end{align*}
The combinator $\times$ preserves rationality, which is proved by finding the separating comma in the input string, and calling $\sem{f_1}$ and $\sem{f_2}$ on the corresponding parts of the input. However, the pairing combinator \emph{does not} preserve rationality. The reason is that in the conclusion of the pairing combinator, the input string is used twice, which is a form of copying that is not available to rational functions.  The failure of pairing is connected to other functions that are not rational, namely duplication and commutativity of products:
    \begin{align*}
\red{ \myunderbrace{\describeterm{\diagterm} {A}{A \times A}}{not a part of our system}
    \qquad 
    \myunderbrace{A \times B \xrightarrow{\mathtt{swap}} B \times A.}{not a part of our system}}
    \end{align*}
A formal proof of nonrationality for these functions can be obtained by running the algorithm for deciding rationality from~\cite[Theorem 2]{FiliotGauwinReynierServais2013}.
Because pairing fails in our system, we will need explicit operations for associativity and unit for products (in the presence of pairing, these could be derived): 
    \begin{atomicterm}
        \label{atomic:monoidal}
\describeterm{\appendoneterm} {A}{A \times \onetype} \qquad
\describeterm{\prependoneterm} {A}{\onetype \times A}
\qquad
 \begin{tikzcd}[ampersand replacement=\&, column sep=large]
A \times (B \times C)
\ar[r,"\assoclterm", shift left=0.6ex]
\&
\ar[l,"\assocrterm", shift left=0.6ex]
(A \times B) \times C.
\end{tikzcd}
    \end{atomicterm}
In categorical terminology, the \kl{denotations} of these terms show that the \kl{category of rational functions} is a monoidal category with respect to the product, with unit $\onetype$. However, it is not a symmetric monoidal category, since the swap function is not rational.\footnote{The coproduct almost gives a symmetric monoidal category, except there is no neutral element. The neutral element for coproduct could be recovered by having a type $0$ with no elements, which would have limited effect on our system.} As a corollary, $\times$ is not a categorical product.

\subsubsection{Distributivity}
\label{sec:atomic-for-distributivity}
    We can distribute product over coproduct.  Because products are not symmetric, we need to have two variants of distribution, on the left and on the right: 
    \begin{atomicterm}
        \label{atomic:distributivity}
\describeterm{\distrterm} {A \times (B + C)}{(A \times B) + (A \times C)} \qquad 
\describeterm{\distlterm} {(A + B) \times C}{(A \times C) + (B \times C)}.
    \end{atomicterm}
    Both  denote \kl{rational functions}, since they  amount to finding the separating comma in the input string, and then shifting $\leftterm$ or $\rightterm$ out to the front of the string.  The \kl{denotation} of each of the two distribution terms is an isomorphism, and hence it has an inverse; these inverses can be derived using the previous constructions (see \cref{sec:theory:isomorphism-axioms}), and therefore they are not included as \kl{atomic terms}.

\subsubsection{Lists}
\label{sec:atomic-for-lists}
The most important \kl{type} constructor for our system is the list constructor $A^*$. We begin with the map combinator:
\begin{align*}
\combinator{A \xrightarrow f B}{A^* \xrightarrow {f^*} B^*}{\ (functoriality of $*$)}
\end{align*}
The combinator $*$ preserves rationality by an argument similar to the one for $\times$: the \kl{string representation} of the input list is split into the representations of its elements, by identifying the separating commas, and then $\sem{f}$ is applied to each element.
Let us  now describe the \kl{atomic rational terms} that operate on lists. All of them denote functions that are easily seen to be rational. There are four terms that describe the inductive definition of lists, since we have both constructors and destructors, and lists can be oriented in two directions:
    \begin{atomicterm}
        \label{atomic:cons-snoc}
    \begin{tikzcd}[ampersand replacement=\&, column sep=large]
\onetype + A \times A^*
\ar[r,"\consterm", shift left=0.6ex]
\&
\ar[l,"\unconsterm", shift left=0.6ex]
A^*
\ar[r,"\unsnocterm", shift left=0.6ex]
\&
\ar[l,"\snocterm", shift left=0.6ex]
\onetype + A^* \times A.
\end{tikzcd}
    \end{atomicterm}
Apart from the constructors and destructors, we also have a term for concatenating lists of lists:
\begin{atomicterm}
    \label{atomic:concat}
A^{**} \xrightarrow{\concatterm} A^*.
\end{atomicterm}
There is an atomic term which clusters the elements of a list over a coproduct type:
\begin{atomicterm}
    \label{atomic:split}
(A + B)^* 
\xrightarrow{\blockterm}
A^* \times (B \times A^*)^*.
\end{atomicterm}
A defining property of this term is that its \kl{denotation} is an isomorphism, whose inverse is defined by injecting $A$ and $B$ into $A+B$, and then concatenating all list elements. This inverse can be derived using the previous terms (see \cref{sec:theory:isomorphism-axioms}), and is hence not included as an atomic term. Finally, we have two terms with mutually inverse denotations:
    \begin{atomicterm}
        \label{atomic:shift}
    \begin{tikzcd}[ampersand replacement=\&, column sep=large]
(A \times B)^* \times A
\ar[r,"\shiftrterm", shift left=0.6ex]
\&
\ar[l,"\shiftlterm", shift left=0.6ex]
A \times (B \times A)^*,
    \end{tikzcd}
    \end{atomicterm}
which shift brackets in the expected way. These two terms are something of a curiosity, since they can be derived from the previous terms. However, the derivation is not natural -- in the technical sense of the word -- and hence it will be more convenient for the equational axiomatisation to have these terms as atomic terms.

\subsubsection{Groups}
\label{sec:atomic-for-groups}
The system defined so far can only describe a fragment of the \kl{rational functions}, namely the \intro{first-order definable rational functions}~\cite[p.~2]{Filiot2016}. In order to escape this fragment, we need an \kl{atomic term} whose \kl{denotation} is not first-order definable (see \cite[Section C.4]{Bojanczyk2026} for more on this topic). This term  is based on group multiplication. 
Consider a finite group $G$ whose underlying set is a \kl{finite alphabet}. (Every finite group can be obtained this way up to group isomorphism.) For example, the underlying set could be $\onetype + \onetype$, and the group could be the cyclic group of order two.\footnote{Group operations (identity, inverse, and multiplication) are all rational functions, since their domains are finite \kl{types}.} For every such finite group and every type $A$, we have a \intro{prefix multiplication term}:
\begin{atomicterm}
    \label{atomic:group-cut}
    \aliaslabel{atomic:last}{atomic:group-cut}
(A \times G)^* \xrightarrow{\prefterm} (A \times G)^*,
\end{atomicterm}
which is defined as follows: the output list has the same length as the input list, the elements from $A$ are not changed,  and the $i$-th group element in the output list is obtained by multiplying the first $i$ group elements of the input list, using the group structure. The denotation of this term is rational, since it can be computed by a transducer that reads the \kl{string representation} of the input list from left to right, and stores in its state the product of the group elements that have been read so far.

\begin{figure}
            \begin{gather*}
\begin{array}{c}
\text{Identity (atomic term \ref{atomic:identity})} \\[2pt]
\describeterm{\idterm}{A}{A}
\end{array}
\qquad\qquad
\begin{array}{c}
\text{Projections (atomic term \ref{atomic:projections})} \\[2pt]
\describeterm{\fstterm} {A \times B}{A}
\qquad 
\describeterm{\sndterm} {A \times B}{B}
\end{array}
\\[10pt]
\text{Coprojections and the codiagonal map (atomic terms \ref{atomic:coprojections} and \ref{atomic:co-diagonal})} \\[2pt]
\describeterm{\leftterm} {A}{A + B} 
\qquad 
\describeterm{\rightterm} {B}{A + B} 
\qquad 
\describeterm{\codiagterm} {A + A}{A}\\[10pt]
\text{Monoidal structure of products (atomic term \ref{atomic:monoidal})} \\[2pt]
\describeterm{\appendoneterm} {A}{A \times \onetype } \qquad 
\describeterm{\prependoneterm} {A}{\onetype  \times A}
\qquad
 \begin{tikzcd}[ampersand replacement=\&, column sep=large]
A \times (B \times C)
\ar[r,"\assoclterm", shift left=0.6ex]
\&
\ar[l,"\assocrterm", shift left=0.6ex]
(A \times B) \times C
\end{tikzcd}
\\[10pt]
\text{Distributivity (atomic terms \ref{atomic:distributivity})} \\[2pt]
\begin{array}{c}
\describeterm{\distrterm} {A \times (B + C)}{(A \times B) + (A \times C)} \qquad 
\describeterm{\distlterm} {(A + B) \times C}{(A \times C) + (B \times C)}  
\end{array}
\\[10pt]
\text{List constructors and destructors (atomic terms \ref{atomic:cons-snoc})} \\[2pt]
    \begin{tikzcd}[ampersand replacement=\&, column sep=large]
\onetype  + A \times A^* 
\ar[r,"\consterm", shift left=0.6ex]
\&
\ar[l,"\unconsterm", shift left=0.6ex]
A^*
\ar[r,"\unsnocterm", shift left=0.6ex]
\&
\ar[l,"\snocterm", shift left=0.6ex]
\onetype  + A^* \times A
\end{tikzcd}
\\[10pt]
\text{List operations (atomic terms  \ref{atomic:concat}, \ref{atomic:split} and \ref{atomic:shift})} \\[2pt]
 \describeterm{\concatterm} {A^{**}}{A^*} 
\qquad
\describeterm{\blockterm} {(A+B)^*}{A^* \times (B \times A^*)^*}  
\qquad 
    \begin{tikzcd}[ampersand replacement=\&, column sep=large]
(A \times B)^* \times A
\ar[r,"\shiftrterm", shift left=0.6ex]
\&
\ar[l,"\shiftlterm", shift left=0.6ex]
A \times (B \times A)^*.
    \end{tikzcd}
\\[10pt]
\text{Finite groups (atomic term \ref{atomic:group-cut})} \\[2pt]
    (A \times G)^* \xrightarrow{\prefterm} (A \times G)^*
\\[6pt] \hline \\[-6pt]
\text{Combinators (composition and functoriality of $+$, $\times$ and $*$)} \\[4pt]
\combinator{A \xrightarrow f B \quad B \xrightarrow g C}{A \xrightarrow {f \, \comp \, g} C}{}
\qquad
\combinator{A_1 \xrightarrow {f_1} B_1 \quad A_2 \xrightarrow {f_2} B_2}{A_1 + A_2 \xrightarrow {f_1 + f_2} B_1 + B_2}{}
\qquad
\combinator{A_1 \xrightarrow {f_1} B_1 \quad A_2 \xrightarrow {f_2} B_2}{A_1 \times A_2 \xrightarrow {f_1 \times f_2} B_1 \times B_2}{}
\qquad
\combinator{A \xrightarrow f B}{A^* \xrightarrow {f^*} B^*}{}
\end{gather*}
\Description[The atomic terms and combinators of the syntax]{A list of the atomic terms, each shown as an arrow from its domain type to its codomain type: identity; coprojections and the codiagonal map; projections; the monoidal structure of products; distributivity; list constructors and destructors; list operations (concatenation, block and shift); and prefix multiplication for finite groups. Below a horizontal line, four inference rules give the combinators: composition, and functoriality of coproducts, products and lists.}
\caption{\label{fig:atomic-rational-terms} The atomic terms and combinators. Observe that the family of atomic terms is infinite, since each term is parameterised by types, and there are infinitely many types. Therefore, in order to be completely formal, we should annotate each term with its types, writing e.g.~$\leftterm_{A,B}$ instead of $\leftterm$. However, we will omit these annotations, since they can be easily inferred from the context.}
\end{figure}

\subsection{Expressive Completeness}
\label{sec:expressive-completeness}
We have now described the whole compositional syntax we will use to denote \kl{rational functions}. The complete list of \kl{atomic terms} and combinators is shown in~\cref{fig:atomic-rational-terms}.

\begin{definition}[Rational terms]
    \label{def:rational-terms}
    A \emph{\intro{rational term}} is any expression built from the \kl{atomic terms}~\ref{atomic:identity}--\ref{atomic:group-cut} by applying the combinators: composition, and functoriality of coproducts, products and lists.
\end{definition}

\AP{}A rational term that does not use the \kl{prefix multiplication term} $\prefterm$ is called an \emph{\intro{aperiodic term}}. Each rational term $f$ represents a rational function $\sem{f}$ of the same \kl{type}, because, as shown earlier in this section, every atomic term is rational and every combinator preserves rationality. 

One can show that all rational functions arise as \kl{denotations} of \kl{rational terms}. This was first proved implicitly in~\cite[Theorems 3.6 and 6.1]{BojanczykDaviaudKrishna2018}. The difference with our system is that the result from~\cite{BojanczykDaviaudKrishna2018} is for a  more powerful system that has nonrational features. However, its part that deals with rational functions does not use any of the nonrational features, although it is not explicitly stated. A formal proof dedicated to the rational fragment can be found in~\cite[Theorem C.5.18]{Bojanczyk2026}.

\begin{theorem}[{Expressive completeness, \cite[Theorem C.5.18]{Bojanczyk2026}}]\label{thm:atomics-and-combinators-completeness}
    Let $A$ and $B$ be  \kl{types}. A function of type $A \to B$ is rational if and only  if it is the \kl{denotation} of some \kl{rational term}.
\end{theorem}

\section{The Equational Axiomatisation}
\label{sec:equational-theory}
\setcounter{equation}{0}

We are now ready to describe the main subject of this paper, which is  the equational axiomatisation of the \kl{rational terms}. It consists of axioms such as 
\begin{align*}
(f\comp{}g)\comp{}h \equiv f\comp(g\comp{}h).
\end{align*}
These axioms are required to be sound, which means that both sides have the same \kl{denotation} for every choice of terms for the variables (in the above axiom, the variables are $f$, $g$ and $h$). 
We also want the axioms to be complete, which means that if two terms have the same denotation, then their equality can be proved using the axioms. When we say that an equality $f \equiv g$ can be proved by some set of axioms, we understand $\equiv$ to be the least equivalence relation that contains the axioms and which is stable under applying the four combinators in our system: composition and functoriality of coproducts,  products, and lists.

To avoid simply adding an axiom $f \equiv g$ for all pairs of terms with the same denotation\footnote{Such a trivial axiomatisation would still be decidable, since as discussed in \cref{sec:intro}, equivalence is decidable for \kl{rational functions}.}, we want to find an axiomatisation which is finitary, as stated in the following main result of this paper.

\begin{theorem}[Finitary axiomatisation]\label{thm:main}
    Two \kl{rational terms} $f$ and $g$ have the same \kl{denotation} $\sem f = \sem g$ if and only if $f \equiv g$ can be proved using Axioms~\ref{ax:associativity-of-composition}--\ref{ax:last} described in Sections~\ref{sec:category-axioms}--\ref{sec:list-axioms}.
\end{theorem}

The statement of the theorem might be a bit misleading, since it suggests that there are 16 axioms. In fact, the referenced axioms are groups of axioms, e.g.~the functoriality axiom (\cref{ax:functoriality-of-type-constructors}) is a group of six axioms which say that the \kl{type} constructors for products, coproducts and lists are all functors. Furthermore, as remarked in \cref{sec:intro}, the axioms are parameterised by types, which means that most axioms represent infinite families of axioms, one for each choice of types. Nevertheless, we think that it is fair to treat each parameterised family of axioms as a single axiom. %

The rest of this paper is devoted to proving \cref{thm:main}. We begin by describing the axioms, and proving that they are sound. For reference, all axioms are also collected as commutative diagrams in \appref{sec:axioms-in-diagrams}, together with explicit definitions of all derived terms that are used in them. We have not made any serious efforts at finding a minimal set of axioms, and hence some of them might be derivable from others.

\subsection{Category Axioms}
\label{sec:category-axioms}

The first group of axioms asserts that \kl{rational terms} form a category with \kl{types} as objects and rational terms quotiented by $\equiv$ as morphisms. We call this category the \intro{category of rational terms}; note that \cref{thm:atomics-and-combinators-completeness,thm:main} imply that the \kl{denotation} map $f \mapsto \sem f$ yields an isomorphism between this category and the \kl{category of rational functions}.
\begin{axiom}
    \label{ax:associativity-of-composition}
    \begin{casesright}
            (f\comp{}g)\comp{}h \equiv f\comp(g\comp{}h) \\
    \idterm\comp{}f \equiv f \equiv f\comp\idterm
    \end{casesright}
\end{axiom}
By associativity of composition, we will often write $f\comp{}g\comp{}h$ instead of $(f\comp{}g)\comp{}h$ or $f\comp(g\comp{}h)$. 

\subsection{Functoriality Axioms} 

The next group of axioms says that the \kl{type} constructors $+$, $\times$, and $*$ are functors. In other words, they preserve identities and composition:
\begin{axiom}
    \label{ax:functoriality-of-type-constructors}
   \begin{casesright}
    \begin{aligned}
        \idterm \times \idterm &\equiv \idterm \qquad (f_1 \times f_2) \comp (g_1 \times g_2) \equiv (f_1 \comp g_1) \times (f_2 \comp g_2) \\[2pt]
        \idterm + \idterm &\equiv \idterm \qquad (f_1 + f_2) \comp (g_1 + g_2) \equiv (f_1 \comp g_1) + (f_2 \comp g_2) \\[2pt]
        \idterm^* &\equiv \idterm \qquad f^* \comp g^* \equiv (f \comp g)^*
    \end{aligned}
  \end{casesright}
\end{axiom}
Observe that when we write equations, we are a bit sloppy with the types. For example, in the equation $\idterm^* \equiv \idterm$ of \cref{ax:functoriality-of-type-constructors}, $\idterm$ on the left has type $A \to A$, while on the right it has type $A^* \to A^*$. We will continue to use this abbreviated notation, as long as the types can be easily inferred from the context.

\subsection{Naturality Axioms}
\label{sec:naturality-axioms}
The next axiom group says that all atomic terms are natural transformations. Let us explain this notion of naturality for the reader unfamiliar with the concept. Each atomic term is parameterised by some choice of types. Naturality says that the operation commutes with the application of morphisms, as in the following example, which says that $\concatterm$ is natural:
\[
\begin{tikzcd}
A^{**} \arrow[r, "f^{**}"] \arrow[d, "\concatterm"'] & B^{**} \arrow[d, "\concatterm"] \\
A^* \arrow[r, "f^*"'] & B^*
\end{tikzcd}
\qquad \text{commutes for every rational term } A \xrightarrow{f} B.
\]
In the diagram, the input functor (which is $(-)^{**}$ in this case) is applied to the term $f$ at the top, and the output functor (which is $(-)^*$ in this case) is applied to the term at the bottom. 
For some \kl{atomic terms}, such as $\assoclterm$, there is more than one \kl{type} parameter. In this case, the naturality axiom involves multiple terms, one for each type parameter, as follows:
\[
\begin{tikzcd}[column sep=huge]
    A_1 \times (A_2 \times A_3) \arrow[r, "f_1 \times (f_2 \times f_3)"] \arrow[d, "\assoclterm"'] & B_1 \times (B_2 \times B_3) \arrow[d, "\assoclterm"] \\
    (A_1 \times A_2) \times A_3 \arrow[r, "(f_1 \times f_2) \times f_3"'] & (B_1 \times B_2) \times B_3 
\end{tikzcd}
\qquad \text{commutes for every rational } A_i \xrightarrow{f_i} B_i.
\]
As it turns out, all atomic terms in our system are natural transformations\footnote{\label{fn:naturality-of-group-prefix-operation}A special case of naturality arises in the case of the \kl{group prefix multiplication}, which uses finite groups in its parameters. This operation is also natural, with a suitable choice of category; however the corresponding naturality property need not be postulated as an axiom, since it can be derived from the other axioms, as we  show in \appref{lem:naturality-of-group-prefix-operation}. Therefore, \cref{ax:naturality} can be used either with or without the group prefix multiplication.
}, and this gives rise to the following axiom, which is part of our finitary axiomatisation:
\begin{axiom}
    \label{ax:naturality}
    \textit{each atomic term is a natural transformation.}
\end{axiom}

\subsection{Isomorphism Axioms}\label{sec:theory:isomorphism-axioms} Another group of axioms states that some of the \kl{atomic terms} are \kl{isomorphisms}, i.e.~they have two-sided inverses. In fact, most terms in our system are isomorphisms, with the only nonisomorphic terms being: the projections and coprojections, the codiagonal, and list concatenations. For the atomic terms that are isomorphisms, in most cases the inverses are also atomic terms, with the exceptions being: distributivity, blocking and the \kl{group prefix multiplication}. Let us explain the inverses for these exceptions. For distributivity on the right, the inverse is defined as follows:
\begin{align*}
(A \times B) + (A \times C) \xrightarrow \undistrterm A \times (B + C)
\quad  \eqdef \quad  \either{(\idterm \times \leftterm)}{(\idterm \times \rightterm)},
\end{align*}
and a similar term $\undistlterm$ is used for distributivity on the left. The inverse of $\blockterm$ is the following term:
\begin{align*}
    A^* \times (B \times A^*)^* \xrightarrowtail{\; \unblockterm \,} (A + B)^*, 
\end{align*}
whose \kl{denotation} injects $A$ and $B$ into the coproduct $A+B$ and then concatenates all resulting lists. The simplest way to define this operation is through the use of \emph{concatenation maps}, which are denoted by tail arrows, and which will also be used to state \cref{ax:shift-concatenation,ax:window} below. Intuitively, if $F(A_1,\ldots,A_n)$ is a \kl{type} built from $A_1,\ldots,A_n$ using the type constructors, then the \kl{concatenation map} $F(A_1,\ldots,A_n) \concmap (A_1 + \cdots + A_n)^*$ collects all elements of the types $A_i$ in its input into a single list, from left to right. The formal definition, by induction on $F$, is given in \cref{sec:concatenation-maps}.

The most interesting case is the inverse of the group prefix multiplication. Rather than deriving this inverse in our system (which we do in \appref{def:unprefterm}), it will be more readable to write the explicit formula:
\begin{align*}
[(a_1,g_1),\ldots,(a_n,g_n)] \in (A \times G)^* \xmapsto{\sem{\unprefterm}}  [(a_1, g_1), (a_2,g_1^{-1} g_2), \ldots, (a_n,g_{n-1}^{-1} g_n)] \in (A \times G)^*.
\end{align*}
This function is rational:  the $i$-th group element in the  output depends on the group elements attached to the $i$-th and $(i-1)$-th input letters. Since the group is finite, the function  can be computed by a transducer which remembers the previous group element in its state. Therefore, by the completeness result from Theorem~\ref{thm:atomics-and-combinators-completeness}, the function $\sem{\unprefterm}$ can be expressed in our system, even without using the group prefix multiplication. 

The isomorphism axiom states that ten pairs of terms are mutual inverses, as follows:
\begin{axiom}
    \label{ax:isomorphisms}
    \begin{casesright}
    (\appendoneterm,\fstterm),
    (\prependoneterm,\sndterm),
    (\assoclterm,\assocrterm),\\
    (\distlterm,\undistlterm),
    (\distrterm,\undistrterm),
    (\consterm,\unconsterm),\\
    (\snocterm,\unsnocterm),
    (\blockterm,\unblockterm), \\
    (\prefterm,\unprefterm), 
    (\shiftlterm,\shiftrterm)
    \end{casesright}
    \text{are mutual inverses}
\end{axiom}

\subsection{Listless Axioms}
\label{sec:listless-axioms}
Let us now present the remaining axioms that do not involve the list \kl{type} constructor. First, since the unit type has only one element, all terms with codomain $\onetype$ are equivalent: 
\begin{axiom}
    \label{ax:unit-type}
    f \equiv g \qquad \text{for every } f,g: A \to \onetype.
\end{axiom}
In other words, the type $\onetype$ is terminal in the \kl{category of rational terms}.

Next, we have an axiom which describes the behaviour of the codiagonal. Its equations correspond to the $\beta$ and $\eta$ rules in \cite[Section 4]{10.1007/BFb0014052}; together, they ensure that \kl{copairing} behaves as expected.
\begin{axiom}
        \label{ax:eval-codiag}
        \begin{casesright}
                 \leftterm\comp \codiagterm \equiv \idterm \equiv \rightterm\comp \codiagterm\\
                 (\leftterm + \rightterm)\comp \codiagterm \equiv \idterm
        \end{casesright}
\end{axiom}

Apart from that, we need axioms about associativity of $\times$. Naturality and the isomorphism axiom say that $\assoclterm$ and $\assocrterm$ are natural and mutually inverse, but this does not yet ensure that all ways of reassociating a product of several factors agree, which is the classical coherence problem for monoidal categories~\cite{MacLane1963}. For $+$, coherence is automatic, since $+$ is a proper coproduct: any two reassociations and reorderings of a coproduct between the same types are provably equivalent, as shown using copairing in~\appref{sec:distributivity}. However, $\times$ is not a categorical product in our system -- there is no pairing combinator, only projections -- and so this coherence does not come for free and must be postulated. We therefore add Mac Lane's pentagon axiom~\cite[VII.1(5)]{MacLane1971}, which says that the two ways of using $\assoclterm$ to reassociate a four-fold product agree:
\begin{axiom}\label{ax:pentagon}
    \myunderbrace{(\idterm \times \assoclterm)\comp \assoclterm\comp (\assoclterm \times \idterm) \equiv \assoclterm\comp \assoclterm}{both sides have type $A \times (B \times (C \times D)) \to ((A \times B) \times C) \times D$.}
\end{axiom}
Together with the pentagon, to make sure that product associativity interacts properly with projections, we also add Mac Lane's standard triangle axiom \cite[VII.1(7)]{MacLane1971}:
\begin{axiom}\label{ax:triangle}
    \begin{tikzcd}[ampersand replacement=\&, column sep=small]
        A \times \left( \onetype \times C \right)
        \ar[rr, "\assoclterm"]
        \ar[dr, "{\idterm \times \sndterm}"']
        \& \& \left(A \times \onetype \right) \times C
        \ar[dl, "{\fstterm \times \idterm}"]
        \\
        \& A \times C
    \end{tikzcd}
\end{axiom}
Axioms \ref{ax:pentagon} and \ref{ax:triangle}, together with naturality and isomorphism axioms for $\assoclterm, \assocrterm$, give exactly the data of a (not necessarily symmetric) monoidal category in the sense of Mac Lane~\cite[VII.1]{MacLane1971}. By a construction in the spirit of the redundancy result of Kelly~\cite{Kelly1964}, we can derive the ``unit on the left or right'' versions of the triangle. Moreover, adding naturality of projections, in \appref{sec:projections-coherence-laws} we are able to strengthen them into what we call \intro{projections coherence laws}, where the discarded unit is replaced with an arbitrary type.

Combined with the isomorphism axioms for distributivity, our listless fragment becomes an instance of what is called a \emph{bimonoidal category}~\cite{Laplaza1972a,Laplaza1972b,Kelly1974} (the case of a rig category in which the multiplicative structure $\times$ is not required to be symmetric), except that in our setting $+$ has no zero object and $\times$ additionally has a terminal (rather than merely two-sided unital) unit, giving the projections $\fstterm,\sndterm$. See also~\cite{Labella2003} for a treatment of categories with sums and a (possibly nonsymmetric) right-distributive tensor product, including normal forms for morphisms, which is close in spirit to parts of our reasoning in \appref{sec:appendix-reduction-to-simple}.

\subsection{List Axioms}
\label{sec:list-axioms}
The last group of axioms is about the list \kl{type} constructor, which is of course the main type constructor in our system. The list type $A^*$ is endowed with two kinds of structure: it is a monoid, and it is a monad. We now describe a series of axioms which capture this structure. In order to state these axioms in a more readable fashion, we introduce the following notation (this notation is best seen using string diagrams, which can be found in \appref{sec:string-diagrams}):
\begin{eqnarray*}
        A \times A^* \xrightarrow{\prependterm} A^* & \eqdef & \rightterm\comp \consterm. \\
        \onetype \xrightarrow {[]} A^* &  \eqdef &  \leftterm\comp \consterm. \\ 
        A \xrightarrow{\pureterm} A^* & \eqdef & \appendoneterm\comp \idterm \times []\comp \prependterm. \\
    A^* \times A^* \xrightarrow{\binconcatterm} A^* & \eqdef &  (\idterm \times \pureterm)\comp \prependterm\comp \concatterm.
\end{eqnarray*}
In the above notation, we have chosen  $\consterm$ as the point of reference. One could consider symmetrically defined notation, based on $\snocterm$ and the term 
\begin{eqnarray*}
        A^* \times A \xrightarrow{\appendterm} A^* & \eqdef & \rightterm\comp \snocterm.
\end{eqnarray*}
We want the symmetrically defined notation to be equivalent, which is stated as an axiom.
\begin{axiom}
    \label{ax:cons-snoc-symmetry}
    \begin{casesright}
        \leftterm\comp \snocterm \equiv \leftterm\comp \consterm \\ 
        \prependoneterm\comp [] \times \idterm\comp \appendterm \equiv \appendoneterm\comp \idterm \times []\comp \prependterm \\
        (\pureterm \times \idterm)\comp \appendterm\comp \concatterm \equiv (\idterm \times \pureterm)\comp \prependterm\comp \concatterm
    \end{casesright}
\end{axiom}
Equipped with the above notation, we state an axiom which says that $A^*$ is a monoid: 
\begin{axiom}
    \label{ax:concatenation-monoid}
   \begin{casesright}
   (\binconcatterm \times \idterm)\comp \binconcatterm \equiv \assocrterm\comp (\idterm \times \binconcatterm)\comp \binconcatterm \\
\appendoneterm\comp (\idterm \times [])\comp \binconcatterm \equiv \idterm \equiv \prependoneterm\comp ([] \times \idterm)\comp \binconcatterm
   \end{casesright} 
\end{axiom}
The following axiom says that $\concatterm$ is a homomorphism between the monoids $A^{**}$ and $A^*$:
\begin{axiom}
    \label{ax:concat-monoid-homomorphism}
    \begin{casesright}
    []\comp \concatterm \equiv []\\
    \binconcatterm\comp \concatterm   \equiv  (\concatterm \times \concatterm)\comp \binconcatterm
    \end{casesright}
\end{axiom}
The list type constructor is also a monad, which is expressed by the following axioms.
\begin{axiom}
    \label{ax:concatenation-unit}
    \begin{casesright}
            \concatterm^*\comp \concatterm  \equiv \concatterm\comp\concatterm \\
            {\pureterm}\comp \concatterm \equiv \idterm \equiv \pureterm^*\comp \concatterm
    \end{casesright}
\end{axiom}
Our axiomatic system prioritises the monoid and monad structure of $A^*$ over its inductive structure. Hence, we have an axiom that defines the latter structure in terms of the former, as follows:
\begin{axiom}
    \label{ax:from-monoid-to-inductive}
    \begin{casesright}
            \prependterm \equiv (\pureterm \times \idterm)\comp \binconcatterm \\
            \appendterm \equiv (\idterm \times \pureterm)\comp \binconcatterm
    \end{casesright}
\end{axiom}
The following axiom says that if we take a list of empty lists and concatenate it, then we could have just as well ignored the input and returned the empty list:
\begin{axiom}
    \label{ax:concatenation-of-empty-lists}
   []^*\comp\concatterm  \equiv 1\comp[] \qquad \text{where $1$ denotes the unique term $\onetype^* \to \onetype$.}
\end{axiom}

\newcommand{\mybin}{\left(\onetype + \onetype\right)}

The last two axioms are about $\shiftrterm$ and the \intro{bundle term} $\bundleterm : A^* \to (A \times A)^* \times (\onetype + A)$, which groups the elements of a list into consecutive pairs, followed by the last element if the length of the list is odd. As for $\unprefterm$, rather than giving this term in our syntax (which we do in \appref{def:bundle-term}, using prefix multiplication for the cyclic group of order two), we write the explicit formula for its \kl{denotation}:
\begin{align*}
    [a_1,\ldots,a_n] \xmapsto{\sem{\bundleterm}}
    \begin{cases}
        \big([(a_1,a_2),(a_3,a_4),\ldots,(a_{n-1},a_n)],\ \leftterm\ 1\big) & \text{if $n$ is even,}\\
        \big([(a_1,a_2),(a_3,a_4),\ldots,(a_{n-2},a_{n-1})],\ \rightterm\ a_n\big) & \text{if $n$ is odd.}
    \end{cases}
\end{align*}
The axioms state that $\shiftrterm$ and $\bundleterm$ do not change the contents of the underlying list, which is expressed by the following commutative diagrams, in which the tail arrows are \kl{concatenation maps} (see \cref{sec:theory:isomorphism-axioms}) that collect all the elements of the input into a single list. We only need an axiom for $\bundleterm$ indexed by the type $2^* = \twotype^*$, since the other instances can be reduced to it (see \appref{sec:bundle-surjectivity}).

\noindent\begin{minipage}{.55\linewidth}
    \begin{axiom}
        \label{ax:shift-concatenation}
        \begin{tikzcd}[
                ampersand replacement=\&,
                column sep=small
            ]
            (A \times B)^* \times A
            \ar[dr, \concar]
            \ar[rr,"\shiftrterm"]
            \& \&
            A \times (B \times A)^*
            \ar[dl, \concar]
            \\
            \&
            (A+B)^*
        \end{tikzcd}
    \end{axiom}
\end{minipage}%
\begin{minipage}{.44\linewidth}
    \begin{axiom}
        \label{ax:window}
        \aliaslabel{ax:last}{ax:window}
        \begin{tikzcd}[
                ampersand replacement=\&,
                column sep=large,
                row sep=large
            ]
            2^{**}
            \ar[dr, "\idterm"']
            \ar[r,"\bundleterm"]
            \&
            (2^* \times 2^*)^* \times (\onetype + 2^*)
            \ar[d, \concar]
            \\
            \&
            2^{**}
        \end{tikzcd}
    \end{axiom}
\end{minipage}

\section{Preliminaries for the Completeness Proof}
\label{sec:proof-overview}
The soundness of our axioms should be clear, and hence the rest of this paper is devoted to proving completeness: if two \kl{rational terms} $f, g$ represent the same function, then  $f \equiv g$ follows from the axioms. The heart of the completeness proof is in the next two sections,  where we follow an approach similar to \cref{sec:rational-functions}, which is to prove completeness for \kl{simple terms} (\cref{sec:simple-rational-equationally}) and then use it to prove the general case (\cref{sec:completeness-for-all-rational-terms}). However, this time, all of our arguments must be formalised in the equational axiomatisation, which  will make the proof considerably more involved. In this section, we present some preliminary notions that will be used in the completeness proof.

\subsection{Concatenation Maps}
\label{sec:concatenation-maps}
Many terms that we consider do nothing but concatenate the input elements into a single list, without deleting or adding any elements. Examples include:
\begin{align*}
\myunderbrace{\consterm, \snocterm, \concatterm}{atomic terms}
\qquad 
\myunderbrace{\prependterm, \appendterm, [], \pureterm, \binconcatterm}{derived in \cref{sec:list-axioms} }.
\end{align*}
To simplify notation, instead of having a separate name for every such operation, we will think of them as instances of a single generic operation called \intro{concatenation map}. We have already used concatenation maps informally in \cref{sec:equational-theory}; here is the formal definition. Consider a \kl{type} $F(A)$ constructed from a type $A$ using type constructors, for example:
\begin{align*}
F(A) = ((A+A)^* \times (A+ \onetype))^*.
\end{align*}
The \emph{concatenation map of type $F(A) \to A^*$}, denoted using the tail arrow notation $F(A) \concmap A^*$, is defined by induction on the structure of $F$ as follows:
\begin{align*}
    A \concmap A^*              & \quad \eqdef \quad  \pureterm                                         \\
    F_1(A) + F_2(A) \concmap A^*      & \quad \eqdef \quad  \either {(F_1(A) \concmap A^*)} {(F_2(A) \concmap A^*)} \\
    F_1(A) \times F_2(A) \concmap A^* & \quad \eqdef \quad {(F_1(A) \concmap A^*)} \fdot {(F_2(A) \concmap A^*)}    \\
    F(A)^* \concmap A^*            & \quad \eqdef \quad \hom{\left( F(A) \concmap A^* \right)}                               \\
    \onetype \concmap A^*       & \quad \eqdef \quad [],
\end{align*}
where $f \odot g$ is syntactic sugar for $(f \times g)\comp \binconcatterm$ and $\hom f$ is syntactic sugar for $f^*\comp\concatterm$, as in the definition of \kl{simple terms} (\cref{def:simple-rational-terms}). From now on, we will prefer the notation of concatenation maps over the \kl{atomic terms}. For example, we will write
\begin{align*}
\onetype + A^* \times A \concmap A^* 
\qquad \text{instead of} \qquad
\onetype + A^* \times A \xrightarrow{\snocterm} A^*.
\end{align*}
The input type of a concatenation map can be seen as a functor, which inputs some type parameter $A$ and outputs a type $F(A)$. Under this perspective, for each choice of $F$, the concatenation map is a natural transformation, which is expressed in the following diagram:
\[
    \begin{tikzcd}
        F(A)
        \ar[r,"F(f)"]
        \ar[d,\concar]
        &
        F(B)
        \ar[d,\concar]
        \\
        A^*
        \ar[r,"{f^*}"']
        &
        B^*
    \end{tikzcd}
    \qquad \text{for every term $f: A \to B$.}
\]

The notion of concatenation maps can be generalised to multiple parameters. An example would be the inverse of $\blockterm$ mentioned in \cref{sec:theory:isomorphism-axioms}, which is of type $A^* \times (B \times A^*)^* \to
(A+B)^*$. We call such terms \intro{multivariate concatenation maps}.
For a formal definition, consider a functor $F(A_1, \dots, A_n)$ that inputs $n$ type parameters and outputs a type, and which is built using the type constructors. Such a functor is called a \intro{multivariate list-polynomial functor}. For such a functor, we define the \reintro{multivariate concatenation map}:
\begin{align*}
F(A_1,\ldots,A_n) \concmap (A_1 + \cdots + A_n)^*
\end{align*}
in the same way as for the single-parameter case, except that in the base case we use
\begin{align*}
A_i \xrightarrow{\text{coprojection}} (A_1 + \cdots + A_n) \xrightarrow \pureterm (A_1 + \cdots + A_n)^*.
\end{align*}
Equivalently, the multivariate concatenation map is the composition of two steps: first, the coprojections inject each type $A_i$ into $A_1 + \cdots + A_n$, and then the single-parameter concatenation map is applied. Multivariate concatenation maps are a crucial tool in our proofs. In particular, in \appref{sec:rebracketings} we use them to formally define rebracketings, i.e.~terms that do nothing except for rebracketing the input, or in other words, terms that commute with concatenation maps. An example is $\shiftrterm$, which is a rebracketing by \cref{ax:shift-concatenation}.

\subsection{Values}
\begin{definition}
    [\intro{Value}] \label{def:value}
    A \kl{value} is a term of type $\onetype  \to A$.
\end{definition}

Values represent elements of \kl{types}. We begin the completeness proof with values, which is also an opportunity to introduce notation used throughout the rest of the proof.

We use the letters $u, v, w, \ldots$ to denote values. To combine two values of types $\onetype \to A_1$ and $\onetype \to A_2$ into a value of type $\onetype \to A_1 \times A_2$, the natural idea would be to use pairing of terms. Although general pairing is not allowed in \kl{rational terms} (see \cref{sec:atomic-for-products}), in this case, it is available thanks to $\unitlterm : \onetype \to \onetype \times \onetype$. Therefore, we can pair two values into a value of product type. In fact, we can do a bit more: we can pair a value with a term that is not necessarily a value, as follows: if $f$ is a term of type $A \to B$  and  $v$ is a value of type $\onetype  \to C$, then we define
\begin{align*}
A \xrightarrow{(v,f)} C \times B \eqdef \prependoneterm\comp (v \times f)  
\qquad 
A \xrightarrow {(f,v)} B \times C \eqdef \appendoneterm\comp (f \times v).
\end{align*}
On the face of it, this notation is ambiguous, since pairing two values $(v,w)$ could be matched to the above definition in two ways to create a new value. However,
\begin{align*}
\prependoneterm\comp(v \times w) \equalbecause{\cref{ax:unit-type} says that $\appendoneterm\comp\sndterm$ is the identity on $\onetype$}\\
\appendoneterm\comp \sndterm\comp \prependoneterm\comp(v \times w) \equalbecause{\cref{ax:isomorphisms} says that $\sndterm\comp \prependoneterm$ is the identity on $\onetype \times \onetype$}\\
\appendoneterm\comp(v \times w),
\end{align*}
so we can pair values $(v,w)$ without any risk of ambiguity.

We now introduce some notation for concatenation with values of list types. To construct singleton lists, we use the combinator  defined by $[v] = v\comp\pureterm$. To concatenate two list values -- in fact more generally, a list value $v$ together with an arbitrary term $f$ that has  list outputs -- we use the following notation: $v \wdot f$ is defined to be the composition of the pair $(v,f)$ with the \kl{concatenation map}, and $f \wdot w$ is defined similarly.
By  the above,   there is no ambiguity when concatenating two list values $v \wdot w$, and the result corresponds to string concatenation.
We now state the relevant monoidality laws, which are the same as in \cref{thm:conjugacy-completeness}. The proof of the following lemma is a straightforward application of our axioms, and is left to the reader.

\begin{lemma}\label{lem:concatenation-monoidality}
    For all \kl{rational terms} $A \xrightarrow{f} B^*$ and $A' \xrightarrow{g} B^*$, and \kl{values} $\onetype \xrightarrow{w,v} B^*$, we have
    \begin{gather*}
        f \wdot [] \equiv f \equiv [] \wdot f \quad \;\;
        (f \wdot w) \wdot v \equiv f \wdot (w \cdot v) \quad \;\;
        w \wdot (v \wdot f) \equiv (w \cdot v) \wdot f \quad \;\;
        (w \wdot f) \wdot v \equiv w \wdot (f \wdot v) \\
        (f \wdot w) \fdot g \equiv f \fdot (w \wdot g) \qquad
        w \wdot (f \fdot g) \equiv (w \wdot f) \fdot g \qquad
        (f \fdot g) \wdot w \equiv f \fdot (g \wdot w).
    \end{gather*}
\end{lemma}

Equipped with the above notation, in \appref{sec:proof-of-normalisation-of-values} we show that every value is provably equivalent to one built using the constructors listed below.
\begin{lemma}[Normalisation of values]\label{lem:value-completeness}
    Every \kl{value} is provably equivalent to one constructed using 
        \begin{align*}
    \myunderbrace{\idterm}{identity on $\onetype $} \qquad \qquad
    \myunderbrace{v\comp\leftterm \quad v\comp\rightterm}{constructors for coproduct values}
    \qquad  \qquad \qquad 
    \myunderbrace{(v,w)}{constructor for product values}
    \qquad \qquad \qquad 
    \myunderbrace{v \wdot w \quad [v] \quad []}{constructors for list values}.
    \end{align*}
\end{lemma}
As a corollary, our equational axiomatisation is complete for values. This is because the only ambiguity in writing a value in normal form is the bracketing of concatenation of values that represent lists, and different bracketings are provably equivalent by \cref{lem:concatenation-monoidality}.

\subsection{Injective, Surjective and Isomorphic Terms}
\label{sec:reduction-to-string-outputs}
We finish this section by introducing some \kl{injective}, \kl{surjective}, and \kl[isomorphism]{isomorphic} terms that will be used throughout the completeness proof.
\begin{definition}[Injective, surjective, and isomorphic terms]\label{def:injective-surjective-terms}
    \AP{}A \kl{rational term} $A \xrightarrow f B$ is called
    \begin{align*}
        \text{\emph{injective}} \label{def:injective} & \quad \text{if } f \comp \inv f \equiv \idterm \text{ for some term } B \xrightarrow{\inv f} A, \tag*{}\\
        \text{\emph{surjective}} \label{def:surjective} & \quad \text{if } \inv f \comp f \equiv \idterm \text{ for some term } B \xrightarrow{\inv f} A, \tag*{}\\
        \text{an \emph{isomorphism}} \label{def:isomorphism} & \quad \text{if } f \comp \inv f \equiv \idterm \text{ and } \inv f \comp f \equiv \idterm \text{ for some term } B \xrightarrow{\inv f} A. \tag*{}
    \end{align*}
\end{definition}
In categorical terms, an injective term is a \emph{section} (i.e.~a split monomorphism) and a surjective term is a \emph{retraction} (i.e.~a split epimorphism) in the \kl{category of rational terms}. However, we prefer the name injective (surjective), since it agrees with the intuition that the function represented by an injective (surjective) term is injective (surjective) in the usual sense.\footnote{As a curiosity, the converse is also true: if $\sem{f}$ is injective (surjective), then $f$ is injective (surjective) in the sense of \cref{def:injective-surjective-terms}. This is a consequence of the fact that the category of rational terms is isomorphic to the \kl{category of rational functions} (a corollary of Theorems~\ref{thm:atomics-and-combinators-completeness} and \ref{thm:main}), in which the notions of injectivity and surjectivity coincide with sections and retractions. For injective functions, this is because the inverse of a \kl{rational relation} is rational, and the image of a \kl{rational function} is a regular set, outside of which the left inverse can output an arbitrary fixed value; for surjective functions, it is a simple consequence of Eilenberg's uniformisation theorem for rational relations~\cite[Proposition IX.8.2]{Eilenberg1974}.} Because of their one-sided inverses, injective and surjective terms can be cancelled out in equivalence proofs, which is an instance of the general fact that sections and retractions are in particular monomorphisms and epimorphisms, respectively. We leave the proof of this fact as an exercise to the reader. 

\begin{lemma}\label{lem:pre-postcompose-with-injective-surjective}
    Let $A \xrightarrow {g_1,g_2} B$ be \kl{rational terms}. Then 
    \begin{align*}
g_1 \equiv g_2 \quad \text{iff} \quad g_1\comp{}f \equiv g_2\comp{}f \qquad &\text{for every \kl{injective} }f : B \to C \\ 
g_1 \equiv g_2 \quad \text{iff} \quad f\comp{}g_1 \equiv f\comp{}g_2 \qquad &\text{for every \kl{surjective}  }f : C \to A.
\end{align*}
\end{lemma}
By a standard argument, a term that is both \kl{injective} and \kl{surjective} is an \kl{isomorphism}, and its inverse is unique up to $\equiv$. As an application of injectivity, we are able to reduce the completeness proof to the special case of terms whose codomains are lists over the \kl{binary alphabet} ${\twotype}$. The reduction uses the following lemma, which is proved in \appref{sec:provably-injective-string-representations}, together with \cref{lem:pre-postcompose-with-injective-surjective} (see \cref{cor:aperiodic-completeness}).

\begin{lemma}\label{lem:provably-injective-string-representation}
    For every \kl{type} $A$ there is an \kl{injective} \kl{aperiodic term} $A \xrightarrow{\stringrep{A}} {\twotype}^*$.
\end{lemma}

\section{Completeness for Simple Rational Terms}
\label{sec:simple-rational-equationally}
In this section, we continue the completeness proof, by showing that our equational axiomatisation is complete for \kl{simple rational terms}. The definition of simple rational terms (\cref{def:simple-rational-terms}), which was presented before we developed our compositional syntax in \cref{sec:syntax}, is meaningful in this syntax. This is thanks to notation developed in the previous sections, which gave meaning to the combinators $\hom f$, $f_1 \fdot f_2$, $w \wdot f$ and $f \wdot w$.

Our proof follows the same lines as the proof of \cref{thm:conjugacy-completeness}, where we showed that any two simple rational terms that define the same function can be proved equivalent using only a handful of simple rules. The difficulty is that two of these rules, case splitting and the conjugacy rule, are implications, so they are not available in our purely equational system. Instead, we need to derive their instances from the equational axioms. This is straightforward for case splitting, whereas deriving the conjugacy rule is the main difficulty of this section.

We begin by introducing some notation. For two \kl{rational terms} $f$ and $g$ that have type $A \to B^*$ and a \kl{value} $w$ of type $\onetype \to B^*$, we  consider three kinds of conjugacy relation:
\begin{enumerate}
    \item $f$ and $g$ are \emph{denotationally $w$-conjugate} if $\sem{f \wdot w} = \sem{w \wdot g}$.
    \item $f$ and $g$ are \emph{provably $w$-conjugate} if $f \wdot w  \equiv    w \wdot g$.
    \item $f$ and $g$ are \emph{syntactically $w$-conjugate} if $f = w \wdot h$ and $g = h \wdot w$ for some term $h$.
\end{enumerate}
These three kinds  are listed in increasing order of strength, with the implication 3 $\Rightarrow 2$ following from \cref{lem:concatenation-monoidality}. Using this terminology, the conjugacy rule from \cref{thm:conjugacy-completeness} says that provable $w$-conjugacy can be lifted via homomorphism:
\begin{align*}
f \wdot w \equiv w \wdot g \quad \text{implies} \quad \hom f \wdot w \equiv w \wdot \hom g.
\end{align*}
If we could prove such an implication in our equational system, then we would be done, which will be demonstrated in the proof of \cref{lem:completeness-for-simple-rational-terms}. Alternatively, we could add the implication as an axiom. However, this would make the system not purely equational, which we want to avoid.

We were unable to derive the conjugacy rule directly, but we are able to do this in mutual recursion with  completeness of the equational axiomatisation for simple rational terms, as stated in the following lemma.

\begin{lemma}[Completeness for simple rational terms]\label{lem:completeness-for-simple-rational-terms}
        For all \kl{simple rational terms} $A \xrightarrow {f,g} B^*$ and \kl{value} $\onetype \xrightarrow w B^*$,
            \begin{align}
            \sem{f} = \sem{g} & \quad \text{implies} \quad f \equiv g
            \label{eq:completeness-for-simple-rational-terms} \\
            f \wdot w \equiv w \wdot g & \quad \text{implies} \quad \hom f \wdot w \equiv w \wdot \hom g.
            \label{eq:conjugacy-rule-for-simple-rational-terms}
        \end{align}
\end{lemma}

\begin{proof}
    We use the name \emph{completeness rule} for the first statement, and \emph{conjugacy rule} for the second statement. The rules are proved in parallel by  induction on the nesting depth of the list constructor in the input \kl{type} $A$.    At each step of the induction, we first prove the completeness rule, and then we prove the conjugacy rule. In particular, in the presence of the completeness rule at this stage of the induction, the conjugacy rule can also be stated in the stronger form, where the assumption is weakened from provable conjugacy to denotational conjugacy.

    For the completeness rule, we use the same argument as in \cref{thm:conjugacy-completeness}. This argument can be applied, since the reasoning rules \ref{item:congruence-rule}--\ref{item:conjugacy-rule} in the statement of \cref{thm:conjugacy-completeness} have already been proved to be admissible in our equational system, as we now explain. The congruence rule~\ref{item:congruence-rule} is one of the rules of equational reasoning. Rule~\ref{item:concatenation-monoidality-rule} for concatenation monoidality is \cref{lem:concatenation-monoidality}. Rules~\ref{item:constants-copairing} and~\ref{item:distributivity-rule} follow from the equivalence:
        \begin{align*}
    (\either {f_1}  {f_2}) \fdot f \equiv \distlterm \comp \either{(f_1 \fdot f)}{(f_2 \fdot f)}
    \end{align*}
    and a similar one for right distributivity $\distrterm$, which are proved in \appref{lem:distributivity-of-concatenation}. Finally, in the proof of \cref{thm:conjugacy-completeness}, the conjugacy rule from item~\ref{item:conjugacy-rule} is only applied for terms that have an input type with strictly smaller nesting of lists than $A$, and hence we use the induction hypothesis of the present lemma. Summing up, we can use \cref{thm:conjugacy-completeness} to prove the induction basis and induction step for the completeness rule.      We are left with doing this for  the conjugacy rule.  We assume that the length of the value $w$, which will be denoted by $k$, is nonzero. In the case of $k=0$, we have $w \equiv []$ by completeness for values (\cref{lem:value-completeness}), and hence the conjugacy rule follows from the unit laws of \cref{lem:concatenation-monoidality}, together with congruence.

    \paragraph*{Induction Basis for the Conjugacy Rule (Illustration).} We only present a special case of the induction basis, where the input type is $A = \onetype$. This special case already illustrates the key ideas, while the general induction basis is proved in the same way as the induction step, which is presented below.

    Already  in this case, the conjugacy rule is nontrivial. To prove this, we will use the following lemma, which shows a variant of the conjugacy rule, where the assumption is strengthened from provable conjugacy to syntactic conjugacy. The proof is deferred to \appref{sec:proof-of-lem:conjugacy-fixed-middle-equational}.
\begin{lemma}\label{lem:conjugacy-fixed-middle-equational}
    Consider a \kl{simple rational term} $A \xrightarrow h  B^*$ and a \kl{value} $\onetype \xrightarrow w B^*$. Then
    \begin{align*}
        \hom(w \wdot h) \wdot w \equiv w \wdot \hom(h \wdot w).
    \end{align*}
\end{lemma}
We can now prove the induction basis of the conjugacy rule. In this case, both $f \wdot w$ and $w \wdot g$ are values. If $f \equiv []$, then also $g \equiv []$ and the rule follows. Otherwise, by Levi's lemma from string combinatorics\footnote{Levi's lemma says that if $u_1 u_2 = v_1 v_2$ for strings $u_1, u_2, v_1, v_2$, then there is a string $t$ such that either $u_1 = v_1 t$ and $v_2 = t u_2$, or $v_1 = u_1 t$ and $u_2 = t v_2$. By induction on the length of $z$, it implies the following fact about conjugate strings, which is the form that we use here: if $u z = z v$ and $u$ is nonempty, then $u = xy$, $v = yx$ and $z = (xy)^\ell x$ for some strings $x, y$ and some $\ell \in \set{0,1,\ldots}$.} and completeness for values, we have
    \begin{align*}
    f \equiv x \wdot y \quad g \equiv y \wdot x \quad w \equiv \myunderbrace{(x \wdot y) \wdot \cdots \wdot (x \wdot y)}{$\ell$ times} \wdot x
    \qquad \text{for some values $\onetype \xrightarrow{x,y} B^*$ and $\ell \in \set{0,1,\ldots}$}.
    \end{align*}
By \cref{lem:conjugacy-fixed-middle-equational}, we deduce that $\hom(x \wdot y) \wdot x \equiv x \wdot \hom(y \wdot x)$. By applying this rule $2\ell+1$ times, we derive $\hom f \wdot w \equiv w \wdot \hom g$, as required in the induction basis.

\paragraph*{Induction Step for the Conjugacy Rule.}  Similarly to the argument above,  the  general idea behind the  induction step is to strengthen the assumption of the rule from provable conjugacy  to  syntactic conjugacy, so that we can apply \cref{lem:conjugacy-fixed-middle-equational}. However, this is not necessarily directly possible, because of short output strings. Indeed, syntactic conjugacy can only hold if all output strings have length at least $k$, and this might not be the case. To overcome this difficulty, we analyse the behaviour of $\sem{f}$ and $\sem{g}$ when the outputs are short. In this analysis, we will  decompose  the input domain  into a finite number of cases, such that in each case, the first and last $k$ letters of the output are syntactically fixed, as in the following definition.

\begin{definition}
    Let $k \in \set{1,2,\ldots}$. 
    A term $A \xrightarrow f B^*$ is called \emph{$k$-guarded} if it is either:
        \begin{enumerate}
        \item \textbf{Constant:} \label{item:provable-constant} a composition of the term $A \to \onetype$ (unique up to equivalence) with some \kl{value}; or
        \item \textbf{Long:} \label{item:guarded} $u \wdot h \wdot v$ for some \kl{values} $u,v$ of length $k$ and a  \kl{simple rational term} $h$.
    \end{enumerate}
\end{definition}

The following lemma shows that, without loss of generality, we can assume that the terms $f$ and $g$ are obtained by \kl{copairing} $k$-guarded terms. The proof is deferred to \appref{sec:proof-of-lem:constants-and-guarded}.

\begin{lemma}\label{lem:constants-and-guarded} Let $k$ be the length of $w$. 
    Without loss of generality we can assume that $f$ and $g$ are 
    \begin{align*}
    \eithermany n {f_1 \cdots f_n} \quad \text{and} \quad 
    \eithermany n {g_1 \cdots g_n} 
    \qquad \text{for some $k$-guarded terms }A_i \xrightarrow{f_i, \, g_i} B^*.
    \end{align*} 
\end{lemma}

From now on, we assume that $A$ and the terms $f$ and $g$ are as in the conclusion of \cref{lem:constants-and-guarded}. In each  case $i \in \set{1,\ldots,n}$, the functions $\sem{f_i}$ and $\sem{g_i}$ must have outputs of the same length on every input. Therefore, if one of the terms $f_i$ and $g_i$ is a constant, then the \kl{denotation} of the other one is a constant function as well, and the lengths of the two constant outputs agree. Empty constants will be inconvenient, so we begin by getting rid of them.
\begin{lemma}\label{lem:nonempty-summands}
    We can assume without loss of generality that all $f_i$ and $g_i$ are nonempty.
\end{lemma}
\begin{proof}
Decompose $A$ as $E+N$ where $E$ is the coproduct of all summands $A_i$ for which both $f_i$ and $g_i$ are empty,  and $N$ is the coproduct of the remaining summands. Let us write $f_E$ and $g_E$ for the corresponding cases of types $E \to B^*$ and $f_N$ and $g_N$ for the corresponding cases of types $N \to B^*$. The following claim, which is proved in \appref{sec:proof-of-claim:ignore-empty-lists} using the axioms about empty lists (Axioms \ref{ax:concatenation-monoid} and \ref{ax:concatenation-of-empty-lists}), shows that $\hom f$ factors through $\hom f_N$.

\begin{claim} 
    \label{claim:ignore-empty-lists}
    The term $\hom f$ is provably equivalent to the following composition:
    \begin{align*}
    (E+N)^*
    \xrightarrow \blockterm
    E^* \times (N \times E^*)^*
    \xrightarrow {\sndterm\comp(\fstterm^*)}
    N^*
    \xrightarrow {\hom f_N}
    B^*
    \end{align*}
\end{claim}
By \cref{claim:ignore-empty-lists}, and an analogous one for $g$,  the conjugacy $\hom f \wdot w \equiv w \wdot \hom g$ reduces to the conjugacy $\hom f_N \wdot w \equiv w \wdot \hom g_N$, in which the nonemptiness assumption is satisfied.
\end{proof}

By \cref{lem:nonempty-summands}, we assume from now on that there are no empty summands. 
Under this assumption, whenever we apply $\sem{f}$ or $\sem{g}$, at least one letter is produced, which is guaranteed syntactically by the form of the summands, and also if the output has unknown length, then it must begin and end with $w$. We will want to strengthen this guarantee so that at least $k$ letters are produced. To achieve this, we will group items into bundles of length $k$, as explained in the following lemma.  (In the lemma, we write $A^i$ for the $i$-fold product of $A$ with itself with $A^0 = \onetype$. Also, we write $A^{< k}$ for the coproduct of all $A^i$ with $0 \leq i <k$.)

\begin{lemma}
    \label{lem:bundles}
    For every \kl{type} $A$, there are arbitrarily large numbers $k$ such that the following \kl{concatenation map} is an \kl{isomorphism}:
    \begin{align*}
        (A^k)^* \times A^{< k} \concmap A^*.
    \end{align*}
\end{lemma}
\begin{proof}[Proof sketch]
    The proof is given in \appref{sec:proof-of-bundles}; here we only give a sketch.  For $A = (\onetype + \onetype)^*$ and $k=2$, we use the term $\bundleterm$ from \cref{ax:window} as the inverse. It uses the cyclic group $\mathbb{Z}_2$ to label each element of the input list by the generator, and then uses the $\prefterm$ \kl{atomic term} to compute the prefixes. As a result, the group elements will alternate in a cyclic pattern, and $\blockterm$ is used to split the list into bundles of length $2$. This is extended to an arbitrary type $A$ using the \kl{binary alphabet} encodings from \cref{lem:provably-injective-string-representation}. By iterating this procedure, we can obtain $k = 2^i$ for every $i \geq 1$. Of course, this argument needs to be formalised in our equational axiomatisation, which requires a substantial amount of work.
\end{proof}
By \kl{surjectivity}, in order to prove the conclusion of the conjugacy rule, i.e.~that
\begin{align*}
\hom f \wdot w \equiv w \wdot \hom g,
\end{align*}
it is enough to prove that this conclusion holds when the two sides are precomposed with the \kl{isomorphism} from \cref{lem:bundles}. By reasoning by cases, it is enough to show that
\begin{align*}
\hom(f^{\fdot k}) \fdot f^{\fdot i} \wdot w \equiv w \wdot \hom(g^{\fdot k}) \fdot g^{\fdot i}
\qquad \text{for every $i \in \set{0,\ldots,k-1}$,}
\end{align*}
where $f^{\fdot i}$ represents the $i$-fold concatenation of $f$ with itself.
By applying the assumption of the conjugacy rule, we know that  $f^{\fdot i}$ and $g^{\fdot i}$ are provably $w$-conjugate, that is, $f^{\fdot i} \wdot w \equiv w \wdot g^{\fdot i}$. We
are left with proving that
\begin{align*}
\hom(f^{\fdot k}) \wdot w \equiv w  \wdot \hom(g^{\fdot k}).
\end{align*}
By having the list elements grouped into bundles of length $k$, we will achieve syntactic conjugacy, as we now explain.
Up to distributivity on the domain, both $f^{\fdot k}$ and $g^{\fdot k}$ are copairings of $k$-guarded terms where all constants have length at least $k$, thanks to the assumption that there are no empty outputs.  Furthermore, by the assumption $f \wdot w \equiv w \wdot g$ of the conjugacy rule, we know that the outputs of each case in the copairing from  $f^{\fdot k}$ must  begin with $w$, and the outputs of each case in $g^{\fdot k}$ must end with $w$. This means that 
\begin{align*}
f^{\fdot k} \equiv w \wdot h_f 
\quad \text{and} \quad 
g^{\fdot k} \equiv h_g \wdot w
\qquad \text{for some simple rational terms $h_f$ and $h_g$.}
\end{align*}
Since these two terms have the same nesting depth of the list constructor as $f$ and $g$, we can use the completeness rule, which has already been proved at this stage of the induction, to conclude that $h_f \equiv h_g$. Therefore, $f^{\fdot k}$ and  $g^{\fdot k}$ must be syntactically $w$-conjugate, and hence we can conclude the proof by \cref{lem:conjugacy-fixed-middle-equational}.
\end{proof}

\section{Completeness for All Rational Terms}
\label{sec:completeness-for-all-rational-terms}
In this section, we finish the proof of \cref{thm:main}, by reducing to the case of \kl{simple rational terms}. To implement this reduction, we introduce two auxiliary classes of terms.

\begin{definition}[Lifts and drops]
    \label{def:lift-drop}
    \intro{Lift} terms (respectively, \intro{drop} terms) are those which can be obtained from the \kl{atomic lift} (respectively, \kl[atomic drop]{drop}) terms
    in \cref{tab:atomic-drops-and-lifts} by applying composition, as well as the functoriality combinators for  coproducts, products, and lists.
\end{definition}

We use arrows $\liftmap$ for lifts and $\dropmap$ for drops. 
Lift terms are meant to denote functions that annotate the input with some structure, while drop terms are meant to denote functions that erase this annotation. Drops are meant to be a mild generalisation of simple rational terms; one can show that simple rational terms are a special case of drops, up to an \kl{isomorphism} on the domain. A key property of  lift terms is that they are isomorphisms, and their inverses are drops. (Drops are not necessarily isomorphisms, and not every drop arises as an inverse of a lift.) 

\begin{table}[t]
    \centering
    \caption{Atomic terms for the \kl{lift} and \kl{drop} classes. Note that we explicitly include atomic lift terms for coproduct commutativity and associativity -- this is because they are derived using coprojections, to which lifts have no access. Apart from that, the notable atomic lift and drop terms are $\shiftterm$ and $\unshiftterm$: they are not atomic rational terms, but are derivable in a natural way. In particular, they are natural transformations by \cref{ax:naturality}. Moreover, they are not exactly each other's inverses, but one can prove that $\shiftterm$ and $\either{[]}{\unshiftterm}$ are (see \cref{lem:lift-provable-inverse}).}
    \label{tab:atomic-drops-and-lifts}
    \renewcommand{\arraystretch}{1.4}
    \begin{tabular}{c | c}
        {\renewcommand\emph[1]{#1}\footnotesize\sffamily\MakeUppercase{\intro{Atomic lift terms}}} & {\renewcommand\emph[1]{#1}\footnotesize\sffamily\MakeUppercase{\intro{Atomic drop terms}}} \\[2pt]
        \hline
        \\[-12pt]
        $A \lift{\idterm} A \qquad A + B \lift{\commterm} B + A$
            & $A \drop{\idterm} A \qquad A \drop{1 \comp v} B$ \\
        $A \lift{\unitrterm} A \times \onetype \qquad A \lift{\unitlterm} \onetype \times A$
            & $A \times B \drop{\fstterm} A \qquad A \times B \drop{\sndterm} B$ \\
        $A \times (B \times C) \lift{\assoclterm} (A \times B) \times C$
            & $A \drop{\leftterm} A + B \qquad B \drop{\rightterm} A + B$ \\
        $(A \times B) \times C \lift{\assocrterm} A \times (B \times C)$
            & $A \times (B \times C) \drop{\assoclterm} (A \times B) \times C$ \\
        $A + (B + C) \lift{\sumassoclterm} (A + B) + C$
            & $(A \times B) \times C \drop{\assocrterm} A \times (B \times C)$ \\
        $(A + B) + C \lift{\sumassocrterm} A + (B + C)$
            & $A + A \drop{\codiagterm} A$ \\
        $(A + B) \times C \lift{\distlterm} (A \times C) + (B \times C)$
            & $A \drop{\pureterm} A^*$ \\
        $A \times (B + C) \lift{\distrterm} (A \times B) + (A \times C)$
            & $A^{**} \drop{\concatterm} A^*$ \\
        $(A + B)^* \lift{\blockterm} A^* \times (B \times A^*)^*$
            & $A^* \times A^* \drop{\binconcatterm} A^*$ \\
        $(A \times B)^* \lift{\shiftterm} \onetype + A \times (B \times A)^* \times B$
            & $A \times (B \times A)^* \times B \drop{\unshiftterm} (A \times B)^*$ \\
    \end{tabular}%
\end{table}

\begin{lemma}\label{lem:lift-provable-inverse}
    Every lift is an \kl{isomorphism}, and its inverse is a \kl{drop}. 
\end{lemma}
\begin{proof}
    Induction on the structure. The induction steps of combinators are straightforward, since drops are closed under the same combinators as lifts, and the same is true for inverses, which is a consequence of the functoriality axiom (\cref{ax:functoriality-of-type-constructors}). We are left with the base cases of the \kl{atomic lift} terms.
    For most of the \kl{atomic lift} terms, the lemma is a direct consequence of the isomorphism axioms. For $\blockterm$, we observe that its inverse $\unblockterm$ (\cref{ax:isomorphisms}) is a \kl{concatenation map}, and all \kl{concatenation maps} are \kl{drops}. For $\shiftterm$, we leave it as an exercise to show that $\either{[]}{\unshiftterm}$ is the inverse.
\end{proof}

As we have remarked above, \kl{drops} are a mild generalisation of \kl{simple rational terms}.  This is formalised in the following lemma, which says that if a \kl{drop} has type $A \to N^*$ for a \kl{finite alphabet} $N$, then it is provably equivalent to a \kl{simple rational term}, up to an isomorphism of its input type. The lemma is proved in \appref{sec:appendix:joint-drops-simplification} using a syntactic analysis of the structure of \kl{drops}.
\begin{lemma}[Provable simplification of drops]\label{lem:joint-drops-normalisation} Let $N$ be a \kl{finite alphabet}, and let  $
    f, g : A \dropmap N^*
$ be drops.
   There is $\pi : A \liftmap C$ and its inverse $\inv \pi$, such  that $\inv \pi\comp{}f$ and $\inv \pi\comp{}g$ are both equivalent to simple rational terms.
\end{lemma}

Since all \kl{lifts} are isomorphisms and our equational axiomatisation is complete for simple rational terms by \cref{lem:completeness-for-simple-rational-terms}, the above lemma shows that it is also complete for drops. We now show how all rational terms can be reduced to drops. This will be proved by factorising rational terms as lifts followed by drops. We begin the reduction with the special case of \kl{aperiodic terms}.

\begin{lemma}[Lift-drop factorisation]\label{lem:lift-then-drop}
	Every aperiodic rational term is provably equivalent to a composition of a lift followed by a drop.
\end{lemma}
\begin{proof}
The  base case is an easy case analysis, which is left to the reader. In the induction step when the rational term is constructed by using one of the functorial combinators, we use the induction hypothesis to factorise the subterms, and then functoriality.

The essential case is the induction step for composition, i.e.~obtaining a lift-drop factorisation for a composition  $f = f_1 \comp f_2$. The induction hypothesis alone is not enough here: applying it gives  a decomposition for $f$ of the form lift-drop-lift-drop. To resolve this issue, we use the following lemma, which will swap the order of the inner drop and lift.
\begin{lemma}[Exchange]\label{lem:exchange} Every drop-lift composition is equivalent to a lift-drop composition.
\end{lemma}
\cref{lem:exchange} is proved in \appref{sec:appendix:exchange-proof}. By applying it to the inner drop-lift in the factorisation of $f$, we get a factorisation of $f$ of the form lift-lift-drop-drop, which can be combined into a single lift and a single drop, using the closure of lifts and drops under composition. 
\end{proof}

Using the lift-drop factorisation, we finish the completeness proof for aperiodic rational terms. 

\begin{corollary}[Aperiodic completeness]\label{cor:aperiodic-completeness}
    If aperiodic rational terms $f$ and $g$ have the same denotation, then $f \equiv g$.
\end{corollary}

\begin{proof}
    By postcomposing with the aperiodic and injective binary encoding term from \cref{lem:provably-injective-string-representation} we can assume without loss of generality that the two terms have type $A \to \twotype^*$.
    We apply \cref{lem:lift-then-drop} to obtain a lift-drop decomposition $f \equiv \gamma \comp f'$. Then, we apply \cref{lem:lift-then-drop} again to decompose $\inv{\gamma} \comp g \equiv \sigma \comp g'$. Since $\inv \sigma$ is a drop, the composition $\inv \sigma \comp f'$ is also a drop. We obtain lift-drop factorisations of $f$ and $g$ with a common lift, as presented in the following diagram:
        \[
    \begin{tikzcd}
        [row sep=small]
        A
        \ar[r, "{f, \, g}"]
        \ar[d, blue, \liftar, "\gamma"']
        \ar[dd, blue, bend right = 90, \liftar, "\text{common lift} \ (\gamma \, \comp \, \sigma)"']
        & \twotype^*
        \\
        \blue{C}
        \ar[ur, blue, Rightarrow, "f'"']
        \ar[d, blue, \liftar, "\sigma"']
        \\
        \blue{D}
        \ar[uur, blue, Rightarrow, bend right, "{(\inv{\sigma} \, \comp \, f'), \, g'}"']
    \end{tikzcd}
    \]
    Consider the two drop terms $g'$ and $\inv{\sigma}\comp f'$ of type $D \to \twotype^*$. These two terms are denotationally equal, and they are drop terms whose codomain is the list type over a \kl{finite alphabet}. Thus, they are provably equivalent by \cref{lem:joint-drops-normalisation}, together with completeness for simple rational terms (\cref{lem:completeness-for-simple-rational-terms}) and the fact that lifts are isomorphisms. We can now use \cref{lem:pre-postcompose-with-injective-surjective} to cancel out the common lift, yielding $f \equiv g$. 
\end{proof}

The aperiodic case is arguably more natural than the general one, since the \kl{atomic terms} for groups are a bit artificial. Unfortunately, its completeness proof does use groups (although only cyclic ones) -- this is hidden in the proof of \cref{lem:bundles}, where a cyclic group is used to prove \kl{surjectivity} of the \kl{concatenation map}
\begin{align*}
(A^k)^* \times A^{<k} \concmap A^*.
\end{align*}
We do not know how to avoid this. The only solution that we see is an implication rule asserting surjectivity of the above concatenation map: if two terms are equivalent when precomposed with it, then they are equivalent. Such a rule would replace \cref{ax:window}, and make the completeness proof for \kl{aperiodic rational terms} purely aperiodic.

\paragraph*{The General Case.} To finish the completeness proof, we need to account for the \kl{group prefix multiplication}. This is done in \appref{sec:completeness-in-the-presence-of-groups}, and we only give a rough description here. The general idea is to extend the lift terms with two kinds of terms: (a) a certain restricted kind of injections into the group \kl{type}, called \emph{group injections}, and (b) the term $\prefterm$ wrapped into a composition chain which has a drop one-sided inverse, which we call a \emph{Simon lift}. The new kinds of lifts are no longer \kl{isomorphisms}, but they are still \kl{injective}, and their one-sided inverses are drops. The crucial step is a version of \cref{lem:exchange} adjusted to the setting of the newly introduced lifts (see \appref{lem:group-exchange}). The completeness proof then proceeds as in the aperiodic case (\cref{cor:aperiodic-completeness}). Using the exact same construction, we obtain the following provable version of \cref{thm:simplification-of-rational-terms}.

\begin{theorem}[Provable simplification of rational terms]\label{thm:provable-simplification}
    Consider \kl{rational terms} $f, g : A \to N^*$, where $N$ is a \kl{finite alphabet}. There is a \kl{type} $B$ and a \kl{surjective} \kl{rational term} $\pi : B \to A$, such that $\pi \comp f$ and $\pi \comp g$ are equivalent to \kl{simple rational terms}.
\end{theorem}

Finally, combining \cref{thm:provable-simplification} with completeness for simple rational terms (\cref{lem:completeness-for-simple-rational-terms}), and arguing exactly as in the aperiodic case (\cref{cor:aperiodic-completeness}), we obtain completeness of the equational axiomatisation, which finishes the proof of \cref{thm:main}.

\section{Conclusions}
\label{sec:conclusions}
Let us  discuss some perspectives for future work. The most natural extension is to consider regular functions. From a transducer perspective, this is a very important class (see the lecture notes~\cite[Part C]{Bojanczyk2026} for an overview of its numerous characterisations). The corresponding compositional syntax of \emph{regular terms}, which was introduced in~\cite{BojanczykDaviaudKrishna2018}, can be obtained from the syntax in \cref{sec:syntax} by adding two new \kl{atomic terms}: list reversal of type $A^* \to A^*$, and a diagonal map $A \to A \times A$.  We would like to extend our equational axiomatisation to cover the regular terms. Since the equivalence problem for regular functions is decidable~\cite[Theorem 1]{Gurari1982}, we see no fundamental reason why such an axiomatisation should not exist, although we expect that designing it and proving its completeness will require substantial new ideas.

A more ambitious goal would be to consider the polyregular functions~\cite{Bojanczyk2022}, which are a transducer class with polynomial output size, as opposed to the linear output size for the rational and regular functions. The corresponding compositional syntax is obtained by adding, on top of the regular terms, one more atomic term of type 
$
A^* \to (A^* \times A^*)^*,
$
which produces all possible partitions of the input list into two parts: a prefix and a suffix. This goal is much harder, since the equivalence problem for polyregular functions is not known to be decidable, and thus finding a (well-behaved) equational axiomatisation would require solving this difficult problem. However, we believe that the syntactic methods and ideas about canonical models that we developed in this paper could  be useful for this goal, and might be a new line of attack on problems where the automata method for transducer equivalence has stalled. A similar situation arises for data structures beyond lists, such as trees or graphs, where the automata method has been applied only with limited success~\cite{BojanczykSchmude2020}. 

The completeness of the theory proved in this paper is about ground terms, i.e.~terms without free variables. One could also ask about equalities that involve free variables, which range over \kl{rational functions}. We do not know about completeness for such a theory, or even decidability. For more expressive variants, in which variables also range over functions but more operations are allowed, the theory can become undecidable~\cite[Theorem 3.1]{BojanczykFabinaskiStefanski2026}.

Another interesting problem is to describe the models of our axiomatisation. Kleene algebra admits many computationally relevant models besides the regular languages, such as the algebra of binary relations on a set and the algebra of square matrices over a Kleene algebra. We do not know whether our axiomatisation has similarly interesting models beyond the rational functions, and we leave this question as a direction for future investigation.

Finally, we state an open problem that is not about the equational axiomatisation of transducers, but about the simpler setting of regular expressions and Kleene algebra. There is a star-free variant of regular expressions, in which the Kleene star is not allowed, but complementation is allowed. This variant has the same expressive power as first-order logic~\cite[Theorem 10.5]{McNaughtonPapert71}. Is there a finite equational axiomatisation of star-free regular expressions? It is unclear to us whether the methods of this paper could be of any use in resolving this problem.

\begin{acks}
This research was supported by the Polish National Science Centre (NCN) grant ``Polynomial finite state computation'' (2022/46/A/ST6/00072).
\end{acks}

\section*{Disclosure of Use of AI in Research}

A large language model (Claude Opus 5.5, Anthropic) was used for editorial purposes during the revision of this paper: correcting grammar, finding typos, rephrasing text, and drawing diagrams. In the research itself, it helped the authors derive the projections coherence laws (see \appref{sec:projections-coherence-laws}) from the axioms of a monoidal category; in an earlier version of the axiomatisation, these laws were postulated as axioms.

We used Aristotle (Harmonic) to produce a Lean formalisation of our results. During the formalisation, the model found several errors in the proofs in \appendixplace. Most of them were straightforward to fix, but one revealed a gap in the list of axioms, which led to the addition of the bundle axiom (\cref{ax:window}). The formalisation was then prepared for submission to the Lax archive by Claude Opus 5.5, and it is publicly available~\cite{lax-149335}.

All content produced with the help of these models was checked by the authors, who take full responsibility for it.

\section*{Data Availability Statement}

\iffullversion\else
Complete proofs of all results are given in the full version of this paper, which is available on arXiv~\cite{fullversion}.
\fi
A Lean formalisation of our results is publicly available in the Lax archive~\cite{lax-149335}. It is provided for reference only, and is not submitted as an official contribution of this paper. As this is purely theoretical work, no other artifacts are associated with this paper.

\bibliographystyle{ACM-Reference-Format}
\bibliography{bib}

\appendix

\section{The Axioms as Commutative Diagrams}
\label{sec:axioms-in-diagrams}

\newcommand{\axcell}[2][0.48]{%
    \begin{minipage}[b]{#1\linewidth}\centering #2\end{minipage}}
\newcommand{\axgroup}[1]{\par\medskip\noindent\rule{\linewidth}{0.4pt}\par\smallskip
    \noindent{\footnotesize\sffamily\MakeUppercase{#1}}\par\medskip}

\axgroup{Derived terms}

{\allowdisplaybreaks
\begin{align*}
    A \xrightarrow{1} \onetype \quad & \eqdef \quad \appendoneterm \comp \sndterm \\
    (A \times B) + (A \times C) \xrightarrow{\undistrterm} A \times (B + C) \quad & \eqdef \quad \either{(\idterm \times \leftterm)}{(\idterm \times \rightterm)} \\
    (A \times C) + (B \times C) \xrightarrow{\undistlterm} (A + B) \times C \quad & \eqdef \quad \either{(\leftterm \times \idterm)}{(\rightterm \times \idterm)} \\
    A \times A^* \xrightarrow{\prependterm} A^* \quad & \eqdef \quad \rightterm \comp \consterm \\
    A^* \times A \xrightarrow{\appendterm} A^* \quad & \eqdef \quad \rightterm \comp \snocterm \\
    \onetype \xrightarrow{[]} A^* \quad & \eqdef \quad \leftterm \comp \consterm \\
    A \xrightarrow{\pureterm} A^* \quad & \eqdef \quad \appendoneterm \comp (\idterm \times []) \comp \prependterm \\
    A^* \times A^* \xrightarrow{\binconcatterm} A^* \quad & \eqdef \quad (\idterm \times \pureterm) \comp \prependterm \comp \concatterm \\
    F(A, B) \xrightarrowtail{\,\text{concatenation map}\,} (A+B)^* \quad & \eqdef \quad \text{\cref{sec:concatenation-maps}} \\
    (A \times G)^* \xrightarrow{\unprefterm} (A \times G)^* \quad & \eqdef \quad \text{\cref{def:unprefterm}} \\
    2^{**} \xrightarrow{\bundleterm} (2^* \times 2^*)^* \times (\onetype + 2^*) \quad & \eqdef \quad \text{\cref{def:bundle-term}} 
\end{align*}
}

\axgroup{Category axioms (\ref{ax:associativity-of-composition})}

\noindent
\axcell[0.32]{$(f \comp g) \comp h \equiv f \comp (g \comp h)$\par\medskip}
\hfill
\axcell[0.32]{\begin{tikzcd}[ampersand replacement=\&]
    A \ar[r,"\idterm"] \ar[dr,"f"'] \& A \ar[d,"f"] \\
    \& B
\end{tikzcd}}
\hfill
\axcell[0.32]{\begin{tikzcd}[ampersand replacement=\&]
    A \ar[r,"f"] \ar[dr,"f"'] \& B \ar[d,"\idterm"] \\
    \& B
\end{tikzcd}}

\axgroup{Functoriality axioms (\ref{ax:functoriality-of-type-constructors})}

\noindent
\axcell[0.32]{\begin{tikzcd}[ampersand replacement=\&]
    A_1 \times A_2 \ar[r, shift left, "\idterm \times \idterm"] \ar[r, shift right, "\idterm"'] \& A_1 \times A_2
\end{tikzcd}}
\hfill
\axcell[0.32]{\begin{tikzcd}[ampersand replacement=\&]
    A_1 + A_2 \ar[r, shift left, "\idterm + \idterm"] \ar[r, shift right, "\idterm"'] \& A_1 + A_2
\end{tikzcd}}
\hfill
\axcell[0.32]{\begin{tikzcd}[ampersand replacement=\&]
    A^* \ar[r, shift left, "\idterm^*"] \ar[r, shift right, "\idterm"'] \& A^*
\end{tikzcd}}

\medskip\noindent
\axcell[0.32]{\begin{tikzcd}[ampersand replacement=\&, column sep=small]
    A_1 \times A_2 \ar[r,"f_1 \times f_2"] \ar[dr,"(f_1 \comp g_1) \times (f_2 \comp g_2)"', near end] \& B_1 \times B_2 \ar[d,"g_1 \times g_2"] \\
    \& C_1 \times C_2
\end{tikzcd}}
\hfill
\axcell[0.32]{\begin{tikzcd}[ampersand replacement=\&, column sep=small]
    A_1 + A_2 \ar[r,"f_1 + f_2"] \ar[dr,"(f_1 \comp g_1) + (f_2 \comp g_2)"', near end] \& B_1 + B_2 \ar[d,"g_1 + g_2"] \\
    \& C_1 + C_2
\end{tikzcd}}
\hfill
\axcell[0.32]{\begin{tikzcd}[ampersand replacement=\&]
    A^* \ar[r,"f^*"] \ar[dr,"(f \comp g)^*"'] \& B^* \ar[d,"g^*"] \\
    \& C^*
\end{tikzcd}}

\axgroup{Naturality axioms (\ref{ax:naturality})}

\noindent
\axcell{\begin{tikzcd}[ampersand replacement=\&]
    A \ar[r,"f"] \ar[d,"\idterm"'] \& B \ar[d,"\idterm"] \\
    A \ar[r,"f"'] \& B
\end{tikzcd}}
\hfill
\axcell{\begin{tikzcd}[ampersand replacement=\&]
    A + A \ar[r,"f + f"] \ar[d,"\codiagterm"'] \& B + B \ar[d,"\codiagterm"] \\
    A \ar[r,"f"'] \& B
\end{tikzcd}}

\medskip\noindent
\axcell{\begin{tikzcd}[ampersand replacement=\&]
    A_1 \ar[r,"f_1"] \ar[d,"\leftterm"'] \& B_1 \ar[d,"\leftterm"] \\
    A_1 + A_2 \ar[r,"f_1 + f_2"'] \& B_1 + B_2
\end{tikzcd}}
\hfill
\axcell{\begin{tikzcd}[ampersand replacement=\&]
    A_2 \ar[r,"f_2"] \ar[d,"\rightterm"'] \& B_2 \ar[d,"\rightterm"] \\
    A_1 + A_2 \ar[r,"f_1 + f_2"'] \& B_1 + B_2
\end{tikzcd}}

\medskip\noindent
\axcell{\begin{tikzcd}[ampersand replacement=\&]
    A_1 \times A_2 \ar[r,"f_1 \times f_2"] \ar[d,"\fstterm"'] \& B_1 \times B_2 \ar[d,"\fstterm"] \\
    A_1 \ar[r,"f_1"'] \& B_1
\end{tikzcd}}
\hfill
\axcell{\begin{tikzcd}[ampersand replacement=\&]
    A_1 \times A_2 \ar[r,"f_1 \times f_2"] \ar[d,"\sndterm"'] \& B_1 \times B_2 \ar[d,"\sndterm"] \\
    A_2 \ar[r,"f_2"'] \& B_2
\end{tikzcd}}

\medskip\noindent
\axcell{\begin{tikzcd}[ampersand replacement=\&]
    A \ar[r,"f"] \ar[d,"{\appendoneterm}"'] \& B \ar[d,"{\appendoneterm}"] \\
    A \times \onetype \ar[r,"f \times \idterm"'] \& B \times \onetype
\end{tikzcd}}
\hfill
\axcell{\begin{tikzcd}[ampersand replacement=\&]
    A \ar[r,"f"] \ar[d,"{\prependoneterm}"'] \& B \ar[d,"{\prependoneterm}"] \\
    \onetype \times A \ar[r,"\idterm \times f"'] \& \onetype \times B
\end{tikzcd}}

\medskip\noindent
\axcell{\begin{tikzcd}[ampersand replacement=\&, column sep=large]
    A_1 \times (A_2 \times A_3) \ar[r,"f_1 \times (f_2 \times f_3)"] \ar[d,"\assoclterm"'] \& B_1 \times (B_2 \times B_3) \ar[d,"\assoclterm"] \\
    (A_1 \times A_2) \times A_3 \ar[r,"(f_1 \times f_2) \times f_3"'] \& (B_1 \times B_2) \times B_3
\end{tikzcd}}
\hfill
\axcell{\begin{tikzcd}[ampersand replacement=\&, column sep=large]
    (A_1 \times A_2) \times A_3 \ar[r,"(f_1 \times f_2) \times f_3"] \ar[d,"\assocrterm"'] \& (B_1 \times B_2) \times B_3 \ar[d,"\assocrterm"] \\
    A_1 \times (A_2 \times A_3) \ar[r,"f_1 \times (f_2 \times f_3)"'] \& B_1 \times (B_2 \times B_3)
\end{tikzcd}}

\medskip\noindent
\axcell[1]{\begin{tikzcd}[ampersand replacement=\&, column sep=huge]
    A_1 \times (A_2 + A_3) \ar[r,"f_1 \times (f_2 + f_3)"] \ar[d,"\distrterm"'] \& B_1 \times (B_2 + B_3) \ar[d,"\distrterm"] \\
    (A_1 \times A_2) + (A_1 \times A_3) \ar[r,"(f_1 \times f_2) + (f_1 \times f_3)"'] \& (B_1 \times B_2) + (B_1 \times B_3)
\end{tikzcd}}

\medskip\noindent
\axcell[1]{\begin{tikzcd}[ampersand replacement=\&, column sep=huge]
    (A_1 + A_2) \times A_3 \ar[r,"(f_1 + f_2) \times f_3"] \ar[d,"\distlterm"'] \& (B_1 + B_2) \times B_3 \ar[d,"\distlterm"] \\
    (A_1 \times A_3) + (A_2 \times A_3) \ar[r,"(f_1 \times f_3) + (f_2 \times f_3)"'] \& (B_1 \times B_3) + (B_2 \times B_3)
\end{tikzcd}}

\medskip\noindent
\axcell{\begin{tikzcd}[ampersand replacement=\&, column sep=large]
    \onetype + A \times A^* \ar[r,"\idterm + f \times f^*"] \ar[d,"\consterm"'] \& \onetype + B \times B^* \ar[d,"\consterm"] \\
    A^* \ar[r,"f^*"'] \& B^*
\end{tikzcd}}
\hfill
\axcell{\begin{tikzcd}[ampersand replacement=\&, column sep=large]
    A^* \ar[r,"f^*"] \ar[d,"\unconsterm"'] \& B^* \ar[d,"\unconsterm"] \\
    \onetype + A \times A^* \ar[r,"\idterm + f \times f^*"'] \& \onetype + B \times B^*
\end{tikzcd}}

\medskip\noindent
\axcell{\begin{tikzcd}[ampersand replacement=\&, column sep=large]
    \onetype + A^* \times A \ar[r,"\idterm + f^* \times f"] \ar[d,"\snocterm"'] \& \onetype + B^* \times B \ar[d,"\snocterm"] \\
    A^* \ar[r,"f^*"'] \& B^*
\end{tikzcd}}
\hfill
\axcell{\begin{tikzcd}[ampersand replacement=\&, column sep=large]
    A^* \ar[r,"f^*"] \ar[d,"\unsnocterm"'] \& B^* \ar[d,"\unsnocterm"] \\
    \onetype + A^* \times A \ar[r,"\idterm + f^* \times f"'] \& \onetype + B^* \times B
\end{tikzcd}}

\medskip\noindent
\axcell{\begin{tikzcd}[ampersand replacement=\&]
    A^{**} \ar[r,"f^{**}"] \ar[d,"\concatterm"'] \& B^{**} \ar[d,"\concatterm"] \\
    A^* \ar[r,"f^*"'] \& B^*
\end{tikzcd}}
\hfill
\axcell{\begin{tikzcd}[ampersand replacement=\&, column sep=large]
    (A_1 + A_2)^* \ar[r,"(f_1 + f_2)^*"] \ar[d,"\blockterm"'] \& (B_1 + B_2)^* \ar[d,"\blockterm"] \\
    A_1^* \times (A_2 \times A_1^*)^* \ar[r,"f_1^* \times (f_2 \times f_1^*)^*"'] \& B_1^* \times (B_2 \times B_1^*)^*
\end{tikzcd}}

\medskip\noindent
\axcell{\begin{tikzcd}[ampersand replacement=\&, column sep=large]
    (A_1 \times A_2)^* \times A_1 \ar[r,"(f_1 \times f_2)^* \times f_1"] \ar[d,"\shiftrterm"'] \& (B_1 \times B_2)^* \times B_1 \ar[d,"\shiftrterm"] \\
    A_1 \times (A_2 \times A_1)^* \ar[r,"f_1 \times (f_2 \times f_1)^*"'] \& B_1 \times (B_2 \times B_1)^*
\end{tikzcd}}
\hfill
\axcell{\begin{tikzcd}[ampersand replacement=\&, column sep=large]
    A_1 \times (A_2 \times A_1)^* \ar[r,"f_1 \times (f_2 \times f_1)^*"] \ar[d,"\shiftlterm"'] \& B_1 \times (B_2 \times B_1)^* \ar[d,"\shiftlterm"] \\
    (A_1 \times A_2)^* \times A_1 \ar[r,"(f_1 \times f_2)^* \times f_1"'] \& (B_1 \times B_2)^* \times B_1
\end{tikzcd}}

\axgroup{Isomorphism axioms (\ref{ax:isomorphisms})}

\noindent
\axcell{\begin{tikzcd}[ampersand replacement=\&]
    A \ar[r,"{\appendoneterm}"] \ar[dr,"\idterm"'] \& A \times \onetype \ar[d,"\fstterm"] \ar[dr,"\idterm"] \& \\
    \& A \ar[r,"{\appendoneterm}"'] \& A \times \onetype
\end{tikzcd}}
\hfill
\axcell{\begin{tikzcd}[ampersand replacement=\&]
    A \ar[r,"{\prependoneterm}"] \ar[dr,"\idterm"'] \& \onetype \times A \ar[d,"\sndterm"] \ar[dr,"\idterm"] \& \\
    \& A \ar[r,"{\prependoneterm}"'] \& \onetype \times A
\end{tikzcd}}

\medskip\noindent
\axcell[1]{\begin{tikzcd}[ampersand replacement=\&]
    A \times (B \times C) \ar[r,"\assoclterm"] \ar[dr,"\idterm"'] \& (A \times B) \times C \ar[d,"\assocrterm"] \ar[dr,"\idterm"] \& \\
    \& A \times (B \times C) \ar[r,"\assoclterm"'] \& (A \times B) \times C
\end{tikzcd}}

\medskip\noindent
\axcell[1]{\begin{tikzcd}[ampersand replacement=\&]
    A \times (B \times A)^* \ar[r,"\shiftlterm"] \ar[dr,"\idterm"'] \& (A \times B)^* \times A \ar[d,"\shiftrterm"] \ar[dr,"\idterm"] \& \\
    \& A \times (B \times A)^* \ar[r,"\shiftlterm"'] \& (A \times B)^* \times A
\end{tikzcd}}

\medskip\noindent
\axcell[1]{\begin{tikzcd}[ampersand replacement=\&]
    (A + B) \times C \ar[r,"\distlterm"] \ar[dr,"\idterm"'] \& (A \times C) + (B \times C) \ar[d,"\undistlterm"] \ar[dr,"\idterm"] \& \\
    \& (A + B) \times C \ar[r,"\distlterm"'] \& (A \times C) + (B \times C)
\end{tikzcd}}

\medskip\noindent
\axcell[1]{\begin{tikzcd}[ampersand replacement=\&]
    A \times (B + C) \ar[r,"\distrterm"] \ar[dr,"\idterm"'] \& (A \times B) + (A \times C) \ar[d,"\undistrterm"] \ar[dr,"\idterm"] \& \\
    \& A \times (B + C) \ar[r,"\distrterm"'] \& (A \times B) + (A \times C)
\end{tikzcd}}

\medskip\noindent
\axcell{\begin{tikzcd}[ampersand replacement=\&]
    \onetype + A \times A^* \ar[r,"\consterm"] \ar[dr,"\idterm"'] \& A^* \ar[d,"\unconsterm"] \ar[dr,"\idterm"] \& \\
    \& \onetype + A \times A^* \ar[r,"\consterm"'] \& A^*
\end{tikzcd}}
\hfill
\axcell{\begin{tikzcd}[ampersand replacement=\&]
    \onetype + A^* \times A \ar[r,"\snocterm"] \ar[dr,"\idterm"'] \& A^* \ar[d,"\unsnocterm"] \ar[dr,"\idterm"] \& \\
    \& \onetype + A^* \times A \ar[r,"\snocterm"'] \& A^*
\end{tikzcd}}

\medskip\noindent
\axcell[1]{\begin{tikzcd}[ampersand replacement=\&, column sep=large]
    (A + B)^* \ar[r,"\blockterm"] \ar[dr,"\idterm"'] \& A^* \times (B \times A^*)^* \ar[d, \concar, "\unblockterm"] \ar[dr,"\idterm"] \& \\
    \& (A + B)^* \ar[r,"\blockterm"'] \& A^* \times (B \times A^*)^*
\end{tikzcd}}

\medskip\noindent
\axcell[1]{\begin{tikzcd}[ampersand replacement=\&, column sep=large]
    (A \times G)^* \ar[r,"\prefterm"] \ar[dr,"\idterm"'] \& (A \times G)^* \ar[d,"\unprefterm"] \ar[dr,"\idterm"] \& \\
    \& (A \times G)^* \ar[r,"\prefterm"'] \& (A \times G)^*
\end{tikzcd}}

\axgroup{Listless axioms (\ref{ax:unit-type}--\ref{ax:triangle})}

\noindent
\axcell[0.18]{\begin{tikzcd}[ampersand replacement=\&]
    A \ar[r, shift left, "f"] \ar[r, shift right, "g"'] \& \onetype
\end{tikzcd}}
\hfill
\axcell[0.36]{\begin{tikzcd}[ampersand replacement=\&, column sep=large, row sep=large]
    A \ar[r,"\leftterm"] \ar[dr,"\idterm"'] \& A + A \ar[d,"\codiagterm" description] \& A \ar[l,"\rightterm"'] \ar[dl,"\idterm"] \\
    \& A \&
\end{tikzcd}}
\hfill
\axcell[0.42]{\begin{tikzcd}[ampersand replacement=\&, column sep=large]
    A + B \ar[r,"\leftterm + \rightterm"] \ar[dr,"\idterm"'] \& (A + B) + (A + B) \ar[d,"\codiagterm"] \\
    \& A + B
\end{tikzcd}}

\medskip\noindent
\axcell[0.56]{\begin{tikzcd}[ampersand replacement=\&, column sep=-1.5em]
    \& A \times (B \times (C \times D)) \ar[dl,"\idterm \times \assoclterm"'] \ar[dr,"\assoclterm"] \& \\
    A \times ((B \times C) \times D) \ar[d,"\assoclterm"'] \& \& (A \times B) \times (C \times D) \ar[d,"\assoclterm"] \\
    (A \times (B \times C)) \times D \ar[rr,"\assoclterm \times \idterm"'] \& \& ((A \times B) \times C) \times D
\end{tikzcd}}
\hfill
\axcell[0.435]{\begin{tikzcd}[ampersand replacement=\&, column sep=small]
    A \times (\onetype \times C) \ar[rr,"\assoclterm"] \ar[dr,"\idterm \times \sndterm"'] \& \& (A \times \onetype) \times C \ar[dl,"\fstterm \times \idterm"] \\
    \& A \times C \&
\end{tikzcd}}

\axgroup{List axioms (\ref{ax:cons-snoc-symmetry}--\ref{ax:window})}

\noindent
\axcell[0.4]{\begin{tikzcd}[ampersand replacement=\&]
    \onetype \ar[r,"\leftterm"] \ar[d,"\leftterm"'] \& \onetype + A \times A^* \ar[d,"\consterm"] \\
    \onetype + A^* \times A \ar[r,"\snocterm"'] \& A^*
\end{tikzcd}}
\hfill
\axcell[0.58]{\begin{tikzcd}[ampersand replacement=\&]
    A \ar[r,"\prependoneterm"] \ar[d,"\appendoneterm"'] \& \onetype \times A \ar[r,"{[] \times \idterm}"] \& A^* \times A \ar[d,"\appendterm"] \\
    A \times \onetype \ar[r,"{\idterm \times []}"'] \& A \times A^* \ar[r,"\prependterm"'] \& A^*
\end{tikzcd}}

\medskip\noindent
\axcell[1]{\begin{tikzcd}[ampersand replacement=\&]
    A^* \times A^* \ar[r,"\pureterm \times \idterm"] \ar[d,"\idterm \times \pureterm"'] \& A^{**} \times A^* \ar[r,"\appendterm"] \& A^{**} \ar[d,"\concatterm"] \\
    A^* \times A^{**} \ar[r,"\prependterm"'] \& A^{**} \ar[r,"\concatterm"'] \& A^*
\end{tikzcd}}

\medskip\noindent
\axcell[0.49]{\begin{tikzcd}[ampersand replacement=\&, column sep=5em]
    (A^* \times A^*) \times A^* \ar[r,"\binconcatterm \times \idterm"] \ar[d,"\assocrterm"'] \& A^* \times A^* \ar[dd,"\binconcatterm"] \\
    A^* \times (A^* \times A^*) \ar[d,"\idterm \times \binconcatterm"'] \& \\
    A^* \times A^* \ar[r,"\binconcatterm"'] \& A^*
\end{tikzcd}}
\hfill
\axcell[0.49]{\begin{tikzcd}[ampersand replacement=\&]
    A^* \ar[r,"\appendoneterm"] \ar[drr,"\idterm"'] \& A^* \times \onetype \ar[r,"{\idterm \times []}"] \& A^* \times A^* \ar[d,"\binconcatterm"] \\
    \& \& A^* \\
    A^* \ar[r,"\prependoneterm"'] \ar[urr,"\idterm"] \& \onetype \times A^* \ar[r,"{[] \times \idterm}"'] \& A^* \times A^* \ar[u,"\binconcatterm"']
\end{tikzcd}}

\medskip\noindent
\axcell[0.35]{\begin{tikzcd}[ampersand replacement=\&]
    \onetype \ar[r,"{[]}"] \ar[dr,"{[]}"'] \& A^{**} \ar[d,"\concatterm"] \\
    \& A^*
\end{tikzcd}}
\hfill
\axcell[0.63]{\begin{tikzcd}[ampersand replacement=\&, column sep=large]
    A^{**} \times A^{**} \ar[r,"\binconcatterm"] \ar[d,"\concatterm \times \concatterm"'] \& A^{**} \ar[d,"\concatterm"] \\
    A^* \times A^* \ar[r,"\binconcatterm"'] \& A^*
\end{tikzcd}}

\medskip\noindent
\axcell[0.35]{\begin{tikzcd}[ampersand replacement=\&]
    A^{***} \ar[r,"\concatterm^*"] \ar[d,"\concatterm"'] \& A^{**} \ar[d,"\concatterm"] \\
    A^{**} \ar[r,"\concatterm"'] \& A^*
\end{tikzcd}}
\hfill
\axcell[0.6]{\begin{tikzcd}[ampersand replacement=\&, column sep=large, row sep=large]
    A^* \ar[r,"\pureterm"] \ar[dr,"\idterm"'] \& A^{**} \ar[d,"\concatterm" description] \& A^* \ar[l,"\pureterm^*"'] \ar[dl,"\idterm"] \\
    \& A^* \&
\end{tikzcd}}

\medskip\noindent
\axcell{\begin{tikzcd}[ampersand replacement=\&]
    A \times A^* \ar[r,"\pureterm \times \idterm"] \ar[dr,"\prependterm"'] \& A^* \times A^* \ar[d,"\binconcatterm"] \\
    \& A^*
\end{tikzcd}}
\hfill
\axcell{\begin{tikzcd}[ampersand replacement=\&]
    A^* \times A \ar[r,"\idterm \times \pureterm"] \ar[dr,"\appendterm"'] \& A^* \times A^* \ar[d,"\binconcatterm"] \\
    \& A^*
\end{tikzcd}}

\medskip\noindent
\axcell[0.4]{\begin{tikzcd}[ampersand replacement=\&]
    \onetype^* \ar[r,"{[]^*}"] \ar[d,"1"'] \& A^{**} \ar[d,"\concatterm"] \\
    \onetype \ar[r,"{[]}"'] \& A^*
\end{tikzcd}}
\hfill
\axcell[0.58]{\begin{tikzcd}[ampersand replacement=\&]
    (A \times B)^* \times A \ar[rr,"\shiftrterm"] \ar[dr,\concar] \& \& A \times (B \times A)^* \ar[dl,\concar] \\
    \& (A + B)^* \&
\end{tikzcd}}

\medskip\noindent
\axcell[1]{\begin{tikzcd}[ampersand replacement=\&, column sep=large]
    2^{**} \ar[r,"{\bundleterm}"] \ar[dr,"\idterm"'] \& (2^* \times 2^*)^* \times (\onetype + 2^*) \ar[d,\concar] \\
    \& 2^{**}
\end{tikzcd}}

\section{String Diagrams}
\label{sec:string-diagrams}
In this section, we illustrate the axioms of concatenation using string diagrams.

This diagram shows \cref{ax:cons-snoc-symmetry} about the symmetry of definitions of $\pureterm$ and $\binconcatterm$.
    \begin{center}
        \includegraphics[page=1,scale=0.2]{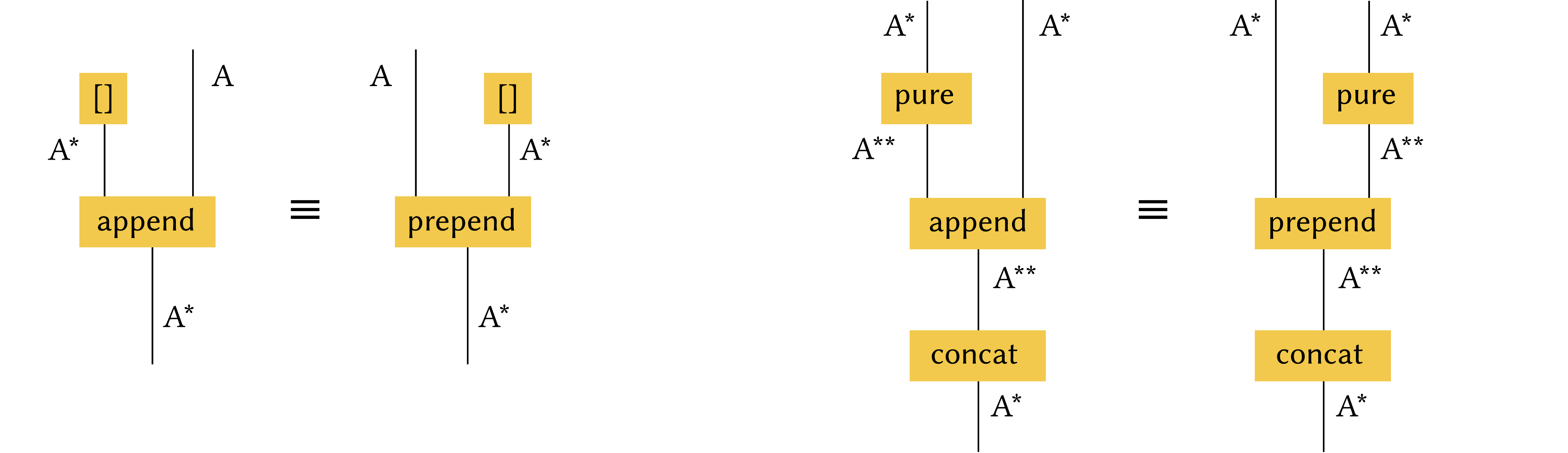}
    \end{center}

This diagram shows     \cref{ax:concatenation-monoid}: $A^*$ is a monoid with respect to $\binconcatterm$ and $[]$.
    \begin{center}
        \includegraphics[page=2,scale=0.2]{B-string-diagrams/equational-theory-rational}
    \end{center}

This diagram shows      \cref{ax:concat-monoid-homomorphism}: $\concatterm$ is a monoid homomorphism from $A^{**}$ to $A^*$.
    \begin{center}
        \includegraphics[page=3,scale=0.2]{B-string-diagrams/equational-theory-rational}
    \end{center}

This diagram shows     \cref{ax:from-monoid-to-inductive}: $\prependterm$ and $\appendterm$ defined in terms of $\pureterm$ and $\binconcatterm$.
    \begin{center}
        \includegraphics[page=4,scale=0.2]{B-string-diagrams/equational-theory-rational}
    \end{center}

\section{Derived Equalities}\label{sec:basic-facts-about-co-products}

In this appendix, we derive auxiliary equivalences used throughout the paper.

\subsection{Projections Coherence Laws}\label{sec:projections-coherence-laws}

In this section, we prove the \kl{projections coherence laws}.

\begin{lemma}\label{lem:projections-coherence-laws}
    \begin{align*}
        \begin{casesright}
        \assoclterm \comp \fstterm \times \idterm \equiv \idterm \times \sndterm \\ 
        \assoclterm \comp \sndterm \times \idterm \equiv \sndterm \\
        \assoclterm \comp \fstterm \equiv \idterm \times \fstterm
        \end{casesright}
        \quad 
        (\text{Projections coherence laws})
    \end{align*}
\end{lemma}
The first equation of \cref{lem:projections-coherence-laws} is a direct strengthening of Mac Lane's triangle (\cref{ax:triangle}), where the unit in the middle is replaced with an arbitrary \kl{type}. Similarly, the second and third equations are analogous strengthenings of the ``unit on the left or right'' versions of the triangle, which we first derive using an argument similar to that of Kelly~\cite{Kelly1964}.

\begin{lemma}[Unit on the left or right]\label{lem:triangle-left-or-right}
    The following diagrams commute provably:
    \[
    \begin{tikzcd}[column sep=small]
        \onetype \times \left( B \times C \right)
        \ar[rr, "\assoclterm"]
        \ar[dr, "\sndterm"']
        && \left(\onetype \times B \right) \times C
        \ar[dl, "{\sndterm \times \idterm}"]
        \\
        & B \times C
    \end{tikzcd}
    \quad 
    \begin{tikzcd}[column sep=small]
        A \times \left( B \times \onetype \right)
        \ar[rr, "\assoclterm"]
        \ar[dr, "{\idterm \times \fstterm}"']
        && \left(A \times B \right) \times \onetype
        \ar[dl, "\fstterm"]
        \\
        & A \times B
    \end{tikzcd}
    \]
\end{lemma}
\begin{proof}
    We only derive the first diagram; the second one is derived using a symmetric argument. Consider the following diagram, where the outer face is an instance of the pentagon (\cref{ax:pentagon}) for $(\onetype, \onetype, B, C)$:
    \[
    \begin{tikzcd}
        &[-3.2em]&[-3.2em]\onetype \times \left( \onetype \times \left( B \times C \right) \right)
        \ar[dll, "{\idterm \times \assoclterm}"']
        \ar[drr, "\assoclterm"]
        \ar[dd, "{\idterm \times \sndterm}"]
        \\ 
        \onetype \times \left( \left( \onetype \times B \right) \times C \right)
        \ar[dddr, "\assoclterm"']
        \ar[drr, "{\idterm \times \left( \sndterm \times \idterm \right)}"']
        &[-3.2em]&[-3.2em]&[-3.2em]&[-3.2em]\left( \onetype \times \onetype \right) \times \left( B \times C \right)
        \ar[dll, "{\fstterm \times \idterm}"]
        \ar[dddl, "\assoclterm"]
        \\ 
        &[-3.2em]&[-3.2em]\onetype \times \left( B \times C \right) 
        \ar[d, "\assoclterm"]
        \\ 
        &[-3.2em]&[-3.2em]\left( \onetype \times B \right) \times C 
        \\ 
        &[-3.2em]\left( \onetype \times \left( \onetype \times B \right) \right) \times C 
        \ar[rr, "{\assoclterm \times \idterm}"']
        \ar[ur, "{\left( \idterm \times \sndterm \right) \times \idterm}"]
        &[-3.2em]&[-3.2em]\left( \left( \onetype \times \onetype \right) \times B \right) \times C
        \ar[ul, "{\left(\fstterm \times \idterm \right) \times \idterm}"']
    \end{tikzcd}
    \]
    Both quadrilateral faces commute by naturality of $\assoclterm$ (\cref{ax:naturality}). The top-right and bottom triangles are the standard triangles (\cref{ax:triangle}). The outer face is the pentagon (\cref{ax:pentagon}). From this we can derive that the top-left triangle also commutes. Chasing around, we have:
    \begin{align*}
        \idterm \times \sndterm \comp \assoclterm \equalbecause{top-right triangle} \\ 
        \assoclterm \comp \fstterm \times \idterm \comp \assoclterm \equalbecause{bottom-right face} \\ 
        \assoclterm \comp \assoclterm \comp \left( \fstterm \times \idterm \right) \times \idterm \equalbecause{outer face} \\ 
        \idterm \times \assoclterm \comp \assoclterm \comp \assoclterm \times \idterm \comp \left( \fstterm \times \idterm \right) \times \idterm \equalbecause{bottom triangle} \\ 
        \idterm \times \assoclterm \comp \assoclterm \comp \left( \idterm \times \sndterm \right) \times \idterm \equalbecause{bottom-left face} \\ 
        \idterm \times \assoclterm \comp \idterm \times \left( \sndterm \times \idterm \right) \comp \assoclterm.
    \end{align*}
    Cancel $\assoclterm$ on both sides by \cref{lem:pre-postcompose-with-injective-surjective} and apply functoriality of $\times$ (\cref{ax:functoriality-of-type-constructors}) to obtain:
    \[
        \idterm \times \sndterm \equiv \idterm \times \left( \assoclterm \comp \sndterm \times \idterm \right).
    \]
    The leading identity on both sides is on $\onetype$, so by precomposing with $\unitlterm$, we get:
    \[
        \unitlterm \comp \idterm \times \sndterm \equiv \unitlterm \comp \idterm \times \left( \assoclterm \comp \sndterm \times \idterm \right).
    \]
    Finally, we apply naturality of $\unitlterm$ (\cref{ax:naturality}) to both sides to obtain the desired equality:
    \[
        \sndterm \equiv \assoclterm \comp \sndterm \times \idterm.
    \]
\end{proof}

To deduce \cref{lem:projections-coherence-laws}, we have to replace the discarded units in \cref{ax:triangle} and \cref{lem:triangle-left-or-right} with an arbitrary type. We achieve this using naturality of projections (\cref{ax:naturality}). Consider the following diagram:
\[
\begin{tikzcd}
    A \times \left(B \times C \right) 
    \ar[rrrr, "\assoclterm"]
    \ar[dr, "{\idterm \times \left( 1 \times \idterm \right)}"']
    &[-3.5em]&[-2em]&[-2em]&[-3.5em]\left( A \times B \right) \times C
    \ar[dl, "{\left( \idterm \times 1 \right) \times \idterm}"] 
    \\ 
    &A \times \left( \onetype \times C \right) 
    \ar[rr, "\assoclterm"]
    \ar[dr, "{\idterm \times \sndterm}"']
    &&\left( A \times \onetype \right) \times C 
    \ar[dl, "{\fstterm \times \idterm}"]
    \\ 
    &&A \times C
\end{tikzcd}
\]
The top face commutes by naturality of $\assoclterm$ (\cref{ax:naturality}), while the bottom triangle is exactly \cref{ax:triangle}. Therefore, the entire diagram commutes. Now, the composition obtained from the left path is:
\begin{align*}
    \idterm \times (1 \times \idterm) \comp \idterm \times \sndterm \equalbecause{functoriality of $\times$, \cref{ax:functoriality-of-type-constructors}} \\ 
    \idterm \times \left( \left(1 \times \idterm \right) \comp \sndterm \right) \equalbecause{naturality of $\sndterm$, \cref{ax:naturality}} \\ 
    \idterm \times \sndterm.
\end{align*}
Similarly, the right path is equivalent to $\assoclterm \comp \fstterm \times \idterm$. This finishes the proof of the first equation in \cref{lem:projections-coherence-laws}. The remaining two laws are derived analogously from triangles in \cref{lem:triangle-left-or-right}.

\subsection{Associativity}

\begin{lemma}\label{lem:assocr-fst-mapped-coherence}
    $\fstterm \comp (f \times g) \equiv \assocrterm \comp (f \times (\fstterm \comp g))$.
\end{lemma}
\begin{proof}
    By \cref{ax:isomorphisms}, it suffices to show that $\assoclterm \comp \fstterm \comp (f \times g) \equiv f \times (\fstterm \comp g)$:
    \begin{align*}
        \assoclterm \comp \fstterm \comp (f \times g) \equalbecause{\cref{lem:projections-coherence-laws}} \\ 
        (\idterm \times \fstterm) \comp (f \times g) \equalbecause{functoriality of $\times$, \cref{ax:functoriality-of-type-constructors}} \\ 
        f \times (\fstterm \comp g).
    \end{align*}
\end{proof}

\begin{lemma}\label{lem:snd-snd-is-assocl-snd}
    $\sndterm \comp \sndterm \equiv \assoclterm \comp \sndterm$.
\end{lemma}
\begin{proof}
    By \cref{ax:isomorphisms}, it suffices to show that $\assocrterm \comp \sndterm \comp \sndterm \equiv \sndterm$:
    \begin{align*}
        \assocrterm \comp \sndterm \comp \sndterm \equalbecause{\cref{lem:projections-coherence-laws}} \\ 
        (\sndterm \times \idterm) \comp \sndterm \equalbecause{naturality of $\sndterm$, \cref{ax:naturality}} \\
        \sndterm \comp \idterm \equalbecause{\cref{ax:associativity-of-composition}} \\
        \sndterm.
    \end{align*}
\end{proof}

\subsection{Copairing}\label{sec:co-pairing-simple-facts}

In this section, we prove some simple facts about the \kl{copairing} combinator $\mathtt{either}$. The following lemma essentially says how copairing is evaluated.

\begin{lemma}\label{lem:case-equivalence}
        $\leftterm\comp \either{f_1}{f_2} \equiv f_1$ and similarly for $\rightterm$.
\end{lemma}
    \begin{proof}
        We only do the proof of the first equation:
        \begin{align*}
            \leftterm\comp \either{f_1}{f_2} &= \why{definition of $\either{f_1}{f_2}$}\\
            \leftterm\comp \codiagterm\comp (f_1 + f_2) \equalbecause{\cref{ax:eval-codiag}}\\
            f_1.
        \end{align*}
    \end{proof}

The following lemma states that any term out of a coproduct is determined by its two restrictions.

\begin{lemma}\label{lem:either-uniqueness}
	$\either{\leftterm\comp{}h}{\rightterm\comp{}h} \equiv h$.
\end{lemma}

\begin{proof}
    \begin{align*}
        \either{\leftterm \comp h }{\rightterm \comp h} &= \why{definition of $\mathtt{either}$} \\ 
        (\leftterm \comp h) + (\rightterm \comp h) \comp \codiagterm \equalbecause{functoriality of $+$, \cref{ax:functoriality-of-type-constructors}} \\ 
        (\leftterm + \rightterm) \comp (h + h) \comp \codiagterm \equalbecause{naturality of $\codiagterm$} \\ 
        (\leftterm + \rightterm) \comp \codiagterm \comp h \equalbecause{\cref{ax:eval-codiag}} \\ 
        h
    \end{align*}
\end{proof}

The following two lemmas show that copairing is functorial.
\begin{lemma}\label{lem:functoriality-of-either}
    $\either{f_1}{f_2}\comp g \equiv \either{(f_1\comp g)}{(f_2\comp g)}$.
\end{lemma}
    \begin{proof}
        \begin{align*}
            \either{f_1}{f_2}\comp g &= \why{definition of $\either{f_1}{f_2}$}\\
             (f_1 + f_2)\comp\codiagterm\comp g \equalbecause{naturality of $\codiagterm$}\\
             (f_1 + f_2)\comp(g+g)\comp\codiagterm \equalbecause{functoriality}\\
             (f_1\comp g) + (f_2\comp g)\comp\codiagterm &= \why{definition of $\either{f_1\comp g}{f_2\comp g}$}\\
            \either{(f_1\comp g)}{(f_2\comp g)}.
        \end{align*}
    \end{proof}

\begin{lemma}\label{lem:functoriality-of-either-left}
    $(g_1 + g_2) \comp \either{f_1}{f_2} \equiv \either{(g_1 \comp f_1)}{(g_2 \comp f_2)}$.
\end{lemma}
    \begin{proof}
        \begin{align*}
            (g_1 + g_2) \comp \either{f_1}{f_2} &= \why{definition of $\either{f_1}{f_2}$} \\ 
            (g_1 + g_2) \comp (f_1 + f_2) \comp \codiagterm \equalbecause{functoriality of $+$, \cref{ax:functoriality-of-type-constructors}} \\ 
            (g_1 \comp f_1) + (g_2 \comp f_2) \comp \codiagterm &= \why{definition of $\either{(g_1 \comp f_1)}{(g_2 \comp f_2)}$} \\ 
            \either{(g_1 \comp f_1)}{(g_2 \comp f_2)}
        \end{align*}
    \end{proof}

\begin{lemma}\label{lem:distributivity-of-either}
    $(\either {f_1}  {f_2}) \times f \equiv \distlterm\comp (\either {f_1 \times f} {f_2 \times f})$ and similarly for $\distrterm$.
\end{lemma}
\begin{proof} We only prove the first equation, the second one is similar.
Before proving the lemma, we prove the following observation about distributivity and the codiagonal.
    \begin{align*}
    \undistlterm\comp  \codiagterm \times \idterm &= \why{definition of $\undistlterm$}\\
\either {(\leftterm \times \idterm)} {(\rightterm \times \idterm)}\comp  \codiagterm \times \idterm \equalbecause{\cref{lem:functoriality-of-either}}\\
\either {((\leftterm\comp\codiagterm) \times \idterm)} {((\rightterm\comp\codiagterm) \times \idterm)} \equalbecause{\cref{ax:eval-codiag} and \cref{ax:functoriality-of-type-constructors}}\\ 
\either{\idterm} {\idterm}  &= \why{definition of copairing}\\ 
(\idterm + \idterm)\comp \codiagterm  \equalbecause{\cref{ax:functoriality-of-type-constructors}}\\
\codiagterm
    \end{align*}
Equipped with the above observation, we can now prove the lemma:
    \begin{align*}
    (\either{f_1}{f_2}) \times f  \equalbecause{isomorphism} \\
    \distlterm\comp \undistlterm\comp (\either{f_1}{f_2}) \times f &= \why{definition of $\either{ }{ }$}\\
        \distlterm\comp \undistlterm\comp ((f_1 + f_2)\comp\codiagterm) \times f \equalbecause{functoriality}\\
                \distlterm\comp \undistlterm\comp ((f_1 + f_2) \times f)\comp \codiagterm \times \idterm \equalbecause{naturality}\\
                \distlterm\comp (f_1 \times f) + (f_2 \times f)\comp \undistlterm\comp  \codiagterm \times \idterm \equalbecause{as proved above}\\
            \distlterm\comp (f_1 \times f) + (f_2 \times f)\comp \codiagterm &= \why{definition of copairing}\\ 
            \distlterm\comp (\either {f_1 \times f} {f_2 \times f})
            \end{align*}
\end{proof}
An immediate consequence of \cref{lem:distributivity-of-either} is the following lemma.
\begin{lemma}\label{lem:distributivity-of-concatenation}
        $(\either {f_1}  {f_2}) \fdot f \equiv \distlterm\comp (\either {f_1 \fdot f} {f_2 \fdot f})$ and similarly for $\distrterm$.
\end{lemma}
\begin{proof} We only prove the first equation, the second one is similar.
        \begin{align*}
    (\either{f_1}{f_2}) \fdot f  &= \why{definition of $\fdot$} \\
    (\either{f_1}{f_2}) \times f\comp \binconcatterm  \equalbecause{\cref{lem:distributivity-of-either}} \\
    \distlterm\comp (\either {f_1 \times f} {f_2 \times f})\comp \binconcatterm 
     \equalbecause{\cref{lem:functoriality-of-either}} \\
    \distlterm\comp (\either {f_1 \times f\comp\binconcatterm } {f_2 \times f\comp\binconcatterm})  
     &= \why{definition of $\fdot$} \\
    \distlterm\comp (\either {f_1 \fdot f} {f_2 \fdot f}).
    \end{align*}
\end{proof}

\subsection{Distributivity}\label{sec:distributivity}

In this section we describe basic properties of distributivity. Since the coproduct is associative and commutative, we can write a coproduct 
\begin{align*}
    \sum_{i \in I} A_i
\end{align*}
without specifying the order of the summands. Such a coproduct is formalised by giving some order on the summands and a bracketing, but all orders and bracketings admit \kl{isomorphisms} with each other. We will also talk about a coprojection
\begin{align*}
A_i 
\xrightarrow {\text{coprojection}}
\sum_{i \in I} A_i,
\end{align*}
which is formalised by using $\leftterm$ and $\rightterm$ a suitable number of times. Similarly, for terms $A_i \xrightarrow{f_i} B$ indexed by $i \in I$, we write
\begin{align*}
\sum_{i \in I} A_i \xrightarrow{\eithermany{i}{f_i}} B
\end{align*}
as an abbreviation for the $n$-fold \kl{copairing} $\eithermany{n}{f_{i_1} \cdots f_{i_n}}$ from \cref{sec:atomic-for-co-products}, where $i_1, \ldots, i_n$ is the order of the summands used to formalise $\sum_{i \in I} A_i$. When the index set is a product, such as $I \times J$, we also write $\eithermany{ij}{f_{ij}}$. Finally, we can extend distributivity and its inverse to arbitrary coproducts
    \[
\begin{tikzcd}
    [column sep=3cm]
\displaystyle{\left(\sum_{i \in I} A_i \right) \times \left(\sum_{j \in J} B_j \right) }
\ar[r,shift left=0.6ex,"\text{distribute}"]
&
\displaystyle{\sum_{(i,j) \in I \times J} A_i \times B_j}
\ar[l,shift left=0.6ex,"\text{undistribute}"],
\end{tikzcd}
\]
which is formalised by using $\distrterm$ and $\distlterm$ a suitable number of times. The following lemma states that relevant properties of distributivity hold provably. The proof is a straightforward induction on the size of the coproducts, using the axioms for distributivity and its inverse.

\begin{lemma}
    \label{lem:distributivity}
    The following diagram commutes provably
    \[
\begin{tikzcd}
    [column sep=3cm]
A_i \times B_j
\ar[r,"{\text{coprojection} \times \text{coprojection}}"]
\ar[dr,"\text{coprojection}"']
&
\displaystyle{\left(\sum_{i \in I} A_i \right) \times \left(\sum_{j \in J} B_j \right) }
\ar[d,"\text{distribute}"]
\\
&
\displaystyle{\sum_{(i,j) \in I \times J} A_i \times B_j}
\end{tikzcd}
\]
\end{lemma}
\begin{proof}
    We first show the special case of the lemma where one index set has two elements and the other index set has one element. By repeatedly applying this special case, we get the general case. The diagram for the special case is as follows:
 \[
\begin{tikzcd}
A_1 \times B
\ar[r,"\leftterm \times \idterm"]
\ar[dr,"\leftterm"']
 & (A_1 + A_2) \times B 
\ar[d,"\distlterm"]
 \\
& (A_1 \times B) + (A_2 \times B)
\end{tikzcd}
\]
By the isomorphism axioms, it is enough to show that the diagram commutes after postcomposing with the inverse of distributivity, which gives the following diagram:
 \[
\begin{tikzcd}
A_1 \times B
\ar[r,"\leftterm \times \idterm"]
\ar[dr,"\leftterm"']
 & (A_1 + A_2) \times B 
 \\
& (A_1 \times B) + (A_2 \times B)
\ar[u,"\undistlterm"']
\end{tikzcd}
\]
The inverse of distributivity is defined by cases, and hence we can apply the reduction from \cref{lem:case-equivalence} to see that the diagram commutes.
\end{proof}

Another fact is the $\distlterm\text{-}\fstterm$ coherence law.

\begin{lemma}\label{lem:distl-fst-coherence}
    $\distlterm \comp (\fstterm + \fstterm) \equiv \fstterm$.
\end{lemma}
\begin{proof}
    By \cref{ax:isomorphisms}, it suffices to show that $(\fstterm + \fstterm) \equiv \undistlterm \comp \fstterm$:
    \begin{align*}
        \undistlterm \comp \fstterm &= \why{definition of $\undistlterm$} \\
        \either{(\leftterm \times \idterm)}{(\rightterm \times \idterm)} \comp \fstterm \equalbecause{\cref{lem:functoriality-of-either}} \\
        \either{\left( \leftterm \times \idterm \right) \comp \fstterm}{\left( \rightterm \times \idterm \right) \comp \fstterm} \equalbecause{naturality of $\fstterm$} \\
        \either{(\fstterm \comp \leftterm)}{(\fstterm \comp \rightterm)} \equalbecause{\cref{lem:functoriality-of-either-left}} \\
        (\fstterm + \fstterm) \comp \either{\leftterm}{\rightterm} \equalbecause{\cref{lem:either-uniqueness}} \\
        (\fstterm + \fstterm).
    \end{align*}
\end{proof}

\subsection{List Facts}\label{sec:basic-facts-about-lists}

In this section we record two basic facts about $\binconcatterm$ and the empty list, which are used when normalising drop terms with list codomain. Recall that $1$ denotes the unique up to equivalence term into $\onetype$ from any domain (\cref{ax:unit-type}), and $[] \eqdef \leftterm\comp\consterm$.

\begin{lemma}\label{lem:binconcat-empty-is-fst}
    $(\idterm \times []) \comp \binconcatterm \equiv \fstterm$.
\end{lemma}
\begin{proof}
    \begin{align*}
        (\idterm \times []) \comp \binconcatterm \equalbecause{\cref{ax:associativity-of-composition}} \\
        \idterm \comp (\idterm \times []) \comp \binconcatterm \equalbecause{$\fstterm$, $\appendoneterm$ mutual inverses, \cref{ax:isomorphisms}} \\
        \fstterm \comp \appendoneterm \comp (\idterm \times []) \comp \binconcatterm \equalbecause{\cref{ax:concatenation-monoid}} \\
        \fstterm \comp \idterm \equalbecause{\cref{ax:associativity-of-composition}} \\
        \fstterm.
    \end{align*}
\end{proof}

\begin{lemma}\label{lem:product-with-empty-list-fst-binconcat}
    For every $f$, $f \times (1 \comp []) \comp \fstterm \equiv f \times (1 \comp []) \comp \binconcatterm$.
\end{lemma}
\begin{proof}
    We compute both sides and show that they are equal to $\fstterm \comp f$. For the left-hand side, this follows directly from naturality of $\fstterm$:
    \begin{align*}
        f \times (1 \comp []) \comp \fstterm \equalbecause{naturality of $\fstterm$} \\
        \fstterm \comp f.
    \end{align*}
    For the right-hand side, we first factor the product using functoriality (\cref{ax:functoriality-of-type-constructors}) and \cref{ax:associativity-of-composition}:
    \begin{align*}
        f \times (1 \comp []) \comp \binconcatterm \equalbecause{\cref{ax:functoriality-of-type-constructors}} \\
        (f \times 1) \comp (\idterm \times []) \comp \binconcatterm \equalbecause{\cref{lem:binconcat-empty-is-fst}} \\
        (f \times 1) \comp \fstterm \equalbecause{naturality of $\fstterm$} \\
        \fstterm \comp f.
    \end{align*}
    Both sides equal $\fstterm \comp f$, which proves the lemma.
\end{proof}

Next, we prove that discarding a pair of inputs, turning them into empty lists and concatenating them, is equivalent to discarding the inputs and returning an empty list.

\begin{lemma}\label{lem:discarding-pair-to-empty-list}
    $(1 \comp []) \times (1 \comp []) \comp \binconcatterm \equiv 1 \comp []$.
\end{lemma} 
\begin{proof}
    \begin{align*}
        (1 \comp []) \times (1 \comp []) \comp \binconcatterm \equalbecause{\cref{lem:product-with-empty-list-fst-binconcat}} \\
        (1 \comp []) \times (1 \comp []) \comp \fstterm \equalbecause{naturality of $\fstterm$, \cref{ax:naturality}} \\
        \fstterm \comp (1 \comp []) \equalbecause{$\onetype$ is terminal, \cref{ax:unit-type}} \\
        1 \comp [].
    \end{align*}
\end{proof}

Recall that $\prependterm \eqdef \rightterm \comp \consterm$ and $[] \eqdef \leftterm \comp \consterm$ are both built from the \kl{atomic term} $\consterm$, and are therefore natural in the following sense.

\begin{lemma}\label{lem:prepend-and-empty-naturality}
    $\prependterm \comp h^* \equiv (h \times h^*) \comp \prependterm$ and $[] \comp h^* \equiv []$.
\end{lemma}
\begin{proof}
    Both equations follow from naturality of $\consterm$ (\cref{ax:naturality}), i.e.~$(\idterm + (h \times h^*)) \comp \consterm \equiv \consterm \comp h^*$, precomposed with $\rightterm$ and $\leftterm$ respectively. For $\prependterm$:
    \begin{align*}
        \prependterm \comp h^* &= \why{definition of $\prependterm$} \\
        \rightterm \comp \consterm \comp h^* \equalbecause{naturality of $\consterm$, \cref{ax:naturality}} \\
        \rightterm \comp (\idterm + (h \times h^*)) \comp \consterm \equalbecause{naturality of $\rightterm$} \\
        (h \times h^*) \comp \rightterm \comp \consterm &= \why{definition of $\prependterm$} \\
        (h \times h^*) \comp \prependterm.
    \end{align*}
    For $[]$:
    \begin{align*}
        [] \comp h^* &= \why{definition of $[]$} \\
        \leftterm \comp \consterm \comp h^* \equalbecause{naturality of $\consterm$, \cref{ax:naturality}} \\
        \leftterm \comp (\idterm + (h \times h^*)) \comp \consterm \equalbecause{naturality of $\leftterm$} \\
        \idterm \comp \leftterm \comp \consterm \equalbecause{\cref{ax:associativity-of-composition}} \\
        \leftterm \comp \consterm &= \why{definition of $[]$} \\
        [].
    \end{align*}
\end{proof}

The following lemma states naturality of $\pureterm$, i.e.~that it is a natural transformation from the identity functor to the list functor.

\begin{lemma}\label{lem:pure-naturality}
    $\pureterm \comp g^* \equiv g \comp \pureterm$.
\end{lemma}
\begin{proof}
    Recall $\pureterm \eqdef \appendoneterm \comp (\idterm \times []) \comp \prependterm$. We compute:
    \begin{align*}
        \pureterm \comp g^* &= \why{definition of $\pureterm$} \\
        \appendoneterm \comp (\idterm \times []) \comp \prependterm \comp g^* \equalbecause{\cref{lem:prepend-and-empty-naturality}} \\
        \appendoneterm \comp (\idterm \times []) \comp (g \times g^*) \comp \prependterm \equalbecause{functoriality of $\times$, \cref{ax:functoriality-of-type-constructors}} \\
        \appendoneterm \comp (g \times ([] \comp g^*)) \comp \prependterm \equalbecause{\cref{lem:prepend-and-empty-naturality}} \\
        \appendoneterm \comp (g \times []) \comp \prependterm \equalbecause{functoriality of $\times$, \cref{ax:functoriality-of-type-constructors}} \\
        \appendoneterm \comp (g \times \idterm) \comp (\idterm \times []) \comp \prependterm \equalbecause{naturality of $\appendoneterm$, \cref{ax:naturality}} \\
        g \comp \appendoneterm \comp (\idterm \times []) \comp \prependterm &= \why{definition of $\pureterm$} \\
        g \comp \pureterm.
    \end{align*}
\end{proof}

The following lemma states that \kl{concatenation maps} are functorial.

\begin{lemma}\label{lem:functoriality-of-concat-maps}
    Suppose that $\vec f$ is a vector of terms with codomain $A^*$ and $g : A \to B$. Then 
    \[
        \left( F(\vec f) \comp \concmap \right) \comp g^* \equiv F(\vec f \comp g^*) \comp \concmap.
    \]
\end{lemma}
\begin{proof}
    The proof proceeds by case analysis on the shape of functor $F$.
    \begin{description}
        \item[Unit node, $F = \onetype$.] In this case, we have $\left( F(\vec f) \comp \concmap \right) \equiv []$ and $[] \comp g^* \equiv []$ by \cref{lem:prepend-and-empty-naturality}.

        \item[Leaf node, $F = X_i$.] In this case, we have $\left( F(\vec f) \comp \concmap \right) \equiv f_i$ and there is nothing to show. 
        
        \item[Sum node, $F = F_1 + F_2$.] In this case, we have 
        \begin{align*}
            \left( F(\vec f) \comp \concmap \right) \comp g^* = \either{\left( F_1(\vec f) \comp \concmap \right)}{\left( F_2(\vec f) \comp \concmap \right)} \comp g^* \equalbecause{\cref{lem:functoriality-of-either}} \\ 
            \either{\left( (F_1(\vec f) \comp \concmap) \comp g^* \right)}{\left( (F_2(\vec f) \comp \concmap) \comp g^*\right)} \equalbecause{induction} \\
            \either{\left( F_1(\vec f \comp g^*) \comp \concmap \right)}{\left( F_2(\vec f \comp g^*) \comp \concmap \right)}.
        \end{align*}

        \item[Product node, $F = F_1 \times F_2$.] In this case, we have 
        \begin{align*}
            \left( F(\vec f) \comp \concmap \right) \comp g^* = \left( (F_1(\vec f) \comp \concmap) \odot (F_2(\vec f) \comp \concmap) \right) \comp g^* \equalbecause{naturality of $\binconcatterm$, \cref{ax:naturality}} \\ 
            \left( (F_1(\vec f) \comp \concmap) \comp g^* \right) \odot \left( (F_2(\vec f) \comp \concmap) \comp g^* \right) \equalbecause{induction} \\
            \left( F_1(\vec f \comp g^*) \comp \concmap \right) \odot \left( F_2(\vec f \comp g^*) \comp \concmap \right).
        \end{align*}

        \item[List node, $F = G^*$.] In this case, we have 
        \begin{align*}
            \left( F(\vec f) \comp \concmap \right) \comp g^* = \left( \hom \left( (G(\vec f) \comp \concmap) \right) \right) \comp g^* \equalbecause{naturality of $\concatterm$, \cref{ax:naturality}} \\
            \hom \left( \left( (G(\vec f) \comp \concmap) \comp g^* \right) \right) \equalbecause{induction} \\
            \hom \left( G(\vec f \comp g^*) \comp \concmap \right).
        \end{align*}
    \end{description}
\end{proof}

The next lemma is similar to \cref{lem:functoriality-of-concat-maps}, but for postcomposition with $\concatterm$.

\begin{lemma}\label{lem:concat-map-concat}
    Suppose that $\vec f$ is a vector of terms with codomain $A^{**}$. Then
    \[
        \left( F(\vec f) \comp \concmap \right) \comp \concatterm \equiv F(\vec f \comp \concatterm) \comp \concmap.
    \]
\end{lemma}
\begin{proof}
    The proof proceeds by case analysis on the shape of functor $F$.
    \begin{description}
        \item[Unit node, $F = \onetype$.] In this case, we have $\left( F(\vec f) \comp \concmap \right) \equiv []$ and $[] \comp \concatterm \equiv []$ by \cref{ax:concat-monoid-homomorphism}.

        \item[Leaf node, $F = X_i$.] In this case, we have $\left( F(\vec f) \comp \concmap \right) \equiv f_i$ and there is nothing to show. 
        
        \item[Sum node, $F = F_1 + F_2$.] In this case, we have 
        \begin{align*}
            \left( F(\vec f) \comp \concmap \right) \comp \concatterm &= \why{definition of $\concmap$} \\ 
            \either{\left( F_1(\vec f) \comp \concmap \right)}{\left( F_2(\vec f) \comp \concmap \right)} \comp \concatterm \equalbecause{\cref{lem:functoriality-of-either}} \\ 
            \either{\left( (F_1(\vec f) \comp \concmap) \comp \concatterm \right)}{\left( (F_2(\vec f) \comp \concmap) \comp \concatterm\right)} \equalbecause{induction} \\
            \either{\left( F_1(\vec f \comp \concatterm) \comp \concmap \right)}{\left( F_2(\vec f \comp \concatterm) \comp \concmap \right)}.
        \end{align*}

        \item[Product node, $F = F_1 \times F_2$.] In this case, we have 
        \begin{align*}
            \left( F(\vec f) \comp \concmap \right) \comp \concatterm &= \why{definition of $\concmap$} \\ 
            \left( (F_1(\vec f) \comp \concmap) \odot (F_2(\vec f) \comp \concmap) \right) \comp \concatterm \equalbecause{\cref{ax:concat-monoid-homomorphism}} \\ 
            \left( (F_1(\vec f) \comp \concmap) \comp \concatterm \right) \odot \left( (F_2(\vec f) \comp \concmap) \comp \concatterm \right) \equalbecause{induction} \\
            \left( F_1(\vec f \comp \concatterm) \comp \concmap \right) \odot \left( F_2(\vec f \comp \concatterm) \comp \concmap \right).
        \end{align*}

        \item[List node, $F = G^*$.] In this case, we have 
        \begin{align*}
            \left( F(\vec f) \comp \concmap \right) \comp \concatterm &= \why{definition of $\concmap$} \\ 
            \hom \left( (G(\vec f) \comp \concmap) \right) \comp \concatterm &= \why{definition of $\hom$} \\ 
            \left( (G(\vec f) \comp \concmap) \right)^* \comp \concatterm \comp \concatterm \equalbecause{\cref{ax:concatenation-unit}} \\
            \left( (G(\vec f) \comp \concmap) \right)^* \comp \concatterm^* \comp \concatterm \equalbecause{functoriality of $*$, \cref{ax:functoriality-of-type-constructors}} \\ 
            \left( \left( G(\vec f) \comp \concmap \right) \comp \concatterm \right)^* \comp \concatterm \equalbecause{induction} \\ 
            \left( G(\vec f \comp \concatterm) \comp \concmap \right)^* \comp \concatterm &= \why{definition of $\hom$} \\
            \hom \left( G(\vec f \comp \concatterm) \comp \concmap \right).
        \end{align*}
    \end{description}
\end{proof}

\section{Rebracketings}
\label{sec:rebracketings}
Many terms in our system do nothing except for rebracketing the input. A prominent example is
\begin{align*}
(A \times B)^* \times A 
\xrightarrow{\shiftrterm}
A \times (B \times A)^*.
\end{align*}
The following definition formalises this concept, by saying that the map commutes with \kl{multivariate concatenation maps} (see \cref{sec:concatenation-maps}).
\begin{definition}
    [Rebracketing] Consider two \kl{multivariate list-polynomial functors} $F$ and $G$. A \intro{rebracketing} is any term $f$ for which  the following diagram commutes provably:
\[
\begin{tikzcd}
F(A_1,\ldots,A_n) 
\ar[rr,"f"] 
\ar[dr,\concar]
& 
& 
G(A_1,\ldots,A_n)
\ar[dl,\concar]
\\
&
(A_1 + \cdots + A_n)^*
\end{tikzcd}
\]
\end{definition}
The following lemma collects some basic facts about rebracketings. The proof is straightforward, and is left as an exercise for the reader.
\begin{lemma}\label{lem:rebracketing-basic-facts}
    All \kl{atomic terms} are \kl{rebracketings}, except for the projections and the \kl{group prefix multiplication} $\prefterm$. Furthermore, \kl{rebracketings} are closed under composition, and applying the \kl{type} constructors.
\end{lemma}
We now present a very useful result, which will allow us to prove equivalence of rebracketings without going into details. The idea is that any two rebracketings of the same type must be provably equivalent, as  long as the output type has a certain  injectivity property, as defined below.
\begin{definition}[Injective type]
    A \kl{type} $F(A_1,\ldots,A_n)$ is called \intro[injective type]{injective} if the \kl{concatenation map}
        \begin{align*}
    F(A_1,\ldots,A_n) \concmap (A_1 + \cdots + A_n)^*
    \end{align*}
    is \kl{injective}.
\end{definition}
An example of an injective type is $(A \times B)^*$, and a nonexample is $A^{**}$. In the latter case, noninjectivity arises because the bracketing cannot be recovered after the list has been flattened. The following lemma is an immediate consequence of the definitions. 
\begin{lemma}[Uniqueness of rebracketings]\label{lem:uniqueness-of-rebracketings}
If
    \begin{align*}
    F(A_1,\ldots,A_n) \xrightarrow{f,g} G(A_1,\ldots,A_n)
    \end{align*}
    are \kl{rebracketings} and the output \kl{type} is \kl[injective type]{injective}, then $f \equiv g$.
\end{lemma}

Let us observe some consequences of the above lemma. 
\begin{lemma}\label{lem:shift-ones}
The following diagram commutes provably
\begin{align*}
\begin{tikzcd}[ampersand replacement=\&]
 \& A^*
    \ar[dl, "{\appendoneterm^*\comp \prependoneterm}"']
    \ar[dr, "{\prependoneterm^*\comp \appendoneterm}"]
    \\
    \onetype \times (A \times\onetype)^* 
    \ar[rr, "\shiftlterm" ]
    \&
    \&
    (\onetype \times A)^* \times \onetype   
\end{tikzcd}
\end{align*}
\end{lemma}
\begin{proof}
        All arrows in the diagram are easily seen to be rebracketings, which can be proved using \cref{lem:rebracketing-basic-facts}. 
 Furthermore, all types in the diagram, in particular the output type  $(\onetype \times A)^* \times \onetype$ are injective. For the output type, the  one-sided inverse of its concatenation map is  the south-east arrow in the above diagram. 
     Therefore, we can apply \cref{lem:uniqueness-of-rebracketings} to conclude that the diagram commutes provably. 
\end{proof}

Another consequence of \cref{lem:uniqueness-of-rebracketings} concerns a symmetric variant of blocking.

\begin{lemma}\label{lem:shift-unsplit}
    The  term
    \begin{align*}
(A+B)^* \xrightarrow{\blocklterm} (A^* \times B)^* \times A^* 
\quad \eqdef \quad 
\blockterm\comp \shiftlterm
\end{align*}
is an \kl{isomorphism}, and its inverse is the \kl{concatenation map}.
\end{lemma}
\begin{proof}
    All terms in the statement of the lemma are easily seen to be rebracketings, which can be proved using \cref{lem:rebracketing-basic-facts}. Therefore, to prove the lemma, it is enough to show that the output type is injective. In fact, the output type is isomorphic to $(A+B)^*$, by composing $\shiftrterm$ with $\unblockterm$. 
    This isomorphism is the same as the concatenation map, which follows from \cref{ax:shift-concatenation}.
\end{proof}

\section{Proof of Normalisation of Values}
\label{sec:proof-of-normalisation-of-values}
In this section, we prove \cref{lem:value-completeness}, which states that every \kl{value} is provably equivalent to a value in \kl[normalised drop term]{normal form}. By induction on the structure of a term $f$, we show that if we take a composition
\begin{align*}
\onetype \xrightarrow v A \xrightarrow f B
\end{align*}
of a value $v$ in normal form with a term $f$, then this composition is provably equivalent to a value in normal form. This will imply the lemma, by taking the special case where $v$ is the identity $\onetype  \to \onetype $.

We begin with the induction step, in which the term $f$ in the application $v\comp{}f$ is constructed using one of the four combinators. 
\begin{description}
    \item[Composition.] Suppose that the term $f$ being applied is  a composition $f_1\comp{}f_2$. Applying the induction hypothesis for $f_1$ allows us to normalise $v\comp{}f_1$, and applying the induction hypothesis for $f_2$ allows us to normalise $(v\comp{}f_1)\comp{}f_2$, which is equivalent to $v\comp(f_1\comp{}f_2)$ by the associativity of composition.
    \item[Product combinator.]      Suppose that the term applied is $f = f_1 \times f_2$.  The argument must be of the form $v = (v_1,v_2)$. We have
    \begin{align*}
    (v_1,v_2)\comp(f_1 \times f_2)&= \why{definition of $(v_1,v_2)$} \\
        \appendoneterm\comp(v_1 \times v_2)\comp(f_1 \times f_2)\equalbecause{functoriality of $\times$}\\
    \appendoneterm\comp((v_1\comp{}f_1) \times (v_2\comp{}f_2))\equalbecause{induction hypothesis} \\
        \text{a value in normal form}
    \end{align*}
    \item[Coproduct combinator.] Suppose that the term being applied is $f = f_1 + f_2$. The value to   which $f$ is applied  must be either $v_1 \comp \leftterm$ or $v_2\comp \rightterm$.  Let us treat only the first case, since the second case is similar:
    \begin{align*}
    v_1 \comp \leftterm\comp (f_1 + f_2)\equalbecause{naturality} \\
    v_1 \comp f_1\comp \leftterm  \equalbecause{induction hypothesis}\\
    \text{a value in normal form}.
    \end{align*}

    \item[Map combinator.] Suppose that the applied term $f$ is of the form $g^*$. The argument must be a list of values. We have 
    \begin{align*}
    ([v_1] \wdot \cdots \wdot [v_n])\comp{}g^* \equalbecause{naturality} \\
    ([v_1\comp{}g]) \wdot \cdots \wdot ([v_n\comp{}g]) \equalbecause{induction hypothesis} \\
    \text{a value in normal form}.
    \end{align*}
   This completes the proof of the induction step for all combinators. 
\end{description}

We now prove the induction basis, in which the applied term $f$ is one of the \kl{atomic terms}.

\begin{description}
    \item[Append and prepend one.] Suppose that the applied term $f$ is one of the two terms $\appendoneterm$ or $\prependoneterm$. The result is equivalent to $(v,1)$ or $(1,v)$, respectively, by definition. 
    \item[Coprojections.] Suppose that the applied term $f$ is a coprojection $\leftterm$ or $\rightterm$. In this case, there is nothing to do, since one of the ways of constructing a value in normal form is to postcompose it with a coprojection.
    \item[Projections.] Suppose that the applied term $f$ is a projection $\fstterm$ or $\sndterm$. By symmetry, we only consider the first projection $\fstterm$. By definition of normal forms, the value $v$ to which $f$ is applied is a pair, and hence it must be of the form $(v_1,v_2)$ for some values $v_1$ and $v_2$.
\begin{align*}
    (v_1,v_2)\comp\fstterm&= \why{definition of $(v_1,v_2)$}\\
\appendoneterm\comp(v_1 \times v_2)\comp\fstterm    \equalbecause{naturality} \\
    \appendoneterm\comp\fstterm\comp{}v_1\equalbecause{isomorphism axiom for $\appendoneterm$} \\
    v_1.
\end{align*}
    \item[Codiagonal.] Suppose that the term $f$ which is applied to the value is  the codiagonal $\codiagterm : A + A \to A$. In this case the value to which $f$ is applied must have coproduct output \kl{type}, and hence it must be  equal to either $v\comp \leftterm$ or $v\comp \rightterm$ for some value $v$. In either of these cases, the composition is provably equivalent to $v$, which is proved  using  \cref{ax:eval-codiag}.

    \item[Distributivity.] Suppose now that the term $f$ is one of the two  distributivity terms. By symmetry, we consider  right distributivity
    \begin{align*}
    \distrterm : A \times (B_1 + B_2) \to (A \times B_1) + (A \times B_2).
    \end{align*}
    The value  $v$ to which $f$ is applied  must have one of two shapes
    \begin{align*}
   v=  (w, (w_1\comp\leftterm)) \qquad \text{or} \qquad v= (w, (w_2\comp\rightterm)),
    \end{align*}
    for some values $w$, $w_1$, and $w_2$. 
    By symmetry, we only treat the first shape.  We will  show
    \begin{align*}
    (w, (w_1\comp\leftterm))\comp \distrterm  \equiv  (w,w_1)\comp \leftterm,
    \end{align*}
    which will complete the proof of this case, since the right-hand side is a value in normal form.
    To prove the above equation, we use an isomorphism trick: since $\undistrterm$ is an \kl{isomorphism} by the isomorphism axiom, in order to prove the above equation, it suffices to prove the variant of the equation in which both sides are postcomposed with $\undistrterm$. This is done below:
    \begin{align*}
        (w,w_1)\comp \leftterm\comp \undistrterm &= \why{definition of $\undistrterm$} \\
    (w,w_1)\comp \leftterm\comp \either{\idterm \times \leftterm}{\idterm \times \rightterm} \equalbecause{see \cref{lem:case-equivalence}}\\ 
    (w,w_1)\comp (\idterm \times \leftterm) \equalbecause{functoriality} \\
    (w, (w_1\comp\leftterm))
    \end{align*}

    \item[List constructors.] Suppose now that the applied term $f$ is one of  the list constructors $\consterm$ or $\snocterm$. By symmetry, we only treat the case of $\consterm$.  The value $v$ to which $f$ is applied must have type $\onetype  \to \onetype  + A \times A^*$, and hence it  must have one of two shapes:
    \begin{align*}
        v = \leftterm \qquad \text{or} \qquad  v = (u,w)\comp \rightterm
    \end{align*}
    for some values $u$ and $w$ of types $\onetype \to A$ and $\onetype \to A^*$ respectively. In the first case, $v\comp{}f$, which is $\leftterm\comp \consterm$, is the definition of $[]$, and hence in  normal form. In the second case,   $v\comp{}f$ which is $(u,w)\comp \prependterm$, is equal to $[u] \wdot w$ thanks to \cref{ax:from-monoid-to-inductive}.
    \item[List destructors.] Suppose that the applied term $f$ is one of the list destructors $\unsnocterm$ or $\unconsterm$. By symmetry, we only treat the case of $\unconsterm$. The value $v$ to which $f$ is applied must have type $\onetype \to A^*$. By definition of values in normal form and \cref{lem:concatenation-monoidality}, the value $v$  must be equivalent to  either $[]$ or $[u] \wdot w$ for some values $u$ and $w$ of types $\onetype \to A$ and $\onetype \to A^*$, respectively. We want to show 
    \begin{align*}
    []\comp \unconsterm \equiv \leftterm  \qquad \text{and} \qquad ([u] \wdot w)\comp\unconsterm \equiv (u,w)\comp \rightterm.
    \end{align*}
    We use the isomorphism trick, by proving new equations obtained from the above ones where both sides are postcomposed with $\consterm$. After this postcomposition, the equations become
    \begin{align*}
    [] \equiv \leftterm\comp \consterm \qquad \text{and} \qquad ([u] \wdot w) \equiv (u,w)\comp \prependterm,
    \end{align*}
    which are exactly the equations that we proved in the previous item. 
    \item[List concatenation.] Assume that the term $f$ is list concatenation $\concatterm$. By  \cref{ax:concat-monoid-homomorphism}, this operation is a monoid homomorphism, and hence we know that 
     \begin{align*}
        (v \wdot w)\comp \concatterm \equiv (v\comp \concatterm) \wdot (w\comp \concatterm).
    \end{align*}
    Therefore, normalisation is proved by a simple induction on  the length of the list represented by $v$. 

    \item[Block.] Suppose now that the applied term $f$ is $\blockterm$.  By the isomorphism trick, it is enough to show that
    \begin{align*}
    v \equiv w\comp \unblockterm \quad\text{for some value $w$ in normal form}.
    \end{align*}
     From this it will follow that $v\comp\blockterm \equiv w$, as required in the induction step. 
    
    To identify the value $w$, we use a syntactic analysis of the value $v$. 
    The value $v$, like any normal form value in type $\onetype \to (A+B)^*$, is obtained by using $\wdot$ to concatenate a list of values, each one having shape 
    \begin{align*}
    \myunderbrace{[x\comp\leftterm] \quad \text{for some value }\onetype \xrightarrow x A}{value of first kind}
    \qquad \text{or} \qquad
    \myunderbrace{[y\comp\rightterm] \quad \text{for some value }\onetype \xrightarrow y B}{value of second kind}.
    \end{align*}
    By identifying the occurrences of values of the second kind, we can decompose $v$ as 
    \begin{align*}
    v \equiv v_0 \wdot [y_1\comp\rightterm] \wdot v_1  \wdot \cdots \wdot [y_n\comp\rightterm] \wdot v_n,
    \end{align*}
    where each $[y_i\comp\rightterm]$ is a value of the second kind, and each $v_i$ is a concatenation of   values of the first kind. This means that each $v_i$ can be obtained as 
    \begin{align*}
    v_i \equiv w_i\comp(\leftterm^*).
    \end{align*}
    This is because of the following property: 
    \begin{align*}
    \binconcatterm\comp f^* \equiv (f^* \times f^*)\comp \binconcatterm,
    \end{align*}
    which says that $\binconcatterm$ is a natural transformation, which in turn holds because it is built by using natural transformations. Summing up, we have shown that 
    \begin{align*}
    v \equiv (w_0\comp\leftterm^*) \wdot ([y_1\comp\rightterm] \wdot (w_1\comp\leftterm^*)) \wdot \cdots \wdot ([y_n\comp\rightterm] \wdot (w_n\comp\leftterm^*)).
    \end{align*}
    We claim that the right-hand side of the above equation is exactly the result of applying $\unblockterm$ to the following value in normal form 
    \begin{align*}
    w \eqdef (w_0, [(y_1,w_1)] \wdot \cdots \wdot [(y_n,w_n)]).
    \end{align*}
    This claim will complete the  proof of the induction step in this case. To prove the claim, we observe that  $\unblockterm$ is a term that is constructed by using the combinators from atomic terms that we have already treated in the previous items (in particular, $\unblockterm$ does not use $\blockterm$). Therefore, by the already proved part of the argument, we know that $w\comp \unblockterm \equiv v$, as required in the claim.

    \item[Prefix multiplication.] By the isomorphism trick, we need to show that if $v$ is a value in normal form of type $\onetype \to (A \times G)^*$ for some type $A$ and finite group $G$, then 
    \begin{align*}
    v  \equiv w\comp \unprefterm \quad \text{for some value $w$ in normal form.}
    \end{align*}
    We proceed as in the previous item. A  candidate for the value $w$ can be easily found by reading the value $v$, and then the above equivalence can be proved by using the part of the argument that we have already treated, since $\unprefterm$ is a term that does not use the \kl{group prefix multiplication}.
\end{description}

\section{Injective String Representations}
\label{sec:provably-injective-string-representations}
In this part of the appendix, we prove \cref{lem:provably-injective-string-representation}, which states that for every \kl{type} there is an \kl{injective} term from this type into lists over the \kl{binary alphabet}. 

The construction follows two steps, first by encoding any type in a \kl{finite alphabet}. Lists over a finite alphabet are then encoded into the binary alphabet.

\begin{claim}\label{lem:provably-injective-string-representation-finite-alphabet}
    For every \kl{type} $A$ there is a \kl{finite alphabet} $N$ and an \kl{injective} \kl{aperiodic term}  $f_A : A \to N^*$.
\end{claim}

\begin{claim}\label{lem:provably-injective-string-representation-finite-alphabet-into-binary}
    For every \kl{finite alphabet} $N$ there is an \kl{injective} \kl{aperiodic term}  $\stringrep N : N^* \to {\twotype}^*$.
\end{claim}

The terms obtained from \cref{lem:provably-injective-string-representation-finite-alphabet} and \cref{lem:provably-injective-string-representation-finite-alphabet-into-binary} can be composed, giving an encoding $\stringrep A$ of arbitrary type $A$ into the lists over the binary alphabet, as required by \cref{lem:provably-injective-string-representation}. This composition is injective, as discussed in \cref{claim:injective-terms-basic-facts}.

\subsection{Injective Terms}

    We begin by establishing some basic facts about \kl{injective} terms. 
    \begin{claim}\label{claim:injective-terms-basic-facts}
        The coprojections are \kl{injective}. Also, \kl{injective} terms are closed under composition and applying the functorial combinators for $+$, $\times$ and $*$. 
    \end{claim}
    \begin{proof}
        We only prove the first statement about coprojections, since the reader can easily verify the closure under combinators -- it is enough to combine the inverses. By \cref{lem:case-equivalence}, a one-sided inverse of the coprojection $A \to A+B$ is   
        \begin{align*}
        \either{\idterm}{g} \qquad \text{where $B \xrightarrow g A$ is any term}.
        \end{align*}
        The term $g$ must exist, since the type $B \to A$, like every \kl{type}, is inhabited by some term, which can be shown by composing $B \to \onetype$ with some \kl{value} of type $\onetype \to A$. (One should also construct a term of type $B \to \onetype$ for every $B$, but this is easily done by induction on the structure of the input type $B$.)
    \end{proof}

\subsection{Proof of \cref{lem:provably-injective-string-representation-finite-alphabet}}

    Equipped with the above claim, we prove \cref{lem:provably-injective-string-representation-finite-alphabet} by induction on the structure of the input \kl{type} $A$.
    \begin{description}
    \item[Base case, $A = \onetype$.] We can use the alphabet $\underline \onetype = \onetype$ and as the string representation we can use any  term of type $\onetype \to \onetype^*$, since all terms with domain $\onetype$ are \kl{injective}, which follows from \cref{ax:unit-type}.

    \item[Induction step for products, $A \times B$.] We define 
    \begin{align*}
    \underline{A \times B} \eqdef \underline A + \underline B.
    \end{align*}
    Since injective  terms are closed under composition, we define the representation as a composition of the following injective  terms:
    \[
    \begin{tikzcd}
    A \times B 
    \ar[d,  "{\stringrep A \times \stringrep B}"] \\
    \underline A^* \times \underline B^* 
    \ar[d,
    "{\idterm \times ((\idterm,[]))^*}"]\\
    \underline A^* \times (\underline B \times \underline A^*)^* 
    \ar[d,"\unblockterm"] \\
    (\underline A + \underline B)^*.
    \end{tikzcd}
    \]
    The first one is injective by the induction hypothesis, and because injective terms are closed under the product combinator. The last one is even an \kl{isomorphism}. The one-sided inverse of the middle one is left as an easy exercise for the reader. 

    \item[Induction step for lists, $A^*$.] We define 
    \begin{align*}
    \underline{A^*} \eqdef \underline A + \onetype.
    \end{align*}
    The representation is defined to be the following composition of injective  terms:
    \[
    \begin{tikzcd}
    A^*
    \ar[d, "{(\stringrep A)^*}"] \\
    \underline A^{**} \ar[d, "{([] , ((\onelement,\idterm))^*)}"] \\
    \underline A^* \times (\onetype \times \underline A^*)^* \ar[d,"\unblockterm"] \\
    (\underline A + \onetype)^*.
    \end{tikzcd}
    \]
    The first one is injective by the induction hypothesis, and because injective terms are closed under the map combinator. Injectivity of the middle one is left as an easy exercise for the reader. As previously, the  last one is  an isomorphism.
    \item[Induction step for coproducts, $A+B$.] This is the most challenging case. Define
\begin{align*}
    \underline{A+B} \eqdef    \underline A + \underline B + \onetype.
\end{align*}
We begin with the composition of the following terms, which is injective by Claim~\ref{claim:injective-terms-basic-facts}: 
\[
\begin{tikzcd}
A + B  
\ar[d,"\stringrep A + \stringrep B"] \\
\underline A^* + \underline B^*  
\ar[d, "(\leftterm)^* + (\rightterm)^*"] \\
(\underline A + \underline B)^* + (\underline A + \underline B)^*.
\end{tikzcd}
\]
To complete the proof, we establish an injective  term of type 
\begin{align*}
 C^* + C^* \to  (\onetype + C)^* \qquad \text{where }C = \underline A + \underline B.
\end{align*}
This is done by using the following composition of injective  terms:
        \[
        \begin{tikzcd}
        C^* + C^* 
        \ar[d,"{\idterm + (\onelement,\idterm)}"] \\
        C^* + \onetype \times C^*
        \ar[d, "{\idterm + ([], (\onelement, \idterm))}"] \\
        C^* + C^* \times (\onetype \times C^*)
        \ar[d, 
        "{\either{(\idterm , [])}{\idterm \times [\idterm]}}"] \\
         C^* \times (\onetype \times C^*)^*
         \ar[d,"\unblockterm"] \\
         (C + \onetype)^*.
        \end{tikzcd}
        \]
    \end{description}

\subsection{Proof of \cref{lem:provably-injective-string-representation-finite-alphabet-into-binary}}

We start with constructing an \kl{injective} term $f : N \to \onetype^*$. Since $N$ is a \kl{finite alphabet}, this can be realised by providing a distinct \kl{value} for each of the summands with the sum operator. 
The inverse is a term that inspects the value with an appropriate number of $\unconsterm$ applications. 
By induction on the size of $N$ it is provably an inverse -- use functoriality of the sum operator and the isomorphism axiom for the pair $\consterm$ used for producing the values and $\unconsterm$ used to inspect it.

We lift $f$ with the star operator to $f^* : N^* \to (\onetype^*)^*$, which preserves injectivity. 

Then, we need to provide a term of the type $(\onetype^*)^* \to \twotype^*$.

We define an injective term $g : (A^*)^* \to (A + \onetype)^*$ by the top path on the upper diagram.
The lower path defines its inverse -- each arrow is an inverse of the corresponding arrow above it. 
For the encoding proof we set $A=\onetype$.

\[
    \begin{tikzcd}[column sep=large]
    (A^*)^* 
    \ar[r, "\unitlterm^*"]
    &
    (\onetype \times A^*)^*
    \ar[r, "{([],\idterm)}"]
    &
    A^* \times (\onetype \times A^*)^*
    \ar[r, "\unblockterm"]
    &
    (A + \onetype)^*
    \ar[dl, "\blockterm", end anchor=10]
    \\
    (A^*)^* 
    \ar[u, equal]
    &
    (\onetype \times A^*)^*
    \ar[l, "\sndterm^*"']
    &
    A^* \times (\onetype \times A^*)^*
    \ar[l, "\sndterm"']
    \end{tikzcd}
\]

\section{Unprefixing}
\label{sec:unprefix-commutes-with-concat}

\newcommand{\functor}{\mathsf T}
\newcommand{\functorplus}{\mathsf T_{\!+}}
\newcommand{\inject}{\mathsf{rep}}
\newcommand{\groupid}{1}
\newcommand{\groupdiag}{\mathsf{repeat}}
\newcommand{\groupdiv}{\mathsf{divide}}

This part of the appendix is devoted to a study of the unprefixing term 
\begin{align*}
(A \times G)^* \xrightarrow \unprefterm (A \times G)^*,
\end{align*}
which is the inverse of the \kl{group prefix multiplication}. We begin by giving an explicit definition of this term, by using the \kl{atomic terms} and their combinators. (In the main body of the paper, we only said that it exists, by the expressive completeness of the system.) In fact, such an explicit definition is the formal content of the isomorphism axiom for $\unprefterm$, since this axiom must contain the source code of the unprefixing term.  As the reader will recall, the unprefixing term is defined denotationally as follows, with $g_0$ being the group identity:
\begin{align*}
[(a_1,g_1),\ldots,(a_n,g_n)] \in (A \times G)^* \xmapsto \unprefterm  [(a_1,g_0^{-1} g_1), (a_2,g_1^{-1} g_2), \ldots, (a_n,g_{n-1}^{-1} g_n)] \in (A \times G)^*
\end{align*}
This process is implemented by a term that is  described in the following definition.

\begin{definition}[Unprefixing term]\label{def:unprefterm}
    For a \kl{type} $A$ and a finite group $G$, the \intro{unprefixing term} $\unprefterm$ is defined to be  the composition of the following terms
    \[
    \begin{tikzcd}
        (A \times G)^* 
        \ar[d,"\text{$\groupdiag^*$ where $\groupdiag$ is defined by $(a,g) \mapsto (g,a,g)$}"]\\
        (G \times A \times G)^*
        \ar[d,"\text{$(\groupid, \idterm)$ where $\groupid : \onetype \to G$ represents the group identity}"]\\
    G \times (G \times A \times G)^*
    \ar[d,"\shiftlterm"]\\
    (G \times G \times A)^* \times G
    \ar[d,"\fstterm"]
    \\
    (G \times G \times A)^*
            \ar[d,"\text{$\groupdiv^*$ where $\groupdiv$ is defined by $(h,g,a) \mapsto (a,h^{-1} \cdot g)$}"]
    \\
    (A \times G)^*
    \end{tikzcd}
    \]
\end{definition}

Before continuing, let us explain the $\groupdiag$ and $\groupdiv$ terms that were used in the above definition. In principle these terms look like they are not admissible in our system, since they use duplication and commutativity. However, duplication and commutativity are applied here in the special case of a \kl{finite alphabet}, and in this special case they are admissible. More formally, we use the \kl{isomorphisms} 
\begin{equation}
    \label{eq:finite-group-isomorphisms}
A \times G \simeq 
\sum_{g \in G} A 
\qquad 
G \times A \times G 
\simeq 
\sum_{g,h \in G} A 
\qquad 
G \times G \times A
\simeq 
\sum_{g,h \in G} A.
\end{equation}
In the presence of these isomorphisms, the terms $\groupdiag$ and $\groupdiv$ can be defined by using case reasoning and coprojections.

\subsection{Naturality of the Group Prefix Multiplication}
\label{sec:naturality-of-group-prefix-operation}
Using the explicit definition of the \kl{unprefixing term}, we can prove that the \kl{group prefix multiplication} is a natural transformation, as was stated in \cref{fn:naturality-of-group-prefix-operation}.

\begin{lemma}\label{lem:naturality-of-group-prefix-operation}
    Let $A \xrightarrow f B$ be a term between \kl{types} and let $G \xrightarrow h H$ be a group homomorphism between groups whose underlying sets are \kl{types}. Then the following diagram commutes: 
\[
\begin{tikzcd}
    (A \times G)^* \arrow[r, "{(f \times h)^*}"] \arrow[d, "\prefterm"'] & (B \times H)^* \arrow[d, "\prefterm"] \\
    (A \times G)^* \arrow[r, "{(f \times h)^*}"'] & (B \times H)^*.
\end{tikzcd}
\]    
\end{lemma}
\begin{proof}
    Since the group prefix multiplication is an \kl{isomorphism}, it suffices to prove that the diagram commutes when $\prefterm$ is replaced by its inverse $\unprefterm$, i.e.
    \[
\begin{tikzcd}
    (A \times G)^* \arrow[r, "{(f \times h)^*}"]  & (B \times H)^*  \\
    (A \times G)^* \arrow[r, "{(f \times h)^*}"']
    \arrow[u, "\unprefterm"] & 
    (B \times H)^* 
    \arrow[u, "\unprefterm"'].
\end{tikzcd}
\]    
To see that the diagram commutes provably, we can use the explicit definition of $\unprefterm$ from \cref{def:unprefterm}, and observe that each of the underlying steps commutes with a group homomorphism. 
\end{proof}

\subsection{Distributing Unprefixing over Concatenation}
In this section we prove  some results in which unprefixing is distributed over concatenation. 
Let us fix the \kl{type} $A$ and a finite group $G$ for the rest of this section. For our distributivity results, we will want to have a stronger typing for lists of pairs $(A \times G)$; namely we will want the last element of the list to have a fixed value. This is implemented by the following notation.
We write 
\begin{align*}
\functor g \eqdef (A \times G)^* \times A \times \onetype.
\end{align*}
Although the group element $g$ does not appear on the right-hand side of the definition, we include it to underline our intention, which is to say that the type represents nonempty lists in $(A \times G)^*$ where the last pair has the group element $g$, and hence there is no need to write this group element explicitly. This intention is formalised by an injection
\begin{align*}
\functor g \hookrightarrow (A \times G)^*,
\end{align*}
which is denoted by $\iota_g$ and is formally defined  to be the following composition
\begin{align*}
\functor g \xrightarrow {\idterm \times (\idterm,g)} (A \times G)^* \times (A \times G) \concmap  (A \times G)^*,
\end{align*}
in which the group element $g$ is treated as a \kl{value} of type $\onetype \to G$. An important special case of this type will arise when $g$ is the identity of the group $G$, which we denote by $\groupid$.

The type $(A \times G)^*$ is isomorphic to 
\begin{align*}
\onetype + \sum_{g \in G} \functor g,
\end{align*}
with the \kl{isomorphism} being the \kl{copairing} of the empty list and the injections defined above.

We are now ready to prove our first distributivity result, which says that unprefixing commutes with binary concatenation, as long as the first list ends with the group identity. 
\begin{lemma}\label{lem:unprefixing-distributed-binary-concat}
    For every $g \in G$, the two paths from the source in the following diagram are provably equivalent (the inner square does not commute on its own):
    \[
\begin{tikzcd}
\functor \groupid \times \functor g 
\ar[d, hookrightarrow, "\iota_\groupid \times \iota_g"']
\\
(A \times G)^* \times (A \times G)^*
\ar[d, "\unprefterm \times \unprefterm"']
\ar[r,\concar]
& 
(A \times G)^*
\ar[d, "\unprefterm"]
\\
(A \times G)^* \times (A \times G)^*
\ar[r,\concar]
& 
(A \times G)^* 
\end{tikzcd}
\]
\end{lemma}
\begin{proof}
    In the proof, it will be more convenient to use a variant of the type $\functor g$, in which the represented list has at least two elements (in $\functor g$ the list is guaranteed to have  at least one element, since the last element is fixed to be $g$). This variant is defined as follows:
    \begin{align*}
    \functorplus g \eqdef (A \times G) \times \functor g.
    \end{align*}
    This type also has an injection into $(A \times G)^*$, which we also denote by $\iota_g$.     We have an isomorphism 
    \begin{align*}
    \functor g \simeq A \times \onetype + \functorplus g
    \end{align*}
The interesting case of the lemma is when both input lists $\functor \groupid$ and $\functor g$ are covered by $\functorplus$, as described in diagram \eqref{eq:unprefixing-distributed-binary-concat-nonempty} below.
The other case, when at least one of these lists is a singleton, is easy to prove and left to the reader as an exercise. 
\begin{equation}
\label{eq:unprefixing-distributed-binary-concat-nonempty}
\begin{tikzcd}[ampersand replacement=\&]
\functorplus \groupid \times \functorplus g
\ar[d, hookrightarrow, "\iota_\groupid \times \iota_g"']
\\
(A \times G)^* \times (A \times G)^*
\ar[d, "\unprefterm \times \unprefterm"']
\ar[r,\concar]
\& 
(A \times G)^*
\ar[d, "\unprefterm"]
\\
(A \times G)^* \times (A \times G)^*
\ar[r,\concar]
\& 
(A \times G)^*
\end{tikzcd}
\end{equation}
We begin by studying the right path in the diagram, in which the lists are first concatenated and then unprefixed.

\begin{claim}\label{claim:right-path-in-binary-unprefixing-distributes-over-concat}
    The right path in the diagram \eqref{eq:unprefixing-distributed-binary-concat-nonempty} is equivalent to the following composition:
    \[
\begin{tikzcd}
\myoverbrace{(A \times G) \times (A \times G)^* \times (A \times \onetype)}{$\functorplus \groupid$} \times 
\myoverbrace{
(A \times G) \times (A \times G)^* \times (A \times \onetype)}{$ \functorplus g$}
\ar[d,"\text{prepend $\onetype$ and duplicate group elements}"]
\\
\onetype \times (G \times A \times G) \times (G \times A \times G)^* \times (\onetype \times A \times \onetype) \times (G \times A \times G) \times (G \times A \times G)^* \times (\onetype \times A \times \onetype)
\ar[d,"\text{rebracket and erase last $\onetype$}"]
\\
(\onetype \times G \times A) \times (G \times G \times A)^* \times (G  \times \onetype  \times A) \times (\onetype \times G \times A) \times (G \times G \times A)^* \times (G \times \onetype \times A)\phantom{\times \onetype}
\ar[d,"\text{apply $\groupid$ to first three $\onetype$'s and $g$ to last  $\onetype$}"]
\\
(G \times G \times A) \times (G \times G \times A)^* \times (G  \times G  \times A) \times (G \times G \times A) \times (G \times G \times A)^* \times (G \times G \times A)
\phantom{ \times G} 
\ar[d,\concar,"\text{concatenate}"]
\\
( G \times G \times A)^* 
\ar[d,"\groupdiv^*"]
\\
(A \times G)^*
\end{tikzcd}
\]
\end{claim}
\begin{proof}[Proof sketch]
    Using naturality and the uniqueness of \kl{rebracketing} from \cref{lem:uniqueness-of-rebracketings}. To apply the latter result, we need to show that all types in the statement of the claim are \kl[injective type]{injective}. The corresponding calculation uses the fact that $\onetype$ appears a constant number of times, and hence each copy of $\onetype$ can be identified uniquely when computing a one-sided inverse.
\end{proof}

A similar argument can be applied to  the left path in the diagram, in which the lists are first unprefixed and then concatenated.
\begin{claim}\label{claim:left-path-in-binary-unprefixing-distributes-over-concat}
    The left path in the diagram \eqref{eq:unprefixing-distributed-binary-concat-nonempty} is equivalent to the following composition (we use colours to help distinguish the two parts of the product):
    \[
\medmuskip=3mu
\begin{tikzcd}
\myoverbrace{(A \times G) \times (A \times G)^* \times (A \times \onetype)}{$\functorplus \groupid$} \times 
\blue{\myoverbrace{(A \times G) \times (A \times G)^* \times (A \times \onetype)}{$ \functorplus g$}}
\ar[d,"\text{in each part, prepend $\onetype$ and duplicate group elements}"]
\\
\onetype \times (G \times A \times G) \times (G \times A \times G)^* \times (\onetype \times A \times \onetype) \times \blue{\onetype \times  (G \times A \times G) \times (G \times A \times G)^* \times (\onetype \times A \times \onetype)}
\ar[d,"\text{in each part, rebracket and remove last $\onetype$}"]
\\
(\onetype \times G \times A) \times (G \times G \times A)^* \times (G  \times \onetype  \times A)  \phantom{ \times \onetype}  \times \blue{(\onetype \times G \times A) \times (G \times G \times A)^* \times (G \times \onetype \times A) \phantom{\times \onetype}}
\ar[d,"\text{apply $\groupid$ to first three $\onetype$'s and $g$ to last  $\onetype$}"]
\\
(G \times G \times A) \times (G \times G \times A)^* \times (G  \times G  \times A) \phantom{\times G} \times \blue{(G \times G \times A) \times (G \times G \times A)^* \times (G \times G \times A)}
\phantom{ \times G}
\ar[d,"\text{in each part, concatenate}"]
\\
( G \times G \times A)^* \times \blue{( G \times G \times A)^*}
\ar[d,"\text{in each part, $\groupdiv^*$}"]
\\
( G \times G \times A)^* \times \blue{( G \times G \times A)^*}
\ar[d,\concar] 
\\
(A \times G)^*
\end{tikzcd}
\]
\end{claim}
\begin{proof}
    Similar to the proof of \cref{claim:right-path-in-binary-unprefixing-distributes-over-concat}.
\end{proof}
To finish the proof of the lemma, it remains to show that the two compositions from \cref{claim:right-path-in-binary-unprefixing-distributes-over-concat,claim:left-path-in-binary-unprefixing-distributes-over-concat} are provably equivalent. By uniqueness of rebracketing, the first two steps are the same in both compositions. The third step (where $\onetype$'s are replaced by group elements) is the same. Finally, the double concatenation in the last three steps of the diagram from \cref{claim:left-path-in-binary-unprefixing-distributes-over-concat} can be replaced by a   single concatenation as in the last two steps of the diagram from \cref{claim:right-path-in-binary-unprefixing-distributes-over-concat}, thanks to
\[
\begin{tikzcd}
A^* \times A^* \ar[r,"f^* \times f^*"]
\ar[d,\concar]
& 
B^* \times B^* 
\ar[d,\concar]
\\
A^* \ar[r,"f^*"]
&
B^*
\end{tikzcd}
\]
which is an instance of the naturality axioms. This finishes the proof of the lemma.
\end{proof}

We now state  a similar result for concatenations of an arbitrary number of lists, as long as each list ends with the group identity. The proof is similar to the proof of \cref{lem:unprefixing-distributed-binary-concat}, and is left to the reader as an exercise.
\begin{lemma}\label{lem:unprefixing-distributed-general-concat}
    The two paths from the source are provably equivalent in the following diagram:
    \[
\begin{tikzcd}
(\functor \groupid)^*
\ar[d, hookrightarrow, "(\iota_\groupid)^*"']
\\
(A \times G)^{**}
\ar[d, "\unprefterm^*"']
\ar[r,\concar]
& 
(A \times G)^*
\ar[d, "\unprefterm"]
\\
(A \times G)^{**}
\ar[r,\concar]
& 
(A \times G)^* 
\end{tikzcd}
\]
\end{lemma}

\section{Proofs for Soundness of the Conjugacy Rule}
In this part of the appendix, we prove the lemmas that we used in the proof of \cref{lem:completeness-for-simple-rational-terms}, whose main goal was to show the soundness of the conjugacy rule for \kl{simple rational terms}.

\subsection{Proof of \cref{lem:conjugacy-fixed-middle-equational}}\label{sec:proof-of-lem:conjugacy-fixed-middle-equational}
We begin with the proof of \cref{lem:conjugacy-fixed-middle-equational}, which says that 
    \begin{align*}
        \hom(w \wdot h) \wdot w \equiv w \wdot \hom(h \wdot w)
        \qquad \text{for every term $A \xrightarrow h B^*$ and value $\onetype \xrightarrow w B^*$.}
    \end{align*}
As we have remarked in the main paper, this can be seen as  a weaker version of the conjugacy rule, where the assumption is strengthened from provable conjugacy  to syntactic conjugacy. 

The proof is based on the following diagram.
\begin{equation}
\label{eq:housie-housie}
    \begin{tikzcd}
        [column sep=huge]
 &  A^*
    \ar[dl, "{\appendoneterm^*\comp \prependoneterm}"']
    \ar[dr, "{\prependoneterm^*\comp \appendoneterm}"]
   \\
    \onetype \times (A \times\onetype)^* 
    \ar[dd,"w \times (f \times w)^*"']
    \ar[r, \concar ]
    &
            (A + \onetype)^*
    \ar[dd,"{\hom{(\either{f}{w})}}"]    
    &
    (\onetype \times A)^* \times \onetype    %
    \ar[dd,"(w \times f)^* \times w"]
    \ar[l, \concar]
    \\
    &
\ 
    \\
    B^* \times (B^* \times B^*)^*
    \ar[r,\concar]
    &
    B^*
    & 
    (B^* \times B^*)^* \times B^*
    \ar[l,\concar]
    \end{tikzcd}
\end{equation}
The top face commutes by \cref{lem:shift-ones}.
To complete the proof of the lemma, we will show that: (a) the lower two rectangular faces commute provably and hence the entire diagram commutes, (b) the leftmost and rightmost paths from $A^*$ to $B^*$ are provably equivalent to the terms in the conclusion of the lemma, which are $w \wdot \hom(f \wdot w)$ and  
$\hom(w \wdot f) \wdot w$, respectively. 
We begin with (b), which is the following calculation (we only do the leftmost path, since the rightmost path is symmetric):
\begin{align*}
\text{leftmost path from $A^*$ to $B^*$ in \eqref{eq:housie-housie}}  & = \\
\appendoneterm^*\comp \prependoneterm\comp w \times (f \times w)^*\comp \concmap  \equalbecause{naturality}\\
\prependoneterm\comp \idterm \times \appendoneterm^*\comp  w \times (f \times w)^*\comp \concmap  \equalbecause{functoriality}\\
\prependoneterm\comp  w \times (\appendoneterm\comp f \times w)^*\comp \concmap  \equalbecause{concatenation axioms}\\
\prependoneterm\comp  w \times (\appendoneterm\comp f \times w\comp \concmap)^*\comp \concmap  &= \why{definition of $f \wdot w$}\\
\prependoneterm\comp  w \times (f \wdot w)^*\comp \concmap  &= \why{definition of $w \wdot \hom(f \wdot w)$}\\
w \wdot \hom(f \wdot w)
\end{align*}
It remains to prove (a). We only consider the lower-left rectangular face, since the lower-right rectangular face is symmetric. This face is expanded in upside-down form in  the following diagram.
    \[
\begin{tikzcd}
    [column sep=5cm]
\onetype \times (A \times \onetype)^*
\ar[dr,"w \times (f \times w)^*"]
\ar[d,"{\rightterm \times (\leftterm \times \rightterm)^*}"']
\\
(A+\onetype) \times ((A + \onetype) \times (A + \onetype))^*
\ar[d,\concar]
\ar[r,"{\either{f}{w} \times (\either{f}{w} \times \either{f}{w})^*}"']
& 
B^* \times (B^* \times  B^*)^*
\ar[d, \concar]
\ar[dd, bend left, \concar]
\\ 
(A + \onetype)^*
\ar[dr,"{\hom(\either{f}{w})}"']
\ar[r,"(\either f w)^*"]
& 
B^{**}
\ar[d,\concar]
\\
& B^*
\end{tikzcd}
\]
The middle rectangular face commutes  provably by naturality of \kl{concatenation maps}. The lower triangular face commutes by definition of $\hom$. The upper triangular face commutes by \cref{lem:case-equivalence}. The rightmost face with three concatenation maps commutes by the list axioms. 
Summing up, the entire diagram commutes provably, which proves (a).

\subsection{Proof of \cref{lem:constants-and-guarded}}
\label{sec:proof-of-lem:constants-and-guarded}
In this part of the appendix, we prove \cref{lem:constants-and-guarded},
which says that in the proof of the induction step for~\eqref{eq:conjugacy-rule-for-simple-rational-terms}, one can assume without loss of generality that the input \kl{type} is a coproduct 
\begin{align*}
A = A_1 + \cdots + A_n,
\end{align*}
and each of the terms $f$ and $g$ is a \kl{copairing} of $k$-guarded terms.  To prove this reduction, it is enough to show the following lemma. 
\begin{lemma} [Value extraction] \label{lem:value-extraction}
    Let $k \in \set{1,2,\ldots}$ and let 
    \begin{align*}
    A \xrightarrow {f,g} B^*
    \end{align*}
    be \kl{simple rational terms} that are  denotationally conjugate. Then there is  an \kl{isomorphism} 
    \begin{align*}
 C_1 + \cdots + C_n \xrightarrow \pi A
    \end{align*}
    such that for every $i \in \set{1,\ldots,n}$, the \kl{type} $C_i$ has the same or smaller nesting of lists as $A$, and both compositions $\pi\comp{}f$ and $\pi\comp{}g$ are provably equivalent to \kl{copairings} of $k$-guarded terms.
\end{lemma}

\begin{proof}
    Induction on the structure of the input type $A$. The induction is on two parameters: (a) more important is the nesting depth of the list constructor; (b) less important is the size of (syntax tree of) the type. This means that we can progress in the induction by decreasing the nesting depth of the list constructor, even if we increase the size of the type. We prove the lemma for only one term $f$, but during the proof we will  observe that the same isomorphism will arise if $f$ is replaced by a conjugate term. 

    \begin{description}
    \item[Induction basis and induction step for coproducts.]
    The induction basis when $A = \onetype$ is trivial, since the term is already a \kl{value}. If the input type is a coproduct $A_1 + A_2$, then the result follows immediately from the induction hypothesis. The more interesting cases are for products and lists. 
    
   \item[Product type.] Suppose now that the input type is a product $A_1 \times A_2$. In this case, we simply apply distributivity. Nevertheless, we present a more detailed argument, to show that all equivalences are provable in this case.  A simple rational term with a product domain $A_1 \times A_2$ is provably equivalent to a concatenation $f_1 \fdot f_2$ of two simple rational terms 
    \begin{align*}
    A_1 \xrightarrow {f_1} B^* \quad \text{and} \quad A_2 \xrightarrow {f_2} B^*.
    \end{align*}
    Apply the induction hypothesis to these two terms, yielding isomorphisms 
    \begin{align*}
    \sum_{i_1 \in I_1} C_{1,i_1} \xrightarrow{\pi_1} A_1 \qquad \text{and} \qquad \sum_{j_2 \in J_2} C_{2,j_2} \xrightarrow{\pi_2} A_2.
    \end{align*}
    Define $\pi$ to be the isomorphism which is the composition
    \begin{align*}
    \sum_{(i_1,j_2) \in I_1 \times J_2} C_{1,i_1} \times C_{2,j_2} \xrightarrow{\undistlterm}
    \left(\sum_{i_1 \in I_1} C_{1,i_1}\right) \times \left(\sum_{j_2 \in J_2} C_{2,j_2}\right)
    \xrightarrow{
    (\pi_1 \times \pi_2)} A_1 \times A_2.
    \end{align*}
    To prove the conclusion of the lemma, we will show that  the composition 
    \begin{align*}
    C_{1,i_1} \times C_{2,j_2} 
    \xrightarrow {\text{coprojection}}
    \sum_{(i_1,j_2) \in I_1 \times J_2} C_{1,i_1} \times C_{2,j_2} 
    \xrightarrow{\pi} A_1 \times A_2 
    \xrightarrow{f_1 \fdot f_2} B^*
    \end{align*}
    is provably equivalent to a $k$-guarded term. 
    By suitable use of distributivity, see \cref{lem:distributivity}, the above composition is provably equivalent to
    \begin{align*}
    C_{1,i_1} \times C_{2,j_2} 
    \xrightarrow{\text{coprojection} \times \text{coprojection}}
    \left(\sum_{i_1 \in I_1} C_{1,i_1}\right) \times \left(\sum_{j_2 \in J_2} C_{2,j_2}\right)
    \xrightarrow{\pi_1 \times \pi_2} A_1 \times A_2 
    \xrightarrow{f_1 \fdot f_2}
    B^*,
    \end{align*}
    which is by induction hypothesis provably equivalent to a concatenation of two $k$-guarded terms. Finally, we observe that  $k$-guarded terms are closed under concatenation, up to provable equivalence, which follows from \cref{lem:concatenation-monoidality}.

    \item[List type.]
    The essential case is when the domain is a list type, and hence the  term is 
    \begin{align*}
    A^* \xrightarrow {w_0 \wdot \hom f \wdot w_1} B^*.
    \end{align*}
    Since $k$-guarded terms are closed under concatenation with values, we can assume without loss of generality that $w_0$ and $w_1$ are both empty, and hence the term is simply 
    \begin{align*}
    A^* \xrightarrow {\hom f} B^*.
    \end{align*}
    Apply the induction hypothesis to $f$, leading to an isomorphism
    \begin{align*}
     C_1 + \cdots + C_n \xrightarrow{\pi} A.
    \end{align*}
    For some cases in $C_1,\ldots,C_n$, the outputs are guarded because they are constant, and for others they are guarded on both sides by strings of length $k$ . Some of the constants are empty, i.e.~the corresponding term always outputs the empty list. Using the same proof as in \cref{lem:nonempty-summands}, we can assume without loss of generality that the constants are always nonempty. (An important observation is that if we take some other term $g$ that is conjugate to $f$, then the same constants will be nonempty, and thus applying the same proof as in \cref{lem:nonempty-summands} will lead to the same isomorphism.) We will now consider a finite number of cases: either the input list from $A^*$ has length $<2k$, or otherwise one can identify a prefix of length $k$ and a suffix of length $k$. This is the content of the following lemma. 

\begin{lemma}\label{lem:fixed-lengths-extracted}
    For every \kl{type} $C$ and $k \in \set{0,1,\ldots}$, the \kl{concatenation map}
        \begin{align*}
    \left(\sum_{\ell < 2k} C^\ell \right) +  C^k \times C^* \times C^k \concmap C^*
    \end{align*}
    is an \kl{isomorphism}.
\end{lemma}
\begin{proof}
    Induction on $k$. For the base case of $k=0$, there is nothing to do, since the concatenation map is the identity. 
    Suppose now that we have proved the lemma for $k$ and we want to prove it for $k+1$. We need to define the inverse of the concatenation map in the statement of the lemma. This is done by using the  list destructors, as in the following composition:
    \[
    \begin{tikzcd}
    C^* 
    \ar[d, "\unconsterm"] 
    \\
    \onetype + C \times C^* 
    \ar[d, "\idterm + \idterm \times (\unsnocterm)"]
    \\
    \onetype  + C \times (\onetype + C^* \times C) 
    \ar[d,"\idterm + \distrterm"]
    \\
    \onetype + C \times \onetype + C \times C^* \times C
    \ar[d,"\text{induction hypothesis}"]
    \\
    \onetype + C \times \onetype + C \times (    (\sum_{\ell < 2k} C^\ell) +  C^k \times C^* \times C^k) \times C
    \ar[d,"\text{distributivity}"]
    \\
    \sum_{\ell < 2k+2} C^\ell + C^{k+1} \times C^* \times C^{k+1}.
    \end{tikzcd}
    \]
    We need to prove that the inverse of the above is the  concatenation map. To see this, we observe that in each step we use a term whose inverse is obtained by  applying the type constructors to concatenation maps. Hence the inverse of the composition of all steps is like this as well, and hence it is a concatenation map.  
\end{proof}

Apply the above lemma and   distributivity, yielding an isomorphism to $C^*$ from the following type:
\begin{align*}
\sum_{\ell < 2k} \sum_{\alpha \in n^\ell} C_{\alpha(1)} \times \cdots \times C_{\alpha(\ell)}
\qquad + \qquad 
\sum_{\alpha, \beta \in n^k} C_{\alpha(1)} \times \cdots \times C_{\alpha(k)} \times C^* \times C_{\beta(1)} \times \cdots \times C_{\beta(k)}.
\end{align*}
When we apply $\hom f$ precomposed with this isomorphism, in each of the cases we get  a $k$-guarded term by the induction hypothesis, and closure of $k$-guarded terms under concatenation, which follows from \cref{lem:concatenation-monoidality}.
    \end{description}
Hence, we have proved the lemma. 
\end{proof}

\subsection{Proof of \cref{claim:ignore-empty-lists}}
\label{sec:proof-of-claim:ignore-empty-lists}
The proof of the claim is based on the following diagram.
            \[
    \begin{tikzcd}
        [column sep=huge]
        (E+N)^* 
        \ar[d,"\blockterm"']
        \ar[drr, bend left = 17, "\hom f \equiv \hom(\either {f_E}  {f_N})"]
        \\
        E^* \times (N \times E^*)^* 
        \ar[rr, "{\hom f_E \fdot \hom ( f_N \fdot \hom f_E)}"] 
        \ar[d,"1 \times (\idterm \times 1)^*"']
        & & 
         B^*
                  \\
         \onetype \times (N \times \onetype)^*
                 \ar[urr, "{[] \fdot \hom ( f_N \fdot [])}"'] 
        \ar[rr,"\sndterm \times \fstterm^*"']
         &  & 
         N^*
         \ar[u,"{\hom f_N}"']
    \end{tikzcd}
    \]
    Let us explain why all faces in this diagram commute. The uppermost face commutes by naturality of $\blockterm$.  The middle face commutes by \cref{ax:concatenation-of-empty-lists}. The lowermost face commutes by \cref{ax:concatenation-monoid}.  Summing up, the entire diagram commutes, which proves the claim.

\subsection{Proof of \cref{lem:bundles}}
\label{sec:proof-of-bundles}

\newcommand{\binrep}{\twotype^*}
\newcommand{\bundle}[2][A]{(#1^{#2})^* \times #1^{<#2}}
\newcommand{\encterm}{\mathtt{enc}}
\newcommand{\decterm}{\mathtt{dec}}

We now prove \cref{lem:bundles}, which says that the \kl{concatenation map}
\begin{align*}
    \bundle[A]{k}\concmap  A^*
\end{align*}
is  an \kl{isomorphism} for arbitrarily large values of $k$. Somewhat surprisingly, this is the most involved  proof for  this part of the appendix. One of the curious parts of this proof is that although the concatenation map itself is \kl{first-order definable}, its inverse is not, and hence proving that it is an isomorphism requires using groups.

We will first prove \cref{lem:bundles} for $k=2$.

\subsubsection{Definition of the Inverse}
\label{sec:inverse-definition}
The inverse of the \kl{concatenation map}  will use
the group $G$ which is the cyclic group of order $k$. Since we already use $1$ a lot, we  describe the elements of this group as
\begin{align*}
    g^{0}, g^1, \ldots, g^{k-1}.
\end{align*}
The group identity is $g^0$ and $g^1=g$ is a generator.

\paragraph*{Semantic Description.}  Since formalising the inverse will be somewhat involved, we begin by explaining how the inverse works semantically, without formalising it in our system.
For an input
\begin{align*}
    [a_1,\ldots,a_n] \in A^*,
\end{align*}
we first annotate each list element with the generator $g^1$  of the group,
leading to an annotated list
\begin{align*}
    [(a_1,g^1),\ldots,(a_n,g^1)] \in (A \times G)^*.
\end{align*}
Next, we apply the \kl{group prefix multiplication}. As a result, the group elements will be arranged in cyclic order: $g^1,g^2,\ldots,g^{k-1},g^0,g^1,g^2,\ldots$. Next, we use the block operation to cut the list after those pairs that use the group identity. As a result, the list will be split into bundles of length exactly $k$, followed by some tail bundle of length $< k$. Finally, we  remove the group annotation, and coerce the lists of length $k$ into tuples of length $k$.

The difficulty in  formalising this  construction in our system arises from the \kl{type} coercion in the last step. Our equational axiomatisation needs to be aware of the fact that this coercion gives the correct output.

\paragraph*{Formal Description.} \label{def:concmap-inverse} We now begin a formal description of the required inverse. We identify a group element  with the corresponding \kl{value} of type $\onetype \to G$. We begin by describing the first step of the construction above, where list elements are annotated by the group. Suppose that we have a list that has already been grouped into bundles of size $k$, and we want to annotate it with the group, as in the following type:
\begin{align*}
    (A^k)^* \times A^{<k}  \to ((A \times G)^k)^* \times (A \times G)^{<k}.
\end{align*}
There are two ways to do this annotation, which we call the \emph{constant annotation} and the \emph{cyclic annotation}. In the constant annotation,  each list element is decorated with the generator $g^1$ of the cyclic group, as in the following term
\begin{align*}
    ((\idterm, g^1) \times \cdots \times  (\idterm, g^1))^* \times (\idterm, g^1)^{<k}.
\end{align*}
In the cyclic annotation, the list elements are decorated with the group elements in cyclic order, as in the following term
\begin{align*}
    ((\idterm, g^1) \times (\idterm, g^2) \times \cdots \times (\idterm,  g^{k-1}) \times  (\idterm, g^0))^* \times f_\text{tail},
\end{align*}
where the tail term continues the cyclic pattern for the last $<k$ elements of the list. The main observation in the proof is that these two annotations are equivalent up to the group prefix multiplication on the output, which is illustrated in the following diagram.
\begin{equation}
    \label{eq:bundles-commutative-diagram}
    \begin{tikzcd}[column sep=small, ]
        & (A^k)^* \times A^{<k}
        \ar[dl,"\text{constant annotation}"']
        \ar[dr,"\text{cyclic annotation}"]
        \\
        ((A \times G)^k)^* \times (A \times G)^{<k}
        \ar[d,\concar]
        & &
        ((A \times G)^k)^* \times (A \times G)^{<k}
        \ar[d,\concar] \\
        (A \times G)^*
        \ar[rr,"\prefterm"']
        & &
        (A \times G)^*
    \end{tikzcd}
\end{equation}
In \cref{sec:bundle-injectivity}, we show that the term
\begin{align*}
    \text{cyclic annotation}\comp \concmap
\end{align*}
in the right path of the diagram is \kl{injective}. In \cref{sec:bundle-diagram-commutes}, we show that the diagram commutes provably. From these two results, it will follow that the
following composition
\begin{multline*}
    (A^k)^* \times A^{<k}
    \xrightarrow{\text{constant annotation}\comp\concmap}
    (A\times G)^*
    \xrightarrow \prefterm
    (A \times G)^*
    \\
    \xrightarrow {\text{one-sided inverse from \cref{sec:bundle-injectivity}}}
    (A^k)^* \times A^{<k},
\end{multline*}
is provably equivalent to the identity on the type $(A^k)^* \times A^{<k}$. By naturality of the concatenation map, the above composition is provably equivalent to
\begin{align*}
    (A^k)^* \times A^{<k}
    \concmap
    \myunderbrace{  A^* \xrightarrow{(\idterm, g^1)^*}
        (A\times G)^*
        \xrightarrow \prefterm
        (A \times G)^*
        \xrightarrow {\text{one-sided inverse from \cref{sec:bundle-injectivity}}}
        (A^k)^* \times A^{<k}
    }{call this composition $f$}.
\end{align*}

We established that if we compose the concatenation map with $f$, then we get the identity on $(A^k)^* \times A^{<k}$. In other words, we know that the concatenation map is injective, as $f$ is its right inverse. It remains to prove that it is also \kl{surjective}, it has a left inverse. We provide such an inverse in \cref{sec:bundle-surjectivity}.

\subsubsection{Proof of Injectivity of the Right-Down Path}
\label{sec:bundle-injectivity}
We begin with the easier part of the lemma, which is that the term on the right  path
\begin{align*}
    \text{cyclic annotation}\comp \concmap
\end{align*}
in diagram~\eqref{eq:bundles-commutative-diagram}
is \kl{injective}. Under the assumption of $k=2$, when we apply this term, we cast even-numbered list items to one copy of $A$ and odd-numbered list items to another copy of $A$. We want to show that this term is injective. This is proved in the following lemma, where $A$ and $B$ should be seen as representing the two copies of $A$ in the cyclic annotation.
\begin{lemma}\label{lem:injectivity-of-cyclic-annotation}
    For all \kl{types} $A$ and $B$, the \kl{concatenation map}
    \begin{align*}
        (A \times B)^* \times A^{<2} \to (A + B)^*
    \end{align*}
    is \kl{injective} and a \kl{rebracketing}.
\end{lemma}
\begin{proof}
    Consider the composition
    \begin{align*}
        (A \times B)^* \times A^{<2}
        \xrightarrow{(\concmap \times \idterm)^* \times \concmap}
        (A^* \times B)^* \times A^*
        \xrightarrow{\unblocklterm}
        (A+B)^*.
    \end{align*}
    The second part of the composition is \kl{injective} as an \kl{isomorphism}. Since injective terms, as well as rebracketings,  are closed under composition and applying the functors, it remains to show that the concatenation maps
    \begin{align*}
        A^{<2}  \concmap A^*
        \qquad
        A \concmap A^*
    \end{align*}
    are  injective. They are clearly rebracketings, as they are concatenation maps. Let us show injectivity.

    The first of these concatenation maps, with domain $A^{<2}$ is equivalent to the  composition
    \[
        \begin{tikzcd}
            \onetype + A
            \ar[r,"{\idterm + (\idterm, [])}"]
            &
            \onetype + A \times A^*
            \ar[r, "\unconsterm"]
            &
            A^*.
        \end{tikzcd}
    \]
    The first term in the composition is injective by \cref{ax:concatenation-monoid} and the second term is injective by the isomorphism axioms. Hence, their composition is injective.

    The more interesting case is showing injectivity of the concatenation map
    \begin{align*}
        A  \concmap A^*,
    \end{align*}
    which is the same as $\pureterm$.
    This is more interesting because the inverse of this map needs to escape the list monad, and this can only be done in a nonnatural way. Let us begin by fixing some \kl{value}
    \begin{align*}
        \onetype \xrightarrow w A.
    \end{align*}
    Such a value exists, since every type is inhabited. Define the \emph{head term for $w$} to be the composition
    \begin{align*}
        A^* \xrightarrow {\headterm_w} A
        \qquad \eqdef \qquad
        A^* \xrightarrow \unconsterm \onetype + A \times A^* \xrightarrow {\either w  \fstterm} A.
    \end{align*}
    This term is a one-sided inverse of $\pureterm$ regardless of the choice of $w$, as shown below:
    \begin{align*}
        \pureterm\comp \headterm_w &= \why{by definition}                                                     \\
        (\idterm, [])\comp \rightterm\comp \consterm\comp \unconsterm\comp \either w \fstterm \equalbecause{isomorphism axioms} \\
        (\idterm, [])\comp \rightterm\comp \either w \fstterm \equalbecause{\cref{lem:case-equivalence}}                \\
        (\idterm, [])\comp \fstterm \equalbecause{naturality}                                                       \\
        \idterm.
    \end{align*}
    This completes the proof of the injectivity of the concatenation map $A \concmap A^*$, and hence the proof of the lemma.
\end{proof}

\subsubsection{The Diagram Commutes}
\label{sec:bundle-diagram-commutes}
To finish the proof in \cref{sec:bundle-injectivity}, we need to show that the diagram~\eqref{eq:bundles-commutative-diagram} commutes provably.  Since $\prefterm$ is an \kl{isomorphism}, it is enough to prove that the diagram commutes when $\prefterm$ is replaced by its inverse, as in the following diagram (where we use the assumption that $k=2$):
\begin{equation}\label{eq:bundles-commutative-diagram-inverse}
    \begin{tikzcd}[column sep=small]
        & (A \times A)^* \times A^{<2}
        \ar[dl,"\text{constant annotation}"']
        \ar[dr,"\text{cyclic annotation}"]
        \\
        ((A \times G) \times (A \times G))^* \times (A \times G)^{<2}
        \ar[d,\concar]
        & &
        ((A \times G) \times (A \times G))^* \times (A \times G)^{<2}
        \ar[d,\concar] \\
        (A \times G)^*
        & &
        (A \times G)^*
        \ar[ll,"\unprefterm"']
    \end{tikzcd}
\end{equation}

The main observation will be the following lemma, which uses the notation and  results from \cref{sec:unprefix-commutes-with-concat} about distributing $\unprefterm$ over concatenation.

\begin{lemma}\label{lem:unpref-commutes-with-concat}
    Recall the notation $\functor g$ from \cref{sec:unprefix-commutes-with-concat}.   The following diagram commutes provably.
    \[
        \begin{tikzcd}[column sep=2cm]
            (\functor g^0)^* \times (A \times G)^*
            \ar[d,"{(\iota_\groupid)^* \times \idterm}"']
            \\
            (A \times G)^{**} \times (A \times G)^{*}
            \ar[r,\concar]
            \ar[d,"\unprefterm^* \times \unprefterm"']
            &
            (A \times G)^{*}
            \ar[d,"\unprefterm"]
            \\
            (A \times G)^{**} \times (A \times G)^{*}
            \ar[r,\concar]
            &
            (A \times G)^*.
        \end{tikzcd}
    \]
\end{lemma}
\begin{proof}
    By \cref{lem:unprefixing-distributed-binary-concat} and \cref{lem:unprefixing-distributed-general-concat}.
\end{proof}

The right path in the diagram is
\begin{align*}
    (A \times A)^* \times A^{<2} \xrightarrow{
        \text{cyclic annotation}\comp \concmap\comp \unprefterm}
    (A \times G)^*.
\end{align*}
Thanks to \cref{lem:unpref-commutes-with-concat}, we can move $\unprefterm$ inside the lists, i.e.~it is provably equivalent to
\begin{align*}
    (A \times A)^* \times A^{<2} \xrightarrow{
        (\text{cyclic annotation}\comp \unprefterm)^* \times (\text{cyclic annotation}\comp \unprefterm)\comp\concmap}
    (A \times G)^*.
\end{align*}
In the new term, both instances of cyclic annotation are used on tuples of bounded length. For such inputs, we have an explicit description of cyclic annotation, namely the above term is equivalent to
\begin{align*}
    (A \times A)^* \times A^{<2} \xrightarrow{
        ((\idterm,g^1) \times (\idterm,g^0))\comp\concmap\comp \unprefterm)^* \times ((\idterm,g^1)^{<2}\comp \concmap\comp \unprefterm)\comp\concmap}
    (A \times G)^*.
\end{align*}
For lists of constant length, we can unfold the definition of $\unprefterm$ to see that the above term is equivalent to
\begin{align*}
    (A \times A)^* \times A^{<2} \xrightarrow{
        ((\idterm,g^1) \times (\idterm,g^1))\comp\concmap)^* \times ((\idterm,g^1)^{<2}\comp \concmap)\comp\concmap}
    (A \times G)^*,
\end{align*}
which is exactly the left path in the diagram~\eqref{eq:bundles-commutative-diagram-inverse}. This proves that the diagram commutes provably, and hence concludes the proof of \kl{injectivity} for $\concmap$ on $\bundle[A]{2}$.

\subsubsection{Surjectivity}
\label{sec:bundle-surjectivity}

We now prove that the \kl{concatenation map} $\bundle{2} \concmap A^*$ is \kl{surjective} for every \kl{type} $A$. From \cref{ax:window} we know that it is the case for $A = \twotype^*$: the inverse is the term $\bundleterm : \twotype^{**} \to (\twotype^* \times \twotype^*)^* \times (\twotype^*)^{< 2}$, of which we now give an explicit definition, based on the description in \cref{sec:inverse-definition}.

\begin{definition}[Bundle term]\label{def:bundle-term}
    The term $\bundleterm$ is defined to be the composition of the following terms:
    \[
    \begin{tikzcd}
        \twotype^{**}
        \ar[d,"{\big(\appendoneterm \comp (\idterm \times \rightterm)\big)^*}"]\\
        (\twotype^* \times \twotype)^*
        \ar[d,"\prefterm_{\mathbb{Z}_2}"]\\
        (\twotype^* \times \twotype)^*
        \ar[d,"{\big(\distrterm \comp \either{(\fstterm \comp \rightterm)}{(\fstterm \comp \leftterm)}\big)^*}"]\\
        (\twotype^* + \twotype^*)^*
        \ar[d,"\blockterm"]\\
        \twotype^{**} \times (\twotype^* \times \twotype^{**})^*
        \ar[d,"\shiftlterm"]\\
        (\twotype^{**} \times \twotype^*)^* \times \twotype^{**}
        \ar[d,"{(\headterm_{[]} \times \idterm)^* \times \big(\unconsterm \comp (\idterm + \fstterm)\big)}"]\\
        (\twotype^* \times \twotype^*)^* \times (\onetype + \twotype^*)
    \end{tikzcd}
    \]
    where $\mathbb{Z}_2$ is the cyclic group of order two whose underlying set is $\twotype$, with identity $\leftterm$, and $\headterm_{[]}$ is the head term for the \kl{value} $\onetype \xrightarrow{[]} \twotype^*$ from the proof of \cref{lem:injectivity-of-cyclic-annotation}.
\end{definition}

Now, we extend the conclusion of \cref{ax:window} to arbitrary type $A$. We take the binary string encoding $\encterm: A \to \twotype^*$ from \cref{lem:provably-injective-string-representation} and its right inverse $\decterm: \twotype^* \to A$. We put 
\[
\bundleterm_A \quad \eqdef \quad \encterm^* \comp \bundleterm \comp F_2(\decterm),
\]
where $F_2(X) = \bundle[X]{2}$ is the functor such that $F_2(A)$ is the input type for $\concmap$ in the lemma.
By naturality we obtain $F_2(\decterm)\comp \concmap_{F_2(A)} \equiv \concmap_{F_2(\twotype^*)} \comp \decterm^*$. We obtain:
\begin{align*}
    \bundleterm_A \comp \concmap_{F_2(A)} &= \why{definition of $\bundleterm_A$} \\
    \encterm^* \comp \bundleterm \comp F_2(\decterm) \comp \concmap_{F_2(A)} & \equiv \why{naturality} \\
    \encterm^* \comp \bundleterm \comp \concmap_{F_2(\twotype^*)} \comp \decterm^* & \equiv \why{$\bundleterm$ is left inverse of $\concmap$ by \cref{ax:window}} \\
    \encterm^* \comp \idterm \comp \decterm^* & \equiv \why{$\decterm$ is right inverse of $\encterm$} \\
    \idterm
\end{align*}
This proves that $\bundleterm_A$ is the left inverse of $\concmap_{F_2(A)}$, so it is a witness of surjectivity.
Together with \cref{sec:bundle-injectivity}, it shows that $\concmap_{\bundle{2}}$ is an \kl{isomorphism} for every $A$.

\subsubsection{Induction}
\label{sec:bundle-induction}

We now show, by induction on $i$, that the \kl{concatenation map} $\bundle[A]{k} \concmap A^*$ is an \kl{isomorphism} for every $k = 2^i$. The case $k = 2$ was shown above. As the induction hypothesis, assume that the claim holds for $k = l$; we show that it holds for $k = 2l$.

Let $F_k(X) = \bundle[X]{k}$, and let $f_2$ and $f_l$ be the two-sided inverses of the concatenation maps on $F_2(A^l)$ and $F_l(A)$, respectively. In particular, both are isomorphisms and \kl{rebracketings}. We define $f_{2l}$ to be the composition
\[
    A^*
    \xrightarrow{f_l}
    (A^l)^* \times A^{<l}
    \xrightarrow{f_2 \times \idterm}
    \big(((A^l)^2)^* \times (A^l)^{<2}\big) \times A^{<l}
    \xrightarrow{h}
    (A^{2l})^* \times A^{<2l},
\]
where $h$ is the rebracketing isomorphism which reassociates the product, and then applies the isomorphisms
\[
    (A^l)^2 \to A^{2l}
    \qquad \text{and} \qquad
    (A^l)^{<2} \times A^{<l} \to A^{<2l},
\]
both of which are obtained by distributing products over coproducts.

The term $f_{2l}$ is built from isomorphisms by using composition and the functors, and hence it is an isomorphism. By the same argument, it is a rebracketing, and hence it is a left inverse of the concatenation map $\bundle[A]{2l} \concmap A^*$. Therefore, this concatenation map is also an isomorphism, and $f_{2l}$ is its inverse, which is unique up to provable equivalence.

By induction, the concatenation map $\bundle[A]{k} \concmap A^*$ is an isomorphism for every $k = 2^i$. Since these values of $k$ form an unbounded set, this completes the proof of \cref{lem:bundles}.

\section{Properties of Drop and Lift Terms}\label{sec:appendix-reduction-to-simple}

The ultimate goal of this section is to prove \cref{lem:exchange} and \cref{lem:joint-drops-normalisation}. In order to do that, we have to develop some technical results about drop and \kl{lift terms}. 

First, we establish auxiliary notation. Throughout this section, we will mostly work with $n$-fold products and coproducts, which by default are right-associative. A big portion of our constructions will involve purely structural term-reshuffling expressible as lift terms: reassociation of products and coproducts, swapping the order of summands in a coproduct, distribution and appending units with $\unitlterm$ and $\unitrterm$. We will denote these transformations by $\lift{\cong}$. 

By $\vec f, \vec A$ we will denote vectors, i.e.~$\vec f = (f_1, \dots, f_n)$ for some natural number $n \geq 1$. If $f_1, \dots, f_n$ all have the same codomain $B$, which is also the domain of a term $g$, then we write $\vec f \comp g$ to denote $(f_1 \comp g, \dots, f_n \comp g)$.

We start with the notion of \intro{normalised drop term} (or \kl{drop terms} in normal form). The definition of normal form depends on the shape of a drop term's codomain.

\begin{definition}[Normalised drop terms]
	Suppose that $f : A \dropmap B$. We say that $f$ is in \emph{normal form} according to the shape of $B$, as follows.
	\begin{description}
		\item[Sum codomain.] If $B = B_1 + B_2$, then $f$ is equal to $\eithermany{i}{(f_i)}$, where each $f_i$ is equal to one of  
		\[
			g_i \comp \leftterm, \qquad g_i \comp \rightterm.
		\]

		\item[Product codomain.] If $B = B_1 \times B_2$, then $f$ is equal to
		\[
			\eithermany{i}{\left( x_i \times y_i \right)}.
		\]

		\item[List codomain.] If $B = C^*$, then $f$ is equal to
		\[
			F(\vec f) \comp \concmap,
		\]
		where $F$ is a \kl{list-polynomial functor} and $\vec f$ is a vector of \kl{drop terms} with codomain $C^*$, such that every $f_i$ is equal to one of
		\[
			(1 \comp v), \qquad [g_i], \qquad g_i^*.
		\]
	\end{description}
\end{definition}

In the following sections we will show that every drop term can be factored as a lift followed by a normalised drop (up to provable equality); we call this ``normalisation'' of drop terms. The cases of sum and product codomain are simpler and can be proved without dependency on any other results, which we do in the next section (\cref{sec:appendix:drops-with-sum-or-product-co-domain}). For the list codomain, we have to first establish some properties of drop terms composed with $\shiftterm$ (\cref{sec:appendix:shift-exchange}), and only then we are able to prove normalisation in \cref{sec:appendix:drops-with-list-co-domain}. 

With the normalisation in hand, we will be in a position to prove \cref{lem:exchange}; we present the proof in \cref{sec:appendix:exchange-proof}. Finally, we present the proof of \cref{lem:joint-drops-normalisation} in \cref{sec:appendix:joint-drops-simplification}.

\subsection{Normalisation with Sum or Product Codomain}\label{sec:appendix:drops-with-sum-or-product-co-domain}

First, we define the \intro{composition rank} of a \kl{drop term}, which will be used to perform induction in the proofs of \cref{lem:drops-with-sum-co-domain} and \cref{lem:drops-with-prod-co-domain}. The rank is defined as the length of the longest chain of compositions found along a single sequential thread of a drop term.

\begin{definition}[Composition rank]
	We define the \emph{rank} $\rk(f) \in \mathbb{N}$ of a \kl{drop term} $f$ as follows. If $f$ is atomic, then $\rk(f) = 0$. Otherwise, we define it inductively:
	\[
		\rk(f_1 \comp f_2) = \rk(f_1) + \rk(f_2) + 1, \quad \rk(f_1 \star f_2) = \max \left( \rk(f_1), \rk(f_2) \right), \quad \rk(f^*) = \rk(f),
	\]
	where $\star$ ranges over $\{ +, \times \}$. We write $g \preceq f$ if $\rk(g) \le \rk(f)$, and $g \prec f$ if $\rk(g) < \rk(f)$.
\end{definition}

Note that $\rk((1\comp{}v)) = 0$, because $(1\comp{}v)$ is an \kl{atomic drop term}, although syntactically it contains a composition. We extend the definition of composition rank to vectors of drop terms. By $\rk(\vec f)$ we mean the maximum rank of any $f_i$ in the vector $\vec f$. This allows us to write $g \preceq \vec f, \vec g \preceq f$, or $\vec g \preceq \vec f$.

The following two lemmas are proved by mutual induction with respect to $\preceq$. 

\begin{lemma}[Sum normalisation]\label{lem:drops-with-sum-co-domain}
	Suppose that $f$ is a \kl{drop term} with sum codomain. There exists a \kl{lift} $\alpha$ and \kl{drop terms} $\vec f$ such that 
	\[
		f \equiv \alpha \comp \eithermany{i}{(f_i)},
	\]
	where every $f_i$ is equal to one of $g_i \comp \leftterm$ or $g_i \comp \rightterm$ for some \kl{drop term} $g_i$ and $\vec g \preceq f$.
\end{lemma}

\begin{lemma}[Product normalisation]\label{lem:drops-with-prod-co-domain}
	Suppose that $f$ is a \kl{drop term} with product codomain. There exists a \kl{lift} $\alpha$ and \kl{drop terms} $\vec x, \vec y \preceq f$ such that 
	\[ 
		f \equiv \alpha \comp \eithermany{i}{\left( x_i \times y_i \right)}.
	\]
\end{lemma}

\begin{proof}[Proof of \cref{lem:drops-with-sum-co-domain}]
	We proceed by induction on $f$ with respect to $\preceq$. The easy cases are when $\rk(f) = 0$ or $f$ is a literal sum of drop terms $f_1 + f_2$. We leave these cases to the reader, as the remaining cases are strictly more involved. What we are left with is the case where $f$ is a composition $f_1 \comp f_2$. Here, we proceed by a nested induction on $f_2$ with respect to $\preceq$. First, suppose that $f_2$ is itself a composition, that is, $f_2 = g_1 \comp g_2$. Then, $f \equiv (f_1 \comp g_1) \comp g_2 \preceq f$ and $g_2 \prec f_2$, so we obtain the desired result by induction. 

	This leaves us with two cases: either $f_2$ is a sum of two drop terms, or $f_2$ is an atomic drop term. We consider these two cases separately. Crucially, in both of them we have $f_1 \prec f$, so we can normalise all terms at most $f_1$ in $\preceq$ by induction, as long as their codomain is a product or coproduct.

	\begin{description}
		\item[Sum, $f_2 = h_1 + h_2$.] By induction we have some $\alpha \in \Lift$ and $\vec f_1 \in \Drop$ such that $f_1 \equiv \alpha \comp \eithermany{i}{(f_i)}$. We get
		\begin{align*}
			f = f_1 \comp f_2 \equalbecause{induction} \\ 
			\alpha \comp \eithermany{i}{(f_i)} \comp (h_1 + h_2) \equalbecause{functoriality of copairing, \cref{lem:functoriality-of-either}} \\ 
			\alpha \comp \eithermany{i}{(f_i \comp h_1 + h_2)}.
		\end{align*}
		Now, for every $i$, we have that $f_i$ is either $g_i \comp \leftterm$ or $g_i \comp \rightterm$. In the former case, we have $f_i \comp (h_1 + h_2) \equiv g_i \comp \leftterm \comp (h_1 + h_2) \equiv (g_i \comp h_1) \comp \leftterm$ by naturality of $\leftterm$ (\cref{ax:naturality}) and $(g_i\comp{}h_1) \preceq f$. The latter case is symmetric.

		\item[Atomic drops.] The atomic drop terms which produce a sum output are $\idterm$, $\leftterm$, $\rightterm$, $(1\comp{}v)$, $\fstterm$, $\sndterm$, and $\codiagterm$. The cases where $f_2 \in \{ \leftterm, \rightterm \}$ are trivial. If $f_2 = \idterm$, then the statement holds by induction on $f_1$. If $f_2 = (1\comp{}v)$, then $v \equiv u\comp\leftterm$ or $v \equiv u\comp\rightterm$ for some \kl{value} $u$ by \cref{lem:value-completeness}. In the former case, we have $f \equiv f_1 \comp (v \comp \leftterm) \equiv (f_1 \comp v) \comp \leftterm$. The latter case is symmetric. 
		
		The remaining cases are $\fstterm$, $\sndterm$, and $\codiagterm$. We consider these three cases separately.

		\begin{description}
			\item[Projection, $f_2 = \fstterm$.] Suppose that $f_2 = \fstterm$ (the case $f_2 = \sndterm$ is symmetric). The construction first applies \cref{lem:drops-with-prod-co-domain} to extract the parts of input producing the left factor $(B_1 \times B_2)$, and then applies \cref{lem:drops-with-sum-co-domain} to normalise the summands. This is illustrated on the following diagram: 
			\[
			\begin{tikzcd}
				A
				\ar[rr, \dropar, "f_1"]
				\ar[d, \liftar, "\text{Lemmas } \ref{lem:drops-with-prod-co-domain} \text{ and } \ref{lem:drops-with-sum-co-domain}"']
				&\hspace*{2cm}&(B_1 + B_2) \times C
				\ar[dd, \dropar, "\fstterm"]
				\\ 
				\sum_i \left( \sum_j X_{i j} \right) \times Y_i
				\ar[urr, \dropar, "{\eithermany{i}{\left( \eithermany{j}{\left( x_{i j} \right)} \times y_i \right)}}"'{xshift=-25pt, yshift=-5pt}]
				\ar[d, \liftar, "\cong"']
				\\ 
				\sum_{i j} X_{i j} \times Y_i
				\ar[rr, \dropar, "{\eithermany{i j}{\left( \fstterm \comp x_{i j} \right)}}"]
				&&B_1 + B_2
			\end{tikzcd}
				\]
			The upper triangle of the diagram commutes by induction. The derivation for the lower face is as follows:
			\begin{align*}
				\eithermany{i}{\left( \eithermany{j}{\left( x_{i j} \right)} \times y_i \right)} \comp \fstterm \equalbecause{\cref{lem:functoriality-of-either}} \\
				\eithermany{i}{\left( \left( \eithermany{j}{\left( x_{i j} \right)} \times y_i \right) \comp \fstterm \right)} \equalbecause{naturality of $\fstterm$, \cref{ax:naturality}} \\
				\eithermany{i}{\left( \fstterm \comp \eithermany{j}{\left( x_{i j} \right)} \right)} \equalbecause{\cref{lem:distl-fst-coherence}} \\
				\eithermany{i}{\left( \distribute \comp \left( \sum \fstterm \right) \comp \eithermany{j}{\left( x_{i j} \right)} \right)} \equalbecause{\cref{lem:functoriality-of-either-left}} \\
				\distribute \comp \eithermany{i}{\left( \left( \sum \fstterm \right) \comp \eithermany{j}{\left( x_{i j} \right)} \right)} \equalbecause{\cref{lem:functoriality-of-either-left}} \\
				\distribute \comp \eithermany{i}{\left( \eithermany{j}{\left(\fstterm \comp x_{i j} \right)} \right)} \equalbecause{\cref{lem:assocr-fst-mapped-coherence}} \\
				\distribute \comp \eithermany{i j}{\left( \assocrterm \comp \left( \fstterm \comp x_{i j} \right) \right)} \equalbecause{\cref{lem:functoriality-of-either-left}} \\
				\distribute \comp \left(\sum \assocrterm \right) \comp \eithermany{i j}{\left( \fstterm \comp x_{i j} \right)}.
			\end{align*}

			\item[Codiagonal, $f_2 = \codiagterm$.] In this case, we apply \cref{lem:drops-with-sum-co-domain} to normalise the outer coproduct of the codomain, and then apply \cref{lem:drops-with-sum-co-domain} again to normalise the inner coproducts. This is illustrated on the following diagram (every $\rho_i$ is either $\leftterm$ or $\rightterm$):
			\[
			\begin{tikzcd}
				A
				\ar[rr, \dropar, "f_1"]
				\ar[d, \liftar, "\cref{lem:drops-with-sum-co-domain}"']
				&\hspace*{2cm}&(B_1 + B_2) + (B_1 + B_2)
				\ar[dd, \dropar, "\codiagterm"]
				\\
				\sum_i \left( \sum_j A_{ij} \right)
				\ar[urr, \dropar, "{\eithermany{i}{\left( \left( \eithermany{j}{f_{ij}} \right) \comp \rho_i \right)}}"'{xshift=-25pt, yshift=-5pt}]
				\ar[d, \liftar, "\cong"']
				\\ 
				\sum_{i j} A_{i j}
				\ar[rr, \dropar, "{\eithermany{i j}{(f_{ij})}}"]
				&&B_1 + B_2
			\end{tikzcd}
			\] 
			The upper triangle commutes by induction. For the lower face, we write:
			\begin{align*}
				\eithermany{i}{\left( \left( \eithermany{j}{(f_{ij})} \right) \comp \rho_i \right)} \comp \codiagterm \equalbecause{\cref{lem:functoriality-of-either}} \\ 
				\eithermany{i}{\left( \left( \eithermany{j}{(f_{ij})} \right) \comp \rho_i \comp \codiagterm \right)} \equalbecause{\cref{ax:eval-codiag}} \\ 
				\eithermany{i}{\left( \left( \eithermany{j}{(f_{ij})} \right) \right)} \equalbecause{coproduct reassociation} \\
				\assoc \comp \eithermany{i j}{(f_{ij})}. 
			\end{align*}
		\end{description}
	\end{description}
\end{proof}

\begin{proof}[Proof of \cref{lem:drops-with-prod-co-domain}]
	We proceed by induction on $f$ with respect to $\preceq$, analogously to the proof of \cref{lem:drops-with-sum-co-domain}. In particular, we reduce to the case where $f = f_1 \comp f_2$ and $f_2$ is either a product or an atomic drop term in the same way as with coproduct codomain. Again, we have $f_1 \prec f$, so we can normalise all terms at most $f_1$ in $\preceq$ by induction, as long as their codomain is a product or coproduct.

	\begin{description}
		\item[Product, $f_2 = h_1 \times h_2$.] By induction, we have some $\alpha \in \Lift$ and collections of terms $x_i, y_i \in \Drop$ such that $f_1 \equiv \alpha \comp \eithermany{i}{(x_i \times y_i)}$. We get
		\begin{align*}
			f = f_1 \comp f_2 \equalbecause{induction} \\ 
			\alpha \comp \left(\eithermany{i}{(x_i \times y_i)} \right) \comp (h_1 \times h_2) \equalbecause{functoriality of copairing, \cref{lem:functoriality-of-either}} \\
			\alpha \comp \eithermany{i}{\left( (x_i \times y_i) \comp (h_1 \times h_2)\right)} \equalbecause{functoriality of $\times$, \cref{ax:functoriality-of-type-constructors}} \\
			\alpha \comp \eithermany{i}{\left( \left( x_i \comp h_1 \right) \times \left( y_i \comp h_2 \right) \right)}.
		\end{align*}

		\item[Atomic drops.] The atomic drop terms which produce a product output are $\idterm$, $(1\comp{}v)$, $\assoclterm$, $\assocrterm$, $\fstterm$, $\sndterm$, and $\codiagterm$. If $f_2 = \idterm$, then the statement holds by induction on $f_1$. If $f_2 = (1 \comp v)$, then $v \equiv (u, w)$ for some values $u, w$ by \cref{lem:value-completeness}. Then the statement holds for $\alpha = \unitlterm$, $x_1 = (1 \comp u)$ and $y_1 = (1 \comp w)$:
		\begin{align*}
			f = f_1 \comp f_2 \equalbecause{the choice of $u, w$} \\ 
			f_1 \comp (1 \comp (u, w)) &= \why{definition of pairing} \\ 
			f_1 \comp (1 \comp \unitlterm \comp (u \times w)) \equalbecause{$\onetype$ is terminal, \cref{ax:unit-type}} \\
			1 \comp \unitlterm \comp (u \times w) \equalbecause{naturality of $\unitlterm$, \cref{ax:naturality}} \\
			\unitlterm \comp (\idterm \times 1) \comp (u \times w) \equalbecause{$\onetype$ is terminal, \cref{ax:unit-type}} \\
			\unitlterm \comp (1 \times 1) \comp (u \times w) \equalbecause{functoriality of $\times$, \cref{ax:functoriality-of-type-constructors}} \\
			\unitlterm \comp (1 \comp u) \times (1 \comp w).
		\end{align*}
		We are left with the cases where $f_2 \in \{ \assoclterm, \assocrterm, \fstterm, \sndterm, \codiagterm \}$. We consider these cases separately.
		
		\begin{description}
			\item[Product associativity, $f_2 = \assoclterm$.] Suppose that $f_2 = \assoclterm$ (the case where $f_2 = \assocrterm$ is symmetric). The construction applies \cref{lem:drops-with-prod-co-domain} to first normalise the outer product of the codomain, and then applies \cref{lem:drops-with-prod-co-domain} again to normalise the inner product. This is illustrated on the following diagram:
			\[
			\begin{tikzcd}
				A
				\ar[rr, \dropar, "f_1"]
				\ar[d, \liftar, "\cref{lem:drops-with-prod-co-domain}"']
				&\hspace*{2cm}&B \times (C \times D)
				\ar[dd, \dropar, "\assoclterm"]
				\\ 
				\sum_i X_i \times \left( \sum_j Y_{ij} \times Z_{ij} \right)
				\ar[urr, \dropar, "{\eithermany{i}{\left( x_i \times \left(  \eithermany{j}{y_{ij} \times z_{ij}} \right) \right)}}"'{xshift=-25pt, yshift=-5pt}]
				\ar[d, \liftar, "\cong"']
				\\ 
				\sum_{ij} (X_i \times Y_{ij}) \times Z_{ij}
				\ar[rr, \dropar, "{\eithermany{ij}{\left((x_i \times y_{ij}) \times z_{ij}\right)}}"]
				&&(B \times C) \times D
			\end{tikzcd}
			\]
			The upper triangle commutes by induction. For the lower face, we write:
			\begin{align*}
				\eithermany{i}{\left( x_i \times \left( \eithermany{j}{y_{ij} \times z_{ij}} \right) \right)} \comp \assoclterm \equalbecause{\cref{lem:distributivity-of-either}} \\
				\distribute \comp \eithermany{ij}{\left( x_i \times \left( y_{ij} \times z_{ij} \right) \right)} \comp \assoclterm \equalbecause{\cref{lem:functoriality-of-either}} \\ 
				\distribute \comp \eithermany{ij}{\left( x_i \times \left( y_{ij} \times z_{ij} \right) \comp \assoclterm \right)} \equalbecause{naturality of $\assoclterm$, \cref{ax:naturality}} \\ 
				\distribute \comp \eithermany{ij}{\left( \assoclterm \comp \left( \left( x_i \times y_{ij} \right) \times z_{ij} \right) \right)} \equalbecause{\cref{lem:functoriality-of-either-left}} \\ 
				\distribute \comp \left(\sum \assoclterm \right) \comp  \eithermany{ij}{\left( \left( x_i \times y_{ij} \right) \times z_{ij} \right)}
			\end{align*}

			\item[Projection, $f_2 = \fstterm$.] Suppose that $f_2 = \fstterm$ (the case where $f_2 = \sndterm$ is symmetric). In this case, we again apply \cref{lem:drops-with-prod-co-domain} to normalise the outer product, and then apply \cref{lem:drops-with-prod-co-domain} again to normalise the inner product. The construction is presented on the following diagram: 
			\[
			\begin{tikzcd}
				A
				\ar[rr, \dropar, "f_1"]
				\ar[d, \liftar, "\cref{lem:drops-with-prod-co-domain}"']
				&\hspace*{2cm}& (B_1 \times B_2) \times C
				\ar[dd, \dropar, "\fstterm"]
				\\ 
				\sum_i \left( \sum_j X_{ij} \times Y_{ij} \right) \times Z_i
				\ar[urr, \dropar, "{\eithermany{i}{\left( \left( \eithermany{j}{(x_{ij} \times y_{ij})} \right) \times z_i \right)}}"'{xshift=-25pt, yshift=-5pt}]
				\ar[d, \liftar, "\cong"']
				\\ 
				\sum_{ij} X_{ij} \times (Y_{ij} \times Z_i)
				\ar[rr, \dropar, "\eithermany{ij}{\left( x_{ij} \times (\fstterm \comp y_{ij}) \right)}"]
				&&B_1 \times B_2
			\end{tikzcd}
			\]
			The upper triangle commutes by induction. For the lower face, we write:
			\begin{align*}
				\eithermany{i}{\left( \left( \eithermany{j}{\left( x_{ij} \times y_{ij} \right)} \right) \times z_i \right)} \comp \fstterm \equalbecause{\cref{lem:functoriality-of-either}} \\ 
				\eithermany{i}{\left( \left( \eithermany{j}{\left( x_{ij} \times y_{ij} \right)} \right) \times z_i \comp \fstterm\right)} \equalbecause{naturality of $\fstterm$, \cref{ax:naturality}} \\ 
				\eithermany{i}{\left( \fstterm \comp \left( \eithermany{j}{\left( x_{ij} \times y_{ij} \right)} \right) \right)} \equalbecause{\cref{lem:distl-fst-coherence}} \\ 
				\eithermany{i}{\left( \distribute \comp \left(\sum \fstterm \right) \comp \left( \eithermany{j}{\left( x_{ij} \times y_{ij} \right)} \right) \right)} \equalbecause{\cref{lem:functoriality-of-either-left}} \\
				\distribute \comp \eithermany{i}{\left( \left( \sum \fstterm \right) \comp \left( \eithermany{j}{\left( x_{ij} \times y_{ij} \right)} \right) \right)} \equalbecause{\cref{lem:functoriality-of-either-left}} \\
				\distribute \comp \eithermany{i}{\left( \eithermany{j}{\left( \fstterm \comp \left( x_{ij} \times y_{ij} \right) \right)} \right)} \equalbecause{\cref{lem:assocr-fst-mapped-coherence}} \\ 
				\distribute \comp \eithermany{ij}{\left( \assocrterm \comp \left( x_{ij} \times (\fstterm \comp y_{ij}) \right) \right)} \equalbecause{\cref{lem:functoriality-of-either-left}} \\ 
				\distribute \comp \left( \sum \assocrterm \right) \comp \eithermany{ij}{\left( \left( x_{ij} \times (\fstterm \comp y_{ij}) \right) \right)}.
			\end{align*}

			\item[Codiagonal, $f_2 = \codiagterm$.] In this case, we apply \cref{lem:drops-with-sum-co-domain} to normalise the outer coproduct of the codomain, and then apply \cref{lem:drops-with-prod-co-domain} again to normalise the inner products. This is illustrated on the following diagram (every $\rho_i$ is either $\leftterm$ or $\rightterm$):
			\[
			\begin{tikzcd}
				A
				\ar[rr, \dropar, "f_1"]
				\ar[d, \liftar, "\text{Lemmas \ref{lem:drops-with-sum-co-domain} and \ref{lem:drops-with-prod-co-domain}}"']
				&\hspace*{2cm}&(B_1 \times B_2) + (B_1 \times B_2)
				\ar[dd, \dropar, "\codiagterm"]
				\\ 
				\sum_i \left( \sum_j X_{ij} \times Y_{ij} \right)
				\ar[d, \liftar, "\cong"']
				\ar[urr, \dropar, "{\eithermany{i}{\left( \left( \eithermany{j}{(x_{ij} \times y_{ij})} \right) \comp \rho_i \right)}}"'{xshift=-30pt, yshift=-5pt}]
				\\ 
				\sum_{ij} X_{ij} \times Y_{ij}
				\ar[rr, \dropar, "{\eithermany{ij}{\left(x_{ij} \times y_{ij} \right)}}"]
				&&B_1 \times B_2
			\end{tikzcd}
			\]
			The upper triangle commutes by induction. For the lower face, we write:
			\begin{align*}
				\eithermany{i}{\left( \left( \eithermany{j}{(x_{ij} \times y_{ij})} \right) \comp \rho_i \right)} \comp \codiagterm \equalbecause{\cref{lem:functoriality-of-either}} \\ 
				\eithermany{i}{\left( \left( \eithermany{j}{(x_{ij} \times y_{ij})} \right) \comp \rho_i \comp \codiagterm \right)} \equalbecause{\cref{ax:eval-codiag}} \\ 
				\eithermany{i}{\left( \left( \eithermany{j}{(x_{ij} \times y_{ij})} \right) \right)} \equalbecause{coproduct reassociation} \\
				\assoc \comp \eithermany{i j}{(x_{ij} \times y_{ij})}. 
			\end{align*}
		\end{description}
	\end{description}
\end{proof}

\subsection{Shift Exchange Lemma}\label{sec:appendix:shift-exchange}

In this section we prove a special case of the exchange lemma (\cref{lem:exchange}) where the drop-lift composition consists of a \kl{drop term} in \kl[normalised drop term]{normal form} followed by $\shiftterm$. This special case will be used in the proof of the normalisation lemma for drops with list codomain (\cref{lem:drops-with-list-co-domain}), which in turn will be used in the general proof of \cref{lem:exchange}. 

We start with the observation that the codomain of $\shiftterm$ is \kl[injective type]{injective}.

\begin{lemma}\label{lem:shift-type-is-injective}
    The \kl{type} $\onetype + S \times (T \times S)^* \times T$ is \kl[injective type]{injective}.
\end{lemma}

\begin{proof}
    Note that the \kl{concatenation map} of type $\onetype + S \times (T \times S)^* \times T \rightarrow (S + T)^*$ is equal to
    \[ 
        \either{[]}{\left( [\leftterm] \times ((T \times S)^* \concmap (S + T)^*) \times [\rightterm] \right)}.
    \] 
    Injectivity translates through type constructors and $\mathtt{either}$ by \cref{ax:eval-codiag}. The term $[] : \onetype \rightarrow (S + T)^*$ is \kl{injective} by \cref{ax:unit-type} with $\unitlterm \comp \fstterm$ as the inverse. Both $\pureterm$ and the coprojections $\leftterm, \rightterm$ are injective, although the inverse must be built in a nonnatural way, i.e.~they involve ``choosing'' a \kl{value} of type $S$ or $T$, which is always possible because all types are inhabited. Injectivity of $\pureterm$ is discussed in the proof of \cref{lem:injectivity-of-cyclic-annotation}. For $\leftterm : S \rightarrow S + T$, choose any value $\onetype \xrightarrow{w} S$ and observe that 
    \begin{align*} 
        \leftterm \comp \idterm + (1\comp{}w) \comp \codiagterm \equalbecause{naturality of $\leftterm$, \cref{ax:naturality}} \\
        \idterm \comp \leftterm \comp \codiagterm \equalbecause{\cref{ax:eval-codiag}} \\ 
        \idterm.
    \end{align*}
    Therefore, it suffices to show that the type $(T \times S)^*$ is injective.
    Due to \cref{lem:uniqueness-of-rebracketings} (the type $(S + T)^*$ is trivially injective), it further suffices to construct an injective \kl{rebracketing} of the type $(T \times S)^* \rightarrow (S + T)^*$. This rebracketing is given by:
    \[ 
        (T \times S)^* \xrightarrow{([], \left( \idterm \times \pureterm \right)^*)} S^* \times (T \times S^*)^* \xrightarrow{\unblockterm} (S + T)^*.
    \]
    The term $([], \idterm)$ is injective by \cref{ax:concatenation-monoid} and $\unblockterm$ is injective by \cref{ax:isomorphisms}. The fact that this term is a rebracketing follows from \cref{lem:rebracketing-basic-facts}.
\end{proof}

We now proceed to the special case of the exchange lemma (\cref{lem:exchange}) with $\shiftterm$. Note that the conclusion of the following statement is stronger than the general conclusion of \cref{lem:exchange} -- this will be needed to wrap the induction step in the proof of \cref{lem:drops-with-list-co-domain}.

\begin{lemma}[Shift exchange]\label{lem:exchange-normal-shift}
    Suppose that $\left( F(\vec f) \comp \concmap \right): F(\vec A) \dropmap (S \times T)^*$ is in \kl[normalised drop term]{normal form}. There exists a \kl{lift} $\alpha$ and \kl{drop terms} $\vec g$ such that 
    \[
        \left( F(\vec f) \comp \concmap \right) \comp \shiftterm \equiv \alpha \comp \eithermany{i}{(g_i)},
    \]
    and every $g_i$ is equal to one of:
    \[
        (1 \comp 1) \comp \leftterm \qquad \text{or} \qquad (\sndterm \comp x_i) \times \left( \myunderbrace{H_i(\vec h_i) \comp \concmap}{\text{normal form}} \right) \times (\fstterm \comp y_i) \comp \rightterm,
    \]
    where in the second case, $H_i$ is a \kl{list-polynomial functor}, $\vec x, \vec y \preceq \vec f$ and every $h_i$ is a \kl{drop term} with codomain $(S \times T)^*$ of one of the following forms:
    \[
        (1 \comp v), \qquad [\fstterm \times \idterm \comp k_i], \qquad \text{or} \qquad (\fstterm \times \idterm \comp k_i)^*,
    \]
    where every $k_i$ is a \kl{drop term} with codomain $(S \times T)$ and $\vec k \preceq \vec f$.
\end{lemma}
\begin{proof}  
    We proceed by induction on the structure of the list-polynomial functor $F$.
    \begin{description}
        \item[Unit node, $F = \onetype$.] Here $\left( F(\vec f) \comp \concmap \right) = []$. The statement holds with $\alpha = \idterm$ and $g = (1\comp1) \comp \leftterm$, because $[] \comp \shiftterm \equiv \leftterm$ by \cref{lem:uniqueness-of-rebracketings}, because the codomain is injective (\cref{lem:shift-type-is-injective}) and both terms are rebracketings by \cref{lem:rebracketing-basic-facts}. Moreover, $\leftterm \equiv (1\comp1) \comp \leftterm$ by terminality of $\onetype$ (\cref{ax:unit-type}).
        
        \item[Leaf node, $F = X_i$.] For simplicity we omit the index $i$ and write $f := f_i$. We consider three subcases depending on the shape of $f$, as described in the definition of normal form.
        
        \begin{description}
            \item[Value, $f = (1 \comp v)$.] In this case, $\left( F(\vec f) \comp \concmap \right) \equiv (1 \comp v)$. If $v$ is equivalent to $[]$, then we can take $\alpha = \idterm$ and $g = (1 \comp 1) \comp \leftterm$, as in the unit case. Otherwise, there are values $s, u, t$ such that $v \comp \shiftterm \equiv (s, u, t) \comp \rightterm$ by \cref{lem:value-completeness}. The statement holds for a lift $\alpha$ which appends units $A \multimap (\onetype \times \onetype) \times A \times (\onetype \times \onetype)$, and $g = (\sndterm \comp s) \times (1\comp{}u) \times (\fstterm \comp t) \comp \rightterm$ by the choice of values $s, u, t$ and the fact that appending units is reversed by projections (\cref{ax:isomorphisms}).
  
            \item[Singleton, $f = {[g]}$.] We have $\left(F(\vec f) \comp \concmap \right) \equiv [g]$, where $g : A \dropmap S \times T$. By \cref{lem:drops-with-prod-co-domain} applied to $g$, there is a lift $\beta$ and drop terms $\vec x, \vec y \preceq f$ such that $g \equiv \beta \comp \eithermany{i}{(x_i \times y_i)}$. We obtain:
            \begin{align*}
                [g] \comp \shiftterm \equalbecause{\cref{lem:drops-with-prod-co-domain}} \\ 
                [\beta \comp \eithermany{i}{(x_i \times y_i)}] \comp \shiftterm \equalbecause{\cref{lem:functoriality-of-either}} \\ 
                \beta \comp \eithermany{i}{\left([x_i \times y_i] \comp \shiftterm \right)} \equalbecause{\cref{lem:functoriality-of-either-left}} \\
                \beta \comp \left( \sum  x_i \times y_i \right) \comp \eithermany{i}{\left([\idterm \times \idterm] \comp \shiftterm \right)} \equalbecause{\cref{lem:uniqueness-of-rebracketings}} \\ 
                \beta \comp \left( \sum  x_i \times y_i \right) \comp \eithermany{i}{\left( (\unitrterm \times \idterm) \comp (\idterm \times [] \times \idterm) \comp \rightterm \right)} \equalbecause{Lemmas \ref{lem:functoriality-of-either} and \ref{lem:functoriality-of-either-left}} \\
                \beta \comp \eithermany{i}{\left( (x_i \times y_i) \comp (\unitrterm \times \idterm) \comp (\idterm \times [] \times \idterm) \right)} \comp \rightterm \equalbecause{functoriality of $\times$, \cref{ax:functoriality-of-type-constructors}} \\ 
                \beta \comp \eithermany{i}{\left( (x_i \comp \unitrterm \comp \idterm \times []) \times y_i \right)} \comp \rightterm \equalbecause{naturality of $\unitrterm$, \cref{ax:naturality}} \\ 
                \beta \comp \eithermany{i}{\left( (\unitrterm \comp x_i \times []) \times y_i \right)} \comp \rightterm \equalbecause{\cref{lem:functoriality-of-either-left}} \\
                \beta \comp \left( \sum \unitrterm \times \idterm \right) \comp \eithermany{i}{\left( x_i \times [] \times y_i \right)} \comp \rightterm \equalbecause{$\onetype$ is terminal, \cref{ax:unit-type}} \\ 
                \beta \comp \left( \sum \unitrterm \times \idterm \right) \comp \eithermany{i}{\left( x_i \times (1 \comp []) \times y_i \right)} \comp \rightterm.
            \end{align*}
            At the end, to formally match the required shape, we can append (by lifts) artificial units with $\unitlterm$ and $\unitrterm$ to have $(\sndterm \comp x_i)$ and $(\fstterm \comp y_i)$, instead of $x_i$ and $y_i$, which is valid by the fact that $\unitlterm \comp \sndterm \equiv \unitrterm \comp \fstterm \equiv \idterm$ by \cref{ax:isomorphisms}.

            \item[List, $f = g^*$.] In this case, $\left(F(\vec f) \comp \concmap \right) \equiv g^*$, where $g : A \dropmap S \times T$. By \cref{lem:drops-with-prod-co-domain} applied to $g$, there is a lift $\beta$ and drop terms $\vec x, \vec y \preceq f$ such that $g \equiv \beta \comp \eithermany{i}{(x_i \times y_i)}$. We can intuitively see that after applying $\beta$, the domain is sufficiently ``granular'' to allow us to extract the leading $S$ and trailing $T$ of the codomain, but the technical details are a bit tricky. This is because of the \kl{copairing} in the normalisation of $g$: we have to proceed by induction on the number of its variants. 
            
            The base case where there is only one variant, i.e.~$g \equiv \beta \comp (x \times y)$, follows by naturality of $\shiftterm$:
            \begin{align*}
                g^* \comp \shiftterm \equalbecause{\cref{lem:drops-with-prod-co-domain}} \\ 
                \beta^* \comp (x \times y)^* \comp \shiftterm \equalbecause{naturality of $\shiftterm$} \\
                \beta^* \comp \shiftterm \comp \idterm + x \times (y \times x)^* \times y \equalbecause{\cref{ax:unit-type} and \cref{lem:either-uniqueness}} \\ 
                \beta^* \comp \shiftterm \comp \either{\left( (1 \comp 1) \comp \leftterm \right)}{\left( x \times (y \times x)^* \times y \comp \rightterm \right)}.
            \end{align*}
            Again, to formally comply with the required shape, we can append (by lifts) artificial units which are then reversed by projections. 

            Moving forward, we describe how to handle the case of two variants -- the general construction will then follow by iterative application of the same argument, which we omit for brevity. The case of two variants is already interesting because it involves $\blockterm$ as part of the lift. Suppose that $g \equiv \beta \comp \either{(x_1 \times y_1)}{(x_2 \times y_2)}$. We first demonstrate a sketch of the construction on the following diagram (let $Z = \onetype + S \times (T \times S)^* \times T$):
            \[ 
            \begin{tikzcd}[column sep = 1cm]
                A^*
                \ar[r, \dropar, "{g^*}"]
                \ar[d, \liftar, "{\beta^*}"]
                &(S \times T)^*
                \ar[r, \liftar, "\shiftterm"]
                &\onetype + S \times (T S)^* \times T
                \\ 
                (X_1 Y_1 + X_2 Y_2)^*
                \ar[r, \dropar, "(x_1 y_1 + x_2 y_2)^*"]
                \ar[d, \liftar, "\blockterm"]
                &(S \times T + S \times T)^*
                \ar[u, \dropar, "\eithermany{i}{\left( \idterm \times \idterm \right)}"']
                \ar[d, \liftar, "\blockterm"]
                \\ 
                \bullet 
                \ar[d, \liftar, "{\shiftterm \times (\idterm \times \shiftterm)^*}"]
                &(S \times T)^* \times ((S \times T) \times (S \times T)^*)^*
                \ar[d, \liftar, "{\shiftterm \times (\idterm \times \shiftterm)^*}"]
                \\ 
                \bullet 
                \ar[d, \liftar, "\cong"]
                &Z \times \left( (S \times T) \times Z \right)^*
                \ar[d, \liftar, "\cong"]
                \\ 
                \bullet 
                \ar[d, \liftar, "\idterm \times \shiftterm"]
                &Z \times \left( S \times (T \times Z) \right)^*
                \ar[d, \liftar, "\idterm \times \shiftterm"]
                \\ 
                \bullet 
                \ar[d, \liftar, "\cong"]
                &Z \times \left( \onetype + S ((T Z) S)^* (T Z) \right)
                \ar[d, \liftar, "\cong"]
                \\
                \bullet 
                \ar[d, \liftar, "\cong"]
                &D_1 + D_2 + D_3 + D_4 + D_5 + D_6
                \ar[d, \liftar, "\cong"]
                \\
                \bullet 
                \ar[r, \dropar, "{x_i, y_i \text{ functorially}}"]
                &E_1 + E_2 + E_3 + E_4 + E_5 + E_6
                \ar[uuuuuuur, \dropar, "\eithermany{i}{(g_i)}"', bend right=40]
            \end{tikzcd}
            \] 
            The last two steps consist of distributing the type and appending units to match the required shape of the drop terms. The types $D_i$ are as follows:
            \begin{align*}
                D_1 &= \onetype \times \onetype \\ 
                D_2 &= (\onetype S) \times ((T \onetype) S)^* \times (T \onetype) \\ 
                D_3 &= (\onetype S) \times (TZS)^* \times (T \onetype) \\ 
                D_4 &= (\onetype S) \times ( (TZS)^* T S (TS)^* ) \times (T \onetype) \\ 
                D_5 &= (\onetype S) \times ((TS)^* TS (TZS)^*) \times (T \onetype) \\ 
                D_6 &= (\onetype S) \times ((TS)^* TS (TZS)^* TS (TS)^*) \times (T \onetype)
            \end{align*}
            The types $E_i$ are similar to types $D_i$, but with slight further lift modifications which we will describe in a moment.
            
            The term $g_1$ is $(1 \comp 1) \comp \leftterm$ (in particular, $E_1 = D_1$), and the other $g_i$ produce nonempty shifted lists and are of the form $(\sndterm \comp \idterm) \times (H_i(\vec h_i) \comp \concmap) \times (\fstterm \comp \idterm) \comp \rightterm$. We describe how $H_3(\vec h_3)$ should look, the other variants are analogous. We want a normal drop term which produces a list of pairs $(TS)^*$ from $(TZS)^*$ and complies with the required shape of the lemma. This is where we need to further modify the type $(TZS)^*$ with lifts to obtain $E_3$ which will be consumed in the end. What we need to do is to distribute the coproduct hidden under $Z$ and append units everywhere between $T$ and $S$. We obtain the following type: 
            \[
                (TZS)^* \xmultimap{\cong} T \onetype S + ((T \onetype) S) ((T \onetype) S)^* ((T \onetype) S).
            \] 
            This is then consumed with the following drop term: 
            \[
                \left( H_3(\vec h_3) \comp \concmap \right) = \either{[\fstterm \times \idterm]}{\left( [\fstterm \times \idterm] \odot (\fstterm \times \idterm)^* \odot [\fstterm \times \idterm] \right)}.
            \]
            There are two things to comment on. First, the bottom drop path of the diagram is a functorial combination of the drop terms $x_i, y_i$ and the drops we have just described, so it satisfies the required shape of the lemma by \cref{lem:functoriality-of-either-left} and naturality of projections (\cref{ax:naturality}). Second, the diagram commutes: the upper-left square commutes by \cref{lem:drops-with-prod-co-domain}, the lower-left face commutes by naturality of used lift terms, and the right face commutes by \cref{lem:uniqueness-of-rebracketings}, because both paths are rebracketings by \cref{lem:rebracketing-basic-facts} and the target type is injective by \cref{lem:shift-type-is-injective}. 
        \end{description}

        \item[Sum node, $F = F_1 + F_2$.] By definition of $\concmap$ on a sum, we have
        \[
            \left( F(\vec f) \comp \concmap \right) = \either{\left(F_1(\vec f) \comp \concmap \right)}{\left(F_2(\vec f) \comp \concmap \right)}.
        \]
        The construction proceeds by applying the induction hypothesis to $F_1$ and $F_2$ separately, and then combining the results by reassociating the nested coproducts. By induction, there are lifts $\alpha_1, \alpha_2$ and drop vectors $\vec g^{(1)}, \vec g^{(2)} \preceq \vec f$, with $F_k(\vec f_k) \comp \concmap \comp \shiftterm \equiv \alpha_k \comp \eithermany{i}{(g_i^{(k)})}$ for $k \in \{1, 2\}$. We get 
        \begin{align*}
            \either{(F_1(\vec f) \comp \concmap)}{(F_2(\vec f) \comp \concmap)} \comp \shiftterm \equalbecause{\cref{lem:functoriality-of-either}} \\
            \either{(F_1(\vec f) \comp \concmap \comp \shiftterm)}{(F_2(\vec f) \comp \concmap \comp \shiftterm)} \equalbecause{induction} \\
            \either{(\alpha_1 \comp \eithermany{i}{(g_i^{(1)})})}{(\alpha_2 \comp \eithermany{i}{(g_i^{(2)})})} \equalbecause{\cref{lem:functoriality-of-either-left}} \\ 
            (\alpha_1 + \alpha_2) \comp \either{(\eithermany{i}{(g_i^{(1)})})}{(\eithermany{i}{(g_i^{(2)})})} \equalbecause{reassociation of coproducts} \\
            (\alpha_1 + \alpha_2) \comp \assoc \comp \eithermany{i j}{(g^{(j)}_i)}.
        \end{align*}

        \item[Product node, $F = F_1 \times F_2$.] By definition of $\concmap$ on a product, we have
        \[
            \left( F(\vec f) \comp \concmap \right) = \left( F_1(\vec f) \comp \concmap \right) \odot \left( F_2(\vec f) \comp \concmap \right).
        \]
        The construction proceeds by applying the induction hypothesis to $F_1$ and $F_2$ separately, and then combining the results by distributing the outer product over the inner coproducts. Let $\mu = (\either{[]}{\unshiftterm})^2 \comp \binconcatterm \comp \shiftterm$.
        
        Now, observe that the following diagram commutes by induction, product functoriality and distribution naturality:
        \[
        \begin{tikzcd}[column sep = 1cm]
            F_1(\vec A) \times F_2(\vec A)
            \ar[rr, \dropar, "{(F_1(\vec f)\comp\concmap) \times (F_2(\vec f)\comp\concmap)}"]
            \ar[d, \liftar, "\text{induction}"]
            &\hspace*{2cm}
            &\left((S \times T)^*\right)^2
            \ar[r, \dropar, "\binconcatterm"]
            \ar[d, \liftar, "{\shiftterm^2}"]
            &(S \times T)^*
            \ar[d, \liftar, "\shiftterm"]
            \\ 
            \sum_i B^{(1)}_i \times \sum_i B^{(2)}_i
            \ar[rr, \dropar, "{\eithermany{i}{(g^{(1)}_i)} \times \eithermany{i}{(g^{(2)}_i)}}"]
            \ar[d, \liftar, "\cong"]
            &&(\onetype + S(TS)^* T)^2 
            \ar[r, "\mu"]
            &\onetype + S(TS)^* T
            \\ 
            \sum_{i j} B^{(1)}_i \times B^{(2)}_j 
            \ar[urr, \dropar, "{\eithermany{i j}{(g^{(1)}_i \times g^{(2)}_j)}}"'{xshift=-0.7cm, yshift=-0.1cm}]
        \end{tikzcd}
        \]
        Therefore, it remains to show every $g^{(1)}_i \times g^{(2)}_j$ can be further lifted by $\beta_{i j}$ into a drop term $g_{i j}$ of desired shape, so that $\beta_{i j} \comp g_{i j} \equiv g^{(1)}_i \times g^{(2)}_j \comp \mu$ -- the statement will then follow by a functorial combination for all $i,j$. We consider three cases (up to symmetry) depending on which coprojection of the coproduct each $g^{(1)}_i$ and $g^{(2)}_j$ is equal to. 

        \begin{description}
            \item[Both empty.] If $g^{(1)}_i = (1 \comp 1) \comp \leftterm$ and $g^{(2)}_j = (1 \comp 1) \comp \leftterm$, then we can take $\beta_{i j} = \idterm$ and $g_{i j} = (1 \comp 1) \comp \leftterm$:
            \begin{align*}
                ((1 \comp 1) \comp \leftterm)^2 \comp \mu \equalbecause{$\onetype$ is terminal, \cref{ax:unit-type}} \\ 
                (1 \comp \leftterm)^2 \comp \mu &= \why{unfold $\mu$} \\
                (1 \comp \leftterm)^2 \comp (\either{[]}{\unshiftterm})^2 \comp \binconcatterm \comp \shiftterm \equalbecause{\cref{lem:case-equivalence}} \\
                (1 \comp [])^2 \comp \binconcatterm \comp \shiftterm \equalbecause{\cref{lem:discarding-pair-to-empty-list}} \\
                (1 \comp []) \comp \shiftterm \equalbecause{\cref{lem:uniqueness-of-rebracketings}} \\
                (1 \comp \leftterm) \equalbecause{$\onetype$ is terminal, \cref{ax:unit-type}} \\
                (1 \comp 1) \comp \leftterm.
            \end{align*}

            \item[One nonempty.] Suppose that $g^{(1)}_i = (1 \comp 1) \comp \leftterm$ and $g^{(2)}_j = (\sndterm\comp{}x) \times (H(\vec h) \comp \concmap) \times (\fstterm\comp{}y) \comp \rightterm$ (the symmetric case is analogous). In this case we have 
            \begin{align*} 
                ((1 \comp 1) \comp \leftterm) \times ((\sndterm\comp{}x) \times (H(\vec h) \comp \concmap) \times (\fstterm\comp{}y)) \comp \mu \equalbecause{\cref{lem:case-equivalence} and \cref{ax:unit-type}} \\ 
                (1 \comp []) \times ((\sndterm\comp{}x) \times (H(\vec h) \comp \concmap) \times (\fstterm\comp{}y) \comp \unshiftterm) \\
                {} \comp\binconcatterm \comp \shiftterm \equalbecause{\cref{lem:product-with-empty-list-fst-binconcat}} \\ 
                (1 \comp []) \times ((\sndterm\comp{}x) \times (H(\vec h) \comp \concmap) \times (\fstterm\comp{}y) \comp \unshiftterm) \comp \sndterm \comp \shiftterm \equalbecause{naturality of $\sndterm$, \cref{ax:naturality}} \\ 
                \sndterm \comp (\sndterm\comp{}x) \times (H(\vec h) \comp \concmap) \times (\fstterm\comp{}y) \comp (\unshiftterm \comp \shiftterm) \equalbecause{\cref{lem:uniqueness-of-rebracketings}} \\
                \sndterm \comp (\sndterm\comp{}x) \times (H(\vec h) \comp \concmap) \times (\fstterm\comp{}y) \comp \rightterm \equalbecause{naturality of associativity} \\ 
                \assoc \comp (\sndterm \comp (\sndterm\comp{}x)) \times (H(\vec h) \comp \concmap) \times (\fstterm\comp{}y) \comp \rightterm \equalbecause{\cref{lem:snd-snd-is-assocl-snd}} \\ 
                \assoc \comp (\sndterm\comp{}x) \times (H(\vec h) \comp \concmap) \times (\fstterm\comp{}y) \comp \rightterm.
            \end{align*}

            \item[Both nonempty.] Suppose that $g^{(1)}_i = (\sndterm\comp{}x_1) \times (H_1(\vec h_1) \comp \concmap) \times (\fstterm\comp{}y_1) \comp \rightterm$ and $g^{(2)}_j = (\sndterm\comp{}x_2) \times (H_2(\vec h_2) \comp \concmap) \times (\fstterm\comp{}y_2) \comp \rightterm$. The construction consists of reassociating products so that the $T$ produced by $y_1$ and the $S$ produced by $x_2$ are combined into one pair $TS$. A sketch of the construction is illustrated on the diagram:
            \[
            \begin{tikzcd}[column sep = 0.6cm]
                (\bullet X_1) H_1(\vec A_1) (Y_1 \bullet) \times (\bullet X_2) H_2(\vec A_2) (Y_2 \bullet)
                \ar[r, \dropar, "{g^{(1)}_i \times g^{(2)}_j}"]
                \ar[d, \liftar, "\cong"]
                &(S(TS)^* T)^2 
                \ar[r, "{\rightterm^2 \comp \mu}"]
                \ar[d, \liftar, "\cong"]
                &\onetype + S(TS)^* T
                \\ 
                (\bullet X_1) \left( H_1(\vec A_1) \times ((Y_1 (\bullet \bullet)) X_2) \times H_2(\vec A_2) \right) (Y_2 \bullet)
                \ar[r, \dropar, "g"]
                &S (TS)^* (TS) (TS)^* T 
                \ar[ur, \dropar]
            \end{tikzcd}
            \]
            The diagonal drop term is the obvious term which collects the $TS$ pairs and then applies the right coprojection. Therefore, the right-hand triangle commutes by \cref{lem:uniqueness-of-rebracketings}. The drop $g$ on the bottom is the functorial application of components of $g_i^{(1)}$ and $g_j^{(2)}$ after the reassociation lift. The left-hand square commutes by naturality of associativity (\cref{ax:naturality}) and \kl{projections coherence laws} (\cref{lem:projections-coherence-laws}). The composed bottom-path drop is equivalent to the following term, which satisfies the required shape of the lemma:
            \[
                (\sndterm \comp x_1) \times \left(H_1(\vec h_1) \times [\fstterm \times \idterm \comp y_1 \times x_2] \times H_2(\vec h_2) \comp \concmap \right) \times (\fstterm \comp y_2) \comp \rightterm.
            \]
        \end{description}

        \item[List node, $F = G^*$.] By definition of $\concmap$ on a list, we have
        \[
            \left( F(\vec f) \comp \concmap \right) = \hom \left( G(\vec f) \comp \concmap \right).
        \]
        The construction proceeds by applying the induction hypothesis to $G$, and then ``folding'' many shifted sublists into one shifted list. First, let $\mu = (\either{[]}{\unshiftterm})^* \comp \concatterm \comp \shiftterm$. The following diagram commutes by induction:
        \[
        \begin{tikzcd}[column sep = 2.5cm]
            G(\vec A)^*
            \ar[r, \dropar, "{(G(\vec f) \comp \concmap)^*}"]
            \ar[d, \liftar, "\text{induction}"]
            &(S \times T)^{**}
            \ar[r, \dropar, "\concatterm"]
            \ar[d, \liftar, "{\shiftterm^*}"]
            &(S \times T)^*
            \ar[d, \liftar, "\shiftterm"]
            \\ 
            \left( \sum_i B_i \right)^*
            \ar[r, \dropar, "{\left( \eithermany{i}{(g_i)} \right)^*}"]
            &(\onetype + S(TS)^* T)^* 
            \ar[r, "\mu"]
            &\onetype + S(TS)^* T
        \end{tikzcd}
        \]
        We now briefly describe how to further lift the bottom path of the diagram to obtain a lift-drop composition of required shape which is equivalent to $(\eithermany{i}{(g_i)})^* \comp \mu$. The idea is similar to the list subcase of the leaf node functor case described earlier: we proceed by induction on the number of variants $g_i$. We use $\blockterm$ to separate the domain list into uniform sublists, then apply $\shiftterm$ to each of them, and finally reassemble the shifted sublists into one with proper reassociating and concatenation maps. One additional technicality which was not present in the leaf node case is that some variants may produce outputs in the left summand (empty lists). The details are tedious but straightforward, and we omit them for brevity.
    \end{description}
\end{proof}

\subsection{Normalisation with List Codomain}\label{sec:appendix:drops-with-list-co-domain}

In this section, we describe the remaining variant of \kl{drops} normalisation for a list codomain.

\begin{lemma}[List normalisation]\label{lem:drops-with-list-co-domain}
	Suppose that $f$ is a \kl{drop term} with list codomain $B^*$. There exists a \kl{lift} $\alpha$, a \kl{list-polynomial functor} $F$ and \kl{drop terms} $\vec f \preceq f$ with codomain $B^*$, such that 
	\[
		f \equiv \alpha \comp \left( F(\vec f) \comp \concmap \right),
	\]
	where every $f_i$ is equal to one of 
	\[
		(1 \comp v), \qquad [g_i], \qquad g_i^*.
	\]
\end{lemma}

\begin{proof}[Proof of \cref{lem:drops-with-list-co-domain}]
	We proceed by induction on $f$ with respect to $\preceq$. The easy cases are when $\rk(f) = 0$ or $f$ is a functorial map of another drop term. We leave these cases to the reader, as the remaining cases are strictly more involved. We are left with the case where $f$ is a composition of two drop terms, i.e.~$f \equiv f_1 \comp f_2$. Similarly to the proofs of \cref{lem:drops-with-sum-co-domain} and \cref{lem:drops-with-prod-co-domain}, we proceed by induction on $f_2$ with respect to $\preceq$. If $f_2$ is itself a composition, that is, $f_2 = g_1 \comp g_2$, then $f \equiv (f_1 \comp g_1) \comp g_2 \preceq f$ and $g_2 \prec f_2$, so we obtain the desired construction by induction. 

	The remaining cases are when $f_2$ is a functorial map of a drop term, or when $f_2$ is an \kl{atomic drop term}. We consider these cases separately. In both of them we have $f_1 \prec f$, so we normalise all terms at most $f_1$ in $\preceq$.
	
	\begin{description}
		\item[List, $f_2 = h^*$.] In this case, by induction there is a lift $\alpha$ and a \kl{normalised drop} $F(\vec f)$ with $\vec f \preceq f_1$, such that $f_1 \equiv \alpha \comp \left( F(\vec f) \comp \concmap \right)$. We obtain:
		\begin{align*}
			f = f_1 \comp f_2 \equalbecause{induction} \\ 
			\alpha \comp \left( F(\vec f) \comp \concmap \right) \comp h^* \equalbecause{functoriality of $\concmap$, \cref{lem:functoriality-of-concat-maps}} \\
			\alpha \comp \left( F(\vec f \comp h^*) \comp \concmap \right).
		\end{align*}
		For every $f_i \in \vec f$, there are three cases depending on its shape. If $f_i = (1\comp{}v)$ for some \kl{value} $v$, we have $(1\comp{}v)\comp{}h^* \equiv (1\comp{}w)$ for some value $w$ by \cref{lem:value-completeness}. If $f_i = [g_i]$, then $f_i \comp h^* \equiv [g_i \comp h]$ by \cref{lem:pure-naturality}. If $f_i = g_i^*$, then $f_i \comp h^* \equiv (g_i \comp h)^*$ by functoriality of $*$ (\cref{ax:functoriality-of-type-constructors}). In each case the \kl{rank} of resulting leaf term is at most $\rk(f_1) + \rk(h) + 1 = \rk(f)$.

		\item[Atomic drops.] The atomic drop terms which produce a list output are values, $\idterm$, $\fstterm$, $\sndterm$, $\codiagterm$, $\pureterm$, $\binconcatterm$, $\concatterm$ and $\unshiftterm$. If $f_2$ is a value, then $\alpha = \idterm, \vec f = (1 \comp f_2)$. If $f_2 = \pureterm$, then $\alpha = \idterm$ and $\vec f = ([f_1])$. If $f_2 = \idterm$, then we obtain the construction by induction on $f_1$. The remaining cases are more involved, and we present them separately.

		\begin{description}
			\item[Projection, $f_2 = \fstterm$.] Suppose that $f_2 = \fstterm$ (the case where $f_2 = \sndterm$ is symmetric). The construction proceeds by applying \cref{lem:drops-with-prod-co-domain} to $f_1$ and then \cref{lem:drops-with-list-co-domain} to the obtained normalisation, which is illustrated on the following diagram:
			\[
			\begin{tikzcd}[column sep=5cm,row sep=1.5cm]
				A
				\ar[r, \dropar, "f_1"]
				\ar[d, \liftar, "\text{Lemmas } \ref{lem:drops-with-prod-co-domain} \text{ and } \ref{lem:drops-with-list-co-domain}"']
				&B^* \times C
				\ar[d, \dropar, "\fstterm"]
				\\
				\sum_i F_i(\vec X_i) \times Y_i
				\ar[ur, \dropar, "{\eithermany{i}{\left( \left( F_i\left( \vec x_i \right) \comp \concmap \right) \times y_i \right)}}"]
				\ar[r, \dropar, "{\eithermany{i}{\left( \left( F_i\left( \vec x_i \right) \comp \concmap \right) \fdot \left( 1 \comp [] \right) \right)}}"]
				&B^*
			\end{tikzcd}
			\]
			The upper triangle commutes by induction. For the lower triangle, we write:
			\begin{align*}
				\eithermany{i}{\left( \left( F_i(\vec x_i) \comp \concmap \right) \times y_i \right)} \comp \fstterm \equalbecause{\cref{lem:functoriality-of-either}} \\ 
				\eithermany{i}{\left( \left( \left( F_i(\vec x_i) \comp \concmap \right) \times  y_i \right) \comp \fstterm \right)} \equalbecause{naturality of $\fstterm$, \cref{ax:naturality}} \\ 
				\eithermany{i}{\left( \left( \fstterm \comp \left( F_i(\vec x_i) \comp \concmap \right) \right) \right)} \equalbecause{naturality of $\fstterm$, \cref{ax:naturality}} \\ 
				\eithermany{i}{\left( \left( \left( F_i(\vec x_i) \comp \concmap \right) \times (1 \comp []) \right) \comp \fstterm \right)} \equalbecause{\cref{lem:product-with-empty-list-fst-binconcat}} \\ 
				\eithermany{i}{\left( \left( F_i(\vec x_i) \comp \concmap \right) \fdot (1 \comp []) \right)}
			\end{align*}

			\item[Codiagonal, $f_2 = \codiagterm$.] Suppose that $f_2 = \codiagterm$. We first apply \cref{lem:drops-with-sum-co-domain} to $f_1$, and then we apply \cref{lem:drops-with-list-co-domain} to the obtained normalisation. This is illustrated on the following diagram (every $\rho_i$ is either $\leftterm$ or $\rightterm$):
			\[
			\begin{tikzcd}[column sep=5cm, row sep=1.5cm]
				A
				\ar[r, \dropar, "f_1"]
				\ar[d, \liftar, "{\text{Lemmas } \ref{lem:drops-with-sum-co-domain} \text{ and } \ref{lem:drops-with-list-co-domain}}"']
				&B^* + B^*
				\ar[d, \dropar, "\codiagterm"]
				\\
				\sum_i F_i(\vec A_i)
				\ar[ur, \dropar, "{\eithermany{i}{\left( \left( F_i(\vec f_i) \comp \concmap \right) \comp \rho_i \right) }}"]
				\ar[r, \dropar, "{\eithermany{i}{\left( F_i(\vec f_i) \comp \concmap \right)}}"]
				&B^*
			\end{tikzcd}
			\]
			The upper triangle commutes by induction. The lower triangle commutes by \cref{lem:functoriality-of-either} and the fact that coprojections and $\codiagterm$ cancel each other out (\cref{ax:eval-codiag}).

			\item[Binary concatenation, $f_2 = \binconcatterm$.] Suppose that $f_2 = \binconcatterm$. The construction proceeds by applying \cref{lem:drops-with-prod-co-domain} to $f_1$ and then \cref{lem:drops-with-list-co-domain} to the obtained normalisation. This is illustrated on the following diagram:
			\[
			\begin{tikzcd}[column sep=4cm, row sep=1cm]
				A
				\ar[r, \dropar, "f_1"]
				\ar[d, \liftar, "\text{Lemmas } \ref{lem:drops-with-prod-co-domain} \text{ and } \ref{lem:drops-with-list-co-domain}"']
				&B^* \times B^*
				\ar[d, \dropar, "\binconcatterm"]
				\\ 
				\sum_i F_i(\vec X_i) \times G_i(\vec Y_i)
				\ar[r, \dropar, "{\eithermany{i}{\left( \left( F_i(\vec x_i) \comp \concmap \right) \, \odot \, \left( G_i(\vec y_i) \comp \concmap \right) \right)}}"]
				&B^*
			\end{tikzcd}
			\]
			In this case, the diagram commutes purely by induction.

			\item[Concatenation, $f_2 = \concatterm$.] Suppose that $f_2 = \concatterm$. Then $f_1 : A \dropmap B^{**}$ and by induction there is $F(\vec f)$ together with a lift $\alpha$ such that $f_1 \equiv \alpha \comp \left( F(\vec f)\comp \concmap \right)$. Again by induction, for every $f_i \in \vec f$ there is $G_i(\vec f_i)$ together with a lift $\alpha_i$ such that $f_i \equiv \alpha_i \comp \left( G_i(\vec f_i) \comp \concmap \right)$. With this in hand, we assemble the construction:
			\[
			\begin{tikzcd}[column sep=6cm]
				A
				\ar[r, \dropar, "f_1"]
				\ar[d, \liftar, "\alpha"']
				&B^{**}
				\ar[dd, \dropar, "\concatterm"]
				\\ 
				F(\vec A)
				\ar[d, \liftar, "{F(\vec \alpha)}"']
				\\ 
				F(G_1(\vec A_1), \dots, G_n(\vec A_n))
				\ar[uur, \dropar, "{F\left((G_1(\vec f_1) \comp \concmap), \dots, (G_n(\vec f_n) \comp \concmap ) \right) \comp \concmap}"]
				\ar[r, \dropar, "{F \left(G_1(\vec f_1 \comp \concatterm), \dots, G_n(\vec f_n \comp \concatterm) \right) \comp \concmap}"]
				&B^*
			\end{tikzcd}
			\]
			The upper triangle commutes by induction, and the lower one by \cref{lem:concat-map-concat} applied once for the outer \kl{concatenation map}, and then for each of the inner concatenation maps. Now, for every $f_{ij} \in \vec f_i$, there are three cases depending on its shape. If $f_{ij} = (1\comp{}v)$ for some value $v$, then $f_{ij} \comp \concatterm \equiv (1\comp{}w)$ for some value~$w$ by \cref{lem:value-completeness}. If $f_{ij} = [g_{ij}]$, then $f_{ij} \comp \concatterm \equiv g_{ij}$ by \cref{ax:concatenation-unit} and $g_{ij}$ can be further normalised since $g_{ij} \preceq f_1$. If $f_{ij} = g_{ij}^*$, then $f_{ij} \comp \concatterm = \hom g_{i j}$, so we can normalise $g_{i j}$ and wrap the resulting subfunctor with a star.

			\item[Unshift, $f_2 = \unshiftterm$.] In this last case, we utilise the tools we developed in \cref{sec:appendix:shift-exchange}. The construction proceeds by first applying \cref{lem:drops-with-prod-co-domain} to $f_1$ to extract parts of the input which produce the leading $S$ and trailing $T$ and the inner list of pairs, and then applying \cref{lem:drops-with-list-co-domain} to normalise the term producing pairs. After that, we can apply \cref{lem:exchange-normal-shift} to push this term through the shift. This is presented on the following diagram (every $\rho_{i j}$ is either $\leftterm$ or $\rightterm$):
			\[
			\begin{tikzcd}
				A
				\ar[rr, \dropar, "f_1"]
				\ar[d, \liftar, "{\text{Lemmas } \ref{lem:drops-with-prod-co-domain}  \text{ and } \ref{lem:drops-with-list-co-domain}}"]
				&\hspace*{1cm}
				&S \times (T \times S)^* \times T
				\ar[r, \dropar, "\unshiftterm"]
				\ar[dd, \liftar, "{\idterm \times \shiftterm \times \idterm}"']
				&(S \times T)^*
				\\ 
				\sum_i L_i \times F_i(\vec A_i) \times R_i
				\ar[d, \liftar, "{\cref{lem:exchange-normal-shift}}"]
				\\ 
				\sum_i L_i \times \left( \sum_j B_{ij} \right) \times R_i
				\ar[rr, \dropar, "{\eithermany{i}{\left( l_i \times \left( \eithermany{j}{(h_{ij}) \comp \rho_{ij}} \right) \times r_i \right)}}"]
				\ar[d, \liftar, "\cong"]
				&&S \left( \onetype + T(ST)^*S \right) T
				\ar[d, \liftar, "\cong"']
				\\ 
				\sum_{i j} L_i \times B_{ij} \times R_i
				\ar[rr, \dropar, "{\eithermany{i j}{\left( l_i \times h_{ij} \times r_i \comp \rho_{ij} \right)}}"]
				\ar[d, \liftar, "\cong"]
				&&\left( S \onetype T \right) + \left( S \times (T(ST)^* S) \times T \right)
				\ar[d, \liftar, "\cong"']
				\\ 
				\sum_{i j} D_{ij}
				\ar[rr, \dropar, "\eithermany{ij}{(g_{ij})}"]
				&&(S (\onetype T)) + ( (S T) \times (S T)^* \times (S T) )
				\ar[uuuur, \dropar, bend right=40, "k"]
			\end{tikzcd}
			\]
			The diagonal drop is the drop term which collects the input into $ST$ pairs: 
			\[
				k = \either{ [\idterm \times \sndterm]}{\left( [\idterm] \odot \idterm \odot [\idterm] \right)}.
			\]
			Before we describe what the \kl{types} $D_{ij}$ and terms $g_{ij}$ are, we note that the diagram commutes (except the bottom square which will commute by the choice of $g_{ij}$): the uppermost-left face commutes by induction, the middle-left face commutes by a straightforward application of \cref{lem:distributivity-of-either} and \cref{lem:distributivity}. The right-hand face commutes by \cref{lem:uniqueness-of-rebracketings}: both paths are \kl{rebracketings} (by \cref{lem:rebracketing-basic-facts} and the fact that the only use of projections removes $\onetype$) and the type $(S \times T)^*$ is \kl[injective type]{injective} (see the proof of \cref{lem:shift-type-is-injective}).

			Now, the types $D_{ij}$ and $g_{ij}$ are defined based on which coprojection $\rho_{ij}$ is. If $\rho_{ij} = \leftterm$, then $h_{ij} = (1 \comp 1)$, which is equivalent to $1$ by \cref{ax:unit-type}. In this case, we simply set $D_{ij} = L_i \times B_{ij} \times R_i$ and $g_{ij} = l_i \times 1 \times r_i \comp \leftterm$. We obtain:
			\begin{align*}
				g_{i j} \comp k &= \why{unfold $g_{i j}$ and $k$} \\ 
				l_i \times 1 \times r_i \comp \leftterm \comp \either{ [\idterm \times \sndterm]}{\left( [\idterm] \odot \idterm \odot [\idterm] \right)} \equalbecause{\cref{lem:case-equivalence}} \\ 
				\left[ l_i \times 1 \times r_i \comp \idterm \times \sndterm \right] \equalbecause{functoriality of $\times$, \cref{ax:functoriality-of-type-constructors}} \\
				\left[ l_i \times \left( 1 \times r_i \comp \sndterm \right) \right] \equalbecause{naturality of $\sndterm$, \cref{ax:naturality}} \\
				\left[ l_i \times \left( \sndterm \comp r_i \right) \right] &= \why{definition of $\concmap$} \\ 
				l_i \times \left( \sndterm \comp r_i \right) \comp \concmap.
			\end{align*} 
			This complies with the rank requirements of the lemma because 
			\[ 
				\rk(l_i \times (\sndterm \comp r_i)) = \max(\rk(l_i), \rk(r_i) + 1) \leq \rk(f_1) + 1 = \rk(f).
			\]

			The remaining case is when $\rho_{ij} = \rightterm$, in which $h_{i j} = (\sndterm \comp x) \times \left( H(\vec h) \comp \concmap \right) \times (\fstterm \comp y)$. We cannot directly compose this form with $k$ -- we have to first perform some lift rebracketing. Given that $B_{i j} = X \times H(\vec A) \times Y$, we set $D_{i j} = (L_i \times X) \times H(\vec A) \times (Y \times R_i)$ and $g_{ij} = \left( l_i \times (\sndterm \comp x) \right) \times \left( H(\vec h) \comp \concmap \right) \times \left( (\fstterm \comp y) \times r_i \right) \comp \rightterm$. We obtain:
			\begin{align*}
				g_{i j} \comp k \equalbecause{\cref{lem:case-equivalence}} \\
				\left( l_i \times (\sndterm \comp x) \right) \times \left( H(\vec h) \comp \concmap \right) \times \left( (\fstterm \comp y) \times r_i \right) \comp \left( [\idterm] \odot \idterm \odot [\idterm] \right) \equalbecause{functoriality of $\times$, \cref{ax:functoriality-of-type-constructors}} \\ 
				\left[ l_i \times (\sndterm \comp x) \right] \odot \left( H(\vec h) \comp \concmap \right) \odot \left[ (\fstterm \comp y) \times r_i \right] &= \why{definition of $\concmap$} \\ 
				\left( l_i \times (\sndterm \comp x) \right) \times H(\vec h) \times \left( (\fstterm \comp y) \times r_i \right) \comp \concmap. 
			\end{align*}
			When it comes to ranks, we have $\vec h \preceq f_1 \preceq f$, $\rk(l_i \times (\sndterm \comp x)) \leq \rk(f_1) + 1 = \rk(f)$ and $\rk((\fstterm \comp y) \times r_i) \leq \rk(f_1) + 1 = \rk(f)$.

			We showed that $g_{i j} \comp k$ is equivalent to a normalised drop term with proper rank bounds. Therefore, so is the entire composition $\eithermany{i j}{(g_{i j})} \comp k$ by \cref{lem:functoriality-of-either}.
		\end{description}
	\end{description}
\end{proof}

\subsection{Proof of the Exchange Lemma}\label{sec:appendix:exchange-proof}

In this section we prove \cref{lem:exchange}. We will use functorial notation for classes of morphisms: given classes of terms $\mathcal{C}_1, \mathcal{C}_2$, we write $\mathcal{C}_1 + \mathcal{C}_2$, $\mathcal{C}_1 \times \mathcal{C}_2$, $\mathcal{C}_1^*$ and $\mathcal{C}_1 \comp \mathcal{C}_2$ to denote the sets of terms constructed by sum, product, list or composition of elements of $\mathcal{C}_1$ and $\mathcal{C}_2$. Moreover, we will write $f$ instead of $\{ f\}$ for brevity, e.g. $\mathcal{C}_1 \comp f$ is the same as $\mathcal{C}_1 \comp \{ f \}$.

\cref{lem:exchange} states that $\Drop \comp \Lift \subseteq \Lift \comp \Drop$, where by $\Lift$ ($\Drop$) we denote the set of all terms provably equivalent to \kl{lift} (drop) terms. We prove this by structural induction on the lift part, i.e.~our induction hypothesis is that if $\delta$ is a lift term, then $\Drop \comp \delta \subseteq \Lift \comp \Drop$. We will first show how to reduce the cases where $\delta$ is built with a combinator, and after that we will consider the cases where $\delta$ is an \kl{atomic lift term}, which are more involved.

\begin{description} 

\item[Composition, $\delta = \delta_1 \comp \delta_2$.] Suppose that $\delta = \delta_1 \comp \delta_2$. By induction, we get
\begin{align*}
	\Drop \comp \delta_1 \comp \delta_2 &\subseteq \why{induction} \\ 
	\Lift \comp \Drop \comp \delta_2             &\subseteq \why{induction} \\ 
	\Lift \comp \Lift \comp \Drop = \Lift \comp \Drop.
\end{align*}

\item[Sum, $\delta = \delta_1 + \delta_2$.] Suppose that $\delta = \delta_1 + \delta_2$. By \cref{lem:drops-with-sum-co-domain} we have that every \kl{drop term} which composes with $\delta$ is in $\Lift \comp (\Drop + \Drop)$. We get
\begin{align*} 
	\Lift \comp (\Drop + \Drop) \comp (\delta_1 + \delta_2) &\subseteq \why{functoriality of $+$, \cref{ax:functoriality-of-type-constructors}} \\ 
	\Lift \comp ((\Drop \comp \delta_1) + (\Drop \comp \delta_2)) &\subseteq \why{induction} \\ 
	\Lift \comp ((\Lift \comp \Drop) + (\Lift \comp \Drop)) &\subseteq \why{functoriality of $+$, \cref{ax:functoriality-of-type-constructors}} \\ 
	\Lift \comp (\Lift + \Lift) \comp (\Drop + \Drop) \subseteq \Lift \comp \Drop.
\end{align*}

\item[Product, $\delta = \delta_1 \times \delta_2$.] Suppose that $\delta = \delta_1 \times \delta_2$. By \cref{lem:drops-with-prod-co-domain} we have that every drop term which composes with $\delta$ is in $\Lift \comp (\eithermany{i}{\left( \Drop \times \Drop \right)})$ for some $i \geq 1$. We get 
\begin{align*}
	\Lift \comp \left(\eithermany{i}{\left( \Drop \times \Drop \right)} \right) \comp (\delta_1 \times \delta_2) &\subseteq \why{\cref{lem:functoriality-of-either}} \\ 
	\Lift \comp \left( \eithermany{i}{\left( \Drop \times \Drop \right) \comp (\delta_1 \times \delta_2)} \right) &\subseteq \why{functoriality of $\times$, \cref{ax:functoriality-of-type-constructors}} \\ 
	\Lift \comp \left( \eithermany{i}{\left( \left( \Drop \comp \delta_1 \right) \times \left( \Drop \comp \delta_2 \right) \right)} \right) &\subseteq \why{induction} \\
	\Lift \comp \left( \eithermany{i}{\left( \left( \Lift \comp \Drop \right) \times \left( \Lift \comp \Drop \right) \right)} \right) &\subseteq \why{functoriality of $\times$, \cref{ax:functoriality-of-type-constructors}} \\ 
	\Lift \comp \left( \eithermany{i}{\left( \left( \Lift \times \Lift \right) \comp \left( \Drop \times \Drop \right) \right)} \right) &\subseteq \why{\cref{lem:functoriality-of-either-left}} \\ 
	\Lift \comp \left( (\Lift \times \Lift) + \dots + (\Lift \times \Lift) \right) \comp \eithermany{i}{\left( \Drop \times \Drop \right)} &\subseteq \Lift \comp \Drop.
\end{align*}

\item[List, $\delta = \gamma^*$.] Suppose that $\delta = \gamma^*$. Let $\mathsf{S} = (1 \comp \Value) \cup [\Drop] \cup \Drop^*$. By \cref{lem:drops-with-list-co-domain} we have that every drop term which composes with $\delta$ is in $\Lift \comp \left( F(\mathsf{S}) \comp \concmap \right)$ for some \kl{list-polynomial functor} $F$. We proceed by nested induction on the shape of $F$.

\begin{description}
	\item[Leaf node, $F = X_i$.] In this case, $\left( F(\mathsf{S}) \comp \concmap \right) = F(\mathsf{S})$. Our goal is to show that $F(\mathsf{S}) \comp \delta \subseteq \Lift \comp \Drop$. We show three inclusions separately:
	\begin{align*}
		(1 \comp \Value) \comp \delta &\subseteq \why{value composed with $\delta$ is a value} \\ 
		(1 \comp \Value) \subseteq \Lift \comp \Drop.
	\end{align*}
	\begin{align*}
		[\Drop] \comp \gamma^* &\subseteq \why{naturality of $\pureterm$, \cref{lem:pure-naturality}} \\ 
		[\Drop \comp \gamma] &\subseteq \why{induction} \\ 
		[\Lift \comp \Drop] \subseteq \Lift \comp \Drop.
	\end{align*}
	\begin{align*}
		\Drop^* \comp \gamma^* &\subseteq \why{functoriality of $*$, \cref{ax:functoriality-of-type-constructors}} \\ 
		(\Drop \comp \gamma)^* &\subseteq \why{induction} \\ 
		(\Lift \comp \Drop)^* &\subseteq \why{functoriality of $*$, \cref{ax:functoriality-of-type-constructors}} \\ 
		\Lift^* \comp \Drop^* \subseteq \Lift \comp \Drop.
	\end{align*}

	\item[Sum node, $F = F_1 + F_2$.] In this case $\left( F(\mathsf{S}) \comp \concmap \right) = \either{\left( F_1(\mathsf{S}) \comp \concmap \right)}{\left( F_2(\mathsf{S}) \comp \concmap \right)}$. We get
	\begin{align*}
		\either{\left( F_1(\mathsf{S}) \comp \concmap \right)}{\left( F_2(\mathsf{S}) \comp \concmap \right)} \comp \delta &\subseteq \why{\cref{lem:functoriality-of-either}} \\ 
		\either{\left( F_1(\mathsf{S}) \comp \concmap \comp \delta \right)}{\left( F_2(\mathsf{S}) \comp \concmap \comp \delta \right)} &\subseteq \why{induction} \\ 
		\either{(\Lift \comp \Drop)}{(\Lift \comp \Drop)} &\subseteq \why{\cref{lem:functoriality-of-either-left}} \\ 
		(\Lift + \Lift) \comp \either{\Drop}{\Drop} \subseteq \Lift \comp \Drop.
	\end{align*}

	\item[Product node, $F = F_1 \times F_2$.] In this case, $\left( F(\mathsf{S}) \comp \concmap \right) = \left( F_1(\mathsf{S}) \comp \concmap \right) \odot \left( F_2(\mathsf{S}) \comp \concmap \right)$. We have 
	\begin{align*}
		\left( F_1(\mathsf{S}) \comp \concmap \right) \odot \left( F_2(\mathsf{S}) \comp \concmap \right) \comp \gamma^* &\subseteq \why{naturality of $\binconcatterm$} \\ 
		\left( F_1(\mathsf{S}) \comp \concmap \comp \gamma^* \right) \odot \left( F_2(\mathsf{S}) \comp \concmap \comp \gamma^* \right) &\subseteq \why{induction} \\ 
		\left( \Lift \comp \Drop \right) \odot \left( \Lift \comp \Drop \right) &= \why{definition of $\odot$} \\
		(\Lift \times \Lift) \comp (\Drop \times \Drop) \comp \binconcatterm \subseteq \Lift \comp \Drop.
	\end{align*}

	\item[List node, $F = G^*$.] In this case $\left( F(\mathsf{S}) \comp \concmap \right) = \left( G(\mathsf{S}) \comp \concmap \right)^* \comp \concatterm$. We get
	\begin{align*}
		\left( G(\mathsf{S}) \comp \concmap \right)^* \comp \concatterm \comp \gamma^* &\subseteq \why{naturality of $\concatterm$} \\ 
		\left( G(\mathsf{S}) \comp \concmap \right)^* \comp \gamma^{**} \comp \concatterm &\subseteq \why{functoriality of $*$, \cref{ax:functoriality-of-type-constructors}} \\ 
		\left( G(\mathsf{S}) \comp \concmap \comp \gamma^* \right)^* \comp \concatterm &\subseteq \why{induction} \\ 
		\left( \Lift \comp \Drop \right)^* \comp \concatterm &\subseteq \why{functoriality of $*$, \cref{ax:functoriality-of-type-constructors}} \\ 
		\Lift^* \comp \Drop^* \comp \concatterm \subseteq \Lift \comp \Drop.
	\end{align*}
\end{description}

\item[Atomic lifts, easy cases.] It now remains to consider cases where $\delta$ is an atomic lift term. If $\delta$ is $\idterm$, $\assoclterm$, $\assocrterm$ or $\commterm$, then the statement holds trivially because $\delta \in \Drop$. If $\delta = \unitlterm$, then
\begin{align*}
	\Drop \comp \unitlterm &\subseteq \why{naturality of $\unitlterm$} \\ 
	\unitlterm \comp (\idterm \times \Drop) \in \Lift \comp \Drop.
\end{align*}
The case $\delta = \unitrterm$ is analogous. 

\item[Distribution, $\delta = \distrterm$.] Suppose that $\delta = \distrterm$ (the case $\delta = \distlterm$ is symmetric) and $f \in \Drop$. The construction is presented on the following diagram:
\[
\begin{tikzcd} 
	A
	\ar[r, \dropar, "f"]
	\ar[d, \liftar, "\cref{lem:drops-with-prod-co-domain}"']
	&B \times (C + D)
	\ar[dd, \liftar, "\distrterm"]
	\\ 
	\sum X_i \times \left( \sum Y_{i,j} + \sum Z_{i,j} \right)
	\ar[ur, \dropar]
	\ar[d, \liftar, "\distrterm"']
	\\ 
	\left( \sum X_i \times \sum Y_{i,j} \right) + \left( \sum X_i \times \sum Z_{i,j} \right)
	\ar[r, \dropar]
	&(B \times C) + (B \times D)
\end{tikzcd}
\]
The upper triangle commutes by \cref{lem:drops-with-prod-co-domain}. The lower rectangle commutes by naturality of $\distrterm$.

\item[Shift, $\delta = \shiftterm$.] Suppose that $\delta = \shiftterm$. This case follows easily by the work we have done in \cref{sec:appendix:shift-exchange}. Let $f$ be a drop term. Then by \cref{lem:drops-with-list-co-domain}, it can be lifted to normal form. This normal form composed with $\delta$ is in $\Lift \comp \Drop$ by \cref{lem:exchange-normal-shift}. Therefore $f \comp \delta \subseteq \Lift \comp \Drop$.

\item[Block, $\delta = \blockterm$.] Suppose that $\delta = \blockterm$. Let $f$ be a drop term. Similarly as in the case of $\shiftterm$, it suffices to show that $\left( F(\vec f) \comp \concmap \right)$ -- the normalised form of $f$ given by \cref{lem:drops-with-list-co-domain} -- postcomposed with $\delta$ is in $\Lift \comp \Drop$. We prove this by induction on the shape of $F$.
\begin{description}
	\item[Leaf node, $F = X_i$.] We omit the index $i$ and write $f := f_i$. We consider three subcases depending on the shape of $f$.
	\begin{description}
	\item[Value, $f = (1 \comp v)$.] There is a \kl{value} $w \equiv v \comp \blockterm$, and therefore $f \comp \blockterm \equiv (1 \comp w) \in \Drop$. 

	\item[Singleton, $f = {[g]}$.] In this case, we lift $g$ to normal form by \cref{lem:drops-with-sum-co-domain}. The construction is illustrated on the diagram:
	\[
	\begin{tikzcd}
		A 
		\ar[r, \dropar, "{g}"]
		\ar[d, \liftar, "\cref{lem:drops-with-sum-co-domain}"'] 
		&S + T 
		\ar[rr, \dropar, "\pureterm"]
		\ar[dd, \liftar, "\cong"]
		&\hspace*{1.5cm}&(S+T)^*
		\ar[dd, \liftar, "\blockterm"] \\ 
		A_1 + A_2
		\ar[ur, \dropar, "{g_1 + g_2}"]
		\ar[d, \liftar, "\cong"']
		\\
		A_1 \onetype + \onetype A_2 \onetype 
		\ar[r, \dropar]
		&S \onetype + \onetype T \onetype 
		\ar[rr, \dropar, "{\left( \pureterm \times [] \right) + \left( [] \times \idterm \times [] \right)}"]
		&&S^* \times (T \times S^*)^*
	\end{tikzcd}
	\]
	The upper triangle commutes by \cref{lem:drops-with-sum-co-domain}. The bottom-left rectangle follows by naturality of $\unitrterm$ and $\unitlterm$. The right-hand rectangle commutes by \cref{lem:uniqueness-of-rebracketings}, because both paths are \kl{rebracketings} by \cref{lem:rebracketing-basic-facts} and the target \kl{type} is \kl[injective type]{injective}.

	\item[List, $f = g^*$.] We further lift $g$ to normal form by \cref{lem:drops-with-sum-co-domain}. The construction is presented on the following diagram:
	\[
	\begin{tikzcd} 
		C_i^*
		\ar[r, \dropar, "g^*"]
		\ar[d, \liftar, "\cref{lem:drops-with-sum-co-domain}"']
		&(S + T)^*
		\ar[dd, \liftar, "\blockterm"]
		\\ 
		\left( A_1 + A_2 \right)^*
		\ar[ur, \dropar]
		\ar[d, \liftar, "\blockterm"']
		\\ 
		A_1^* \times (A_2 \times A_1^*)^*
		\ar[r, \dropar]
		&S^* \times (T \times S^*)^*
	\end{tikzcd}
	\]
	The upper triangle commutes by \cref{lem:drops-with-sum-co-domain}. The lower rectangle commutes by naturality of $\blockterm$.
	\end{description}

	\item[Sum node, $F = F_1 + F_2$.] By induction on $F_1$ and $F_2$ we get
	\begin{align*}
		\left( F(\vec f) \comp \concmap \right) \comp \blockterm = \either{F_1(\vec f)}{F_2(\vec f)} \comp \blockterm \equalbecause{\cref{lem:functoriality-of-either}} \\ 
		\either{\left( F_1(\vec f) \comp \blockterm \right)}{\left( F_2(\vec f) \comp \blockterm \right)} &\subseteq \why{induction} \\ 
		\either{(\Lift \comp \Drop)}{\Lift \comp \Drop} &\subseteq \why{\cref{lem:functoriality-of-either-left}} \\ 
		(\Lift + \Lift) \comp \either{\Drop}{\Drop} \subseteq \Lift \comp \Drop.
	\end{align*}

	\item[Product node, $F = F_1 \times F_2$.] We have 
	\[
		\left( F(\vec f) \comp \concmap \right) = (F_1(\vec f) \comp \concmap) \times (F_2(\vec f) \comp \concmap) \comp \binconcatterm.
	\] 
	The construction uses induction on $F_1,F_2$ and the instances of \cref{lem:exchange} for lifts which we have already handled: $\unitrterm$, $\shiftterm$ and distribution. The construction is illustrated on the following diagram, where $Z := T \times S^*$:
	\[
	\begin{tikzcd} 
		\bullet
		\ar[r, \dropar, "{(F_1(\vec f)\comp\concmap) \times (F_2(\vec f)\comp\concmap)}"]
		\ar[d, \liftar, "\text{induction}"]
		&\left((S + T)^*\right)^2
		\ar[r, \dropar, "\binconcatterm"]
		\ar[d, \liftar, "{\blockterm^2}"]
		&[-5em](S+T)^*
		\ar[d, \liftar, "\blockterm"]
		\\ 
		\bullet
		\ar[dd, \liftar, "\cref{lem:exchange}"]
		\ar[r, \dropar]
		&(S^* \times Z^*)^2 
		\ar[d, \liftar, "{(\idterm \times (\unitrterm^* \comp \shiftterm))^2}"]
		&S^* \times Z^* 
		\\ 
		&\left( S^* \left( \onetype + Z(\onetype Z)^* \onetype \right) \right)^2
		\ar[d, \liftar, "\cong"]
		\\ 
		\bullet
		\ar[r, \dropar]
		&(S^*\onetype)^2 + S^*\onetype S^* Z(\onetype Z)^*\onetype + S^* Z (\onetype Z)^* \onetype S^* \onetype + (S^* Z (\onetype Z)^* \onetype)^2
		\ar[uur, \dropar, start anchor={[xshift=40pt, yshift=0pt]}]
	\end{tikzcd}
	\]
	The top-left square commutes from induction. The bottom-left square commutes by \cref{lem:exchange} for $\shiftterm$, $\unitrterm$, distribution, and their functorial combinations. The right face of the diagram commutes by \cref{lem:uniqueness-of-rebracketings}, because both paths are rebracketings by \cref{lem:rebracketing-basic-facts} and the type $S^* \times (T \times S^*)^*$ is injective.

	\item[List node, $F = G^*$.] We have $\left( F(\vec f) \comp \concmap \right) = \hom (G(\vec f) \comp \concmap)$. First, consider the following diagram, where $Z := T \times S^*$:
	\[
	\begin{tikzcd}
		(S + T)^{**}
		\ar[r, \dropar, "\concatterm"]
		\ar[d, \liftar, "{\blockterm^*}"]
		&(S + T)^*
		\ar[ddddddd, \liftar, "\blockterm"]
		\\ 
		(S^* Z^*)^*
		\ar[d, blue, \liftar, "{(\idterm \times \unitlterm^*)^*}"]
		\\ 
		(S^* \times (\onetype Z)^*)^*
		\ar[d, blue, \liftar, "(\idterm \times \shiftterm)^*"] 
		\\ 
		(S^* \times (\onetype + \onetype(Z \onetype)^* Z))^*
		\ar[d, blue, \liftar, "{\distrterm^*}"] 
		\\ 
		(S^* \onetype + S^* \onetype(Z \onetype)^* Z)^*
		\ar[d, blue, \liftar, "\blockterm"] 
		\\ 
		(S^* \onetype)^* \times (S^* \onetype(Z \onetype)^* Z (S^* \onetype)^*)^* 
		\ar[d, \liftar, "{\idterm \times \shiftterm}"] 
		\\
		(S^* \onetype)^* \times (\onetype + S^* (\onetype(Z \onetype)^* Z (S^* \onetype)^* S^*)^* \onetype(Z \onetype)^* Z (S^* \onetype)^*)
		\ar[d, \liftar, "\distrterm"]
		\\ 
		(S^* \onetype)^*\onetype + (S^* \onetype)^* S^* (\onetype(Z \onetype)^* Z (S^* \onetype)^* S^*)^* \onetype(Z \onetype)^* Z (S^* \onetype)^*
		\ar[r, \dropar]
		&S^* Z^*
	\end{tikzcd}
	\]
	The diagram above commutes by \cref{lem:rebracketing-basic-facts} and \cref{lem:uniqueness-of-rebracketings}: both paths are rebracketings and $S^* Z^*$ is an injective type.

	Let $\gamma$ be the lift obtained by composing 4 consecutive lifts on the above diagram, starting from the second from the top (the lifts highlighted with blue colour). First, we claim that $\gamma$ admits exchange for every drop constructed with list combinator.

	\begin{claim}\label{claim:exchange-for-split-aux}
		Suppose that $g^*$ is a \kl{drop term}. Then $g^* \comp \gamma \in \Lift\comp \Drop$.
	\end{claim}
	\begin{proof}[Proof of \cref{claim:exchange-for-split-aux}]
		We write
		\[
			\gamma \equiv \left( \myunderbrace{\left( \idterm \times \unitlterm^* \right) \comp \left( \idterm \times \shiftterm \right) \comp \distrterm}{$\zeta$} \right)^* \comp \blockterm.
		\]
		The lift $\zeta$ is composed from terms for which we have already proved \cref{lem:exchange}, therefore there is a lift $\alpha$ and drop $h$ such that $g \comp \zeta \equiv \alpha \comp h$. We obtain the following diagram:
		\[
		\begin{tikzcd}
			\bullet 
			\ar[r, \dropar, "{g^*}"]
			\ar[d, \liftar, "{\alpha^*}"']
			&(S^*Z^*)^*
			\ar[d, \liftar, "{\zeta^*}"]
			\\ 
			\bullet
			\ar[r, \dropar, "h^*"]
			\ar[d, \liftar, "\cref{lem:drops-with-sum-co-domain}"']
			&(S^* \onetype + S^* \onetype(Z \onetype)^*Z)^*
			\ar[dd, \liftar, "\blockterm"]
			\\ 
			(A + B)^*
			\ar[ur, \dropar] 
			\ar[d, \liftar, "\blockterm"']
			\\
			A^* \times (B \times A^*)^* 
			\ar[r, \dropar] 
			&(S^* \onetype)^* \times (S^* \onetype(Z \onetype)^* Z (S^* \onetype)^*)^*
		\end{tikzcd}
		\]
		The upper rectangle commutes as we discussed. The triangle commutes by \cref{lem:drops-with-sum-co-domain}. The lower rectangle commutes by naturality of $\blockterm$.
	\end{proof}
	We are now in position to assemble the entire construction, which consists of inductive exchange of $G(\vec f)\comp\concmap \comp \blockterm$, \cref{claim:exchange-for-split-aux}, and exchange for lifts that we have already handled. We illustrate it on the following diagram:
	\[
	\begin{tikzcd}
		\bullet
		\ar[rr, \dropar, "{\left( G(\vec f) \comp \concmap \right)^*}"]
		\ar[d, \liftar, "\text{induction}"']
		&\hspace{1.5cm}&(S + T)^{**}
		\ar[r, \dropar, "\concatterm"]
		\ar[d, \liftar, "{\blockterm^*}"']
		&(S + T)^*
		\ar[d, \liftar, "\blockterm"]
		\\ 
		\bullet
		\ar[rr, \dropar, "g^*"]
		\ar[d, \liftar, "\cref{claim:exchange-for-split-aux}"']
		&&(S^* \times (T \times S^*)^*)^*
		\ar[d, \liftar, "\gamma"']
		&S^* \times (T \times S^*)^*
		\\ 
		\bullet
		\ar[rr, \dropar] 
		\ar[d, \liftar, "\cref{lem:exchange}"']
		&&\bullet
		\ar[d, \liftar, "\idterm \times \shiftterm \comp \distrterm"']
		\\ 
		\bullet 
		\ar[rr, \dropar]
		&&\bullet 
		\ar[uur, \dropar]
	\end{tikzcd}
	\]
	On the left side of the diagram, the rectangles commute by induction, \cref{claim:exchange-for-split-aux} and \cref{lem:exchange} for a lift constructed out of lifts for which we have already proved it. The right face of the diagram is exactly the diagram presented at the beginning of this case. 
\end{description}

\end{description}

\subsection{Proof of Joint Simplification}\label{sec:appendix:joint-drops-simplification}

We will represent a simple drop term $f$ as a functor $F$ and a vector of \kl{values}  $\vec v$ in $N^*$,
where $f = F(\vec v) \comp \concmap $ is of type $F(\vec \onetype) \drop{} N^*$. In particular, values can only be produced in the leaves, not in arbitrary places, and all those leaves are $\onetype$.
This form is equivalent to the definition of \kl{simple rational terms} (\cref{def:simple-rational-terms}), up to introducing some additional $\onetype$s to realise values as explicit subterms rather than appending them.

We begin with a simplification lemma for a single term:

\begin{lemma}[Drop simplification]\label{lem:drop-normalization}
    Suppose that $N$ is a \kl{finite alphabet} and $A \drop{f} N^*$. There exists a functor $F(X_1, \dots, X_n)$, a \kl{lift term} $ A \lift{\delta} F(\vec \onetype)$ and \kl{values} $\vec v_i$ such that $f \equiv \delta \comp F(\vec v_i) \comp \concmap$.
\end{lemma}
\begin{proof}
    First, apply \cref{lem:drops-with-list-co-domain}. This deconstructs $f$ into constant terms of the form $1\comp v_i$, singletons of the form $[h_i]$ and terms of the form $(g_i)^*$, where all $g_i$'s are simpler than $f$.

    Apply \cref{lem:drops-with-sum-co-domain} to the singletons, since each $h_i$ has output in $N$ which is a finite sum \kl{type}.
    We obtain a decomposition of $h_i$ into a case-split of constant terms.

    In the terms of the form $(g_i)^*$, we simplify $g_i$'s by induction, since the lemma ensures that they are simpler than $f$. 
    From this we obtain functors $G_1, ...., G_k$ and a vector of values for each of them. Merge those into the main functor.

    We finish by pushing values down the type tree, so that they are produced on $\onetype$ instead of any other type. 
    This is possible for sum type (produce the value in all cases separately) and in the product case (produce the value in first component, and empty in the other). In the case of a list there is an issue, as we cannot push nonempty value down -- it would be repeated many times by the map combinator. Instead, in the place of this list we make a product of it with $\onetype$, push the value to this $\onetype$ and empty value to the list -- this is obtained by lifting the input with $\unitlterm$. 
\end{proof}

Now we prove \cref{lem:joint-drops-normalisation}. 

\begin{proof}
    Simplify the term $f$ first. By precomposing with $\delta$ we can assume that $g$ is defined on $F(\onetype)$ -- the lifted type on which $f$ becomes simple. 
    We will now simplify $g$, and prove that this procedure does not interfere with simplicity of $f$.
    
    This process is summarised on the diagram below. When simplifying $g$ we obtain a new lifted type $T$.
    We represent it by two functors: $T = G(\vec\onetype) = G'(\vec\onetype)$, depending on our perspective. The functor $G$ is obtained by manipulating $F$ with lift terms, it uses the same input parameters as $F$. The functor $G'$ is the functor given by simplification procedure for $g$. It differs from $G$ in this way, that some additional input parameters were added. We can extend the vector $\vec v$ to $\vec v'$ by following the origin of each added parameter -- if it was added to represent a $\onetype$ created by $G$ without parameter, then apply empty value and where multiple new parameters are used in place of single one, they all inherit the same value.

    This difference between $G$ and $G'$ can be illustrated on the type $\onetype^*$ to which $\unconsterm$ is applied resulting in $T = \onetype  + \onetype \times \onetype^*$. We have $G(A) = \onetype + A \times A^*$ and $G'(A, B, C) = A + B \times C$, so $G(\onetype) = T = G(\onetype, \onetype, \onetype)$. A term $f$ that was simple on $\onetype^*$ with representation $F(v)$ has representation $G(v)$ and a corresponding representation $G([], v, v)$ -- this is because the first parameter of $G'$ corresponds to the empty list case, on which $f$ produces empty output, and the second and third to elements of the list, on which $f$ produces $v$. The term $g$ has access to more specific information, so it can make separate decision for each input parameter.

    The diagram commutes: face I because the lift is a \kl{rebracketing}, faces II and III are two instances of naturality of $\concmap$, and faces $IV$ and $V$ because the corresponding arrows describe the same terms, but with different notation.

    In this way on $G'(\vec \onetype)$ we have a simple representation of both terms $f, g$, proving the lemma.
      
    \[
    \begin{tikzcd}
    F(\vec \onetype)
    \ar[rr, Rightarrow, "F(\vec v) = f"]
    \ar[d, \concar]
    \ar[dd, bend right=90, \liftar, ""{name=U1}]
    \ar[drr, phantom, "\blue{II}"]
    & \ 
    & F(\vec N^*) 
    \ar[d, \concar]
    \\
    |[alias=U2]| \left(\sum_1^k\onetype\right)^*
    \ar[r, Rightarrow, "\casesterm (\vec v)"]
    \ar[from=U1, to=U2, phantom, "\blue{I}"]
    & (\sum_1^kN^*)^*
    \ar[d, phantom, "\blue{III}"]
    \ar[r, \concar] 
    & N^*
    \\
    G(\vec \onetype)
    \ar[u, \concar]
    \ar[d, equal]
    \ar[r, Rightarrow, "G(\vec v)"]
    \ar[drr, phantom, "\blue{IV}"]
    & |[alias=U4]| G(\vec N^*)
    \ar[ur, \concar]
    \ar[dr, equal]
    & \ 
    \\
    G'(\vec \onetype)
    \ar[rr, Rightarrow,"G'(\vec v')"]
    & \
    & 
    G'(\vec N^*)
    \ar[uu, \concar, ""{name=U3}]
    \ar[from=U4, to=U3, phantom, "\blue{V}"]
    \end{tikzcd}
    \]

\end{proof}

\section{Completeness in the Presence of Groups}
\label{sec:completeness-in-the-presence-of-groups}

\newcommand{\checkterm}{\mathsf{check}}
\newcommand{\precheckterm}{\mathsf{precheck}}
\newcommand{\repairterm}{\mathsf{repair}}
\newcommand{\postrepairterm}{\mathsf{postrepair}}

In \cref{cor:aperiodic-completeness}, we have shown that our equational axiomatisation is complete for \kl{aperiodic terms}, i.e.~if two aperiodic terms are denotationally equal, then their equality can be proved in our theory. 
In this part of the appendix, we finish the completeness proof, by taking into account terms that  use the \kl{group prefix multiplication} $\prefterm$. Since we already have completeness for aperiodic terms, we will no longer need to justify provability for equivalences on aperiodic terms, as long as they are denotationally true. This will spare us from the kind of low-level equational reasoning that we had to do in the aperiodic case.

\subsection{A Single Group Suffices}

\label{sec:single-group-suffices}
We begin by showing that, without loss of generality for the completeness proof, there is only a single finite group in the system. 
\begin{lemma}\label{lem:just-one-group}
    Assume that for every finite group $G$, our equational axiomatisation is complete for terms that use $\prefterm$ only for $G$. Then our equational axiomatisation is complete for all terms.
\end{lemma}
\begin{proof}
    Suppose that we want to prove equality $f \equiv g$ for two terms, which might use \kl{group prefix multiplication} for different groups. Take the product 
    \begin{align*}
    G = G_1 \times \cdots \times G_n.
    \end{align*}
    of all the groups that appear in $f$ and $g$. For each $i$ consider the two group homomorphisms 
    \begin{align*}
    G_i \xhookrightarrow {\text{$i$-th injection}} 
    G \xtwoheadrightarrow {\text{$i$-th projection}}
    G_i,
    \end{align*}
    which are the identity on $G_i$ when composed. By naturality of the group prefix multiplication in terms of the finite group, which we have shown in \cref{lem:naturality-of-group-prefix-operation},  the following diagram commutes provably for every \kl{type} $A$:
    \[
        \begin{tikzcd}[column sep=3cm]
        (A \times G_i)^*
        \ar[r,hookrightarrow, "{\text{$i$-th injection}}"]
        \ar[d, "\prefterm"']
        &
        (A \times G)^*
        \ar[r,twoheadrightarrow, "{\text{$i$-th projection}}"]
        \ar[d, "\prefterm"]
        &
        (A \times G_i)^* 
        \ar[d, "\prefterm"]
        \\
        (A \times G_i)^*
        \ar[r,hookrightarrow, "{\text{$i$-th injection}}"]
        &
        (A \times G)^*
        \ar[r,twoheadrightarrow, "{\text{$i$-th projection}}"]
        &
        (A \times G_i)^*. 
        \end{tikzcd}
    \]
Therefore, the group prefix multiplication for $G_i$ can be expressed in terms of the group prefix multiplication for $G$, as well as some \kl{aperiodic terms} (the horizontal arrows). This means that we can find terms that are provably equivalent to $f$ and $g$, respectively, and which use the group prefix multiplication only for $G$. By the  assumption of the lemma, we can then prove equality of these new terms, and therefore also of $f$ and $g$.
\end{proof}

Thanks to the above lemma, in the rest of this section we will assume that there is a single finite group $G$ in the system, and we will show that the equational axiomatisation is complete for terms that use $\prefterm$ only for this group $G$. Fix the group $G$ for the rest of \cref{sec:completeness-in-the-presence-of-groups}.

\subsection{Coproduct Notation}

When defining the \kl{group prefix multiplication}, we used the input \kl{type} $(A \times G)^*$. However, in this section it will be more convenient to use coproducts ranging over $G$ instead of products with $G$. In this section, we present the corresponding notation. The type $A \times G$ is isomorphic to 
\begin{align*}
\Sigma A \eqdef \sum_{g \in G}A.
\end{align*}
The idea is that $\Sigma A$ is a type annotated with values in the group.
Under this \kl{isomorphism}, the input and output types of the group prefix multiplication are isomorphic to $(\Sigma A)^*$, and therefore we can use the group prefix multiplication over this type. For the rest of this section, we assume that the group prefix multiplication is defined as 
\[
    \left( \Sigma A \right)^* \xrightarrow{\prefterm} \left( \Sigma A \right)^*.
\]

\subsection{Partial Prefix Multiplication}

Suppose that $H \subseteq G$ is a (nonempty) subset of elements of the group $G$ and that $\sum_H A$ is a coproduct \kl{type}, where the variants are indexed by elements of $H$. Formally, unless $H = G$, we are not able to apply the prefix multiplication to the type $(\sum_H A)^*$, since the \kl{group prefix multiplication} is only defined for the full coproduct $(\Sigma A)^*$. However, we can first inject the coproduct into the full type $(\Sigma A)^*$, and then apply the prefix multiplication: 
\[
    \left( \sum_H A \right)^* \drop{\rho^*} (\Sigma A)^* \xrightarrow{\prefterm} (\Sigma A)^*.
\]
After that, we would like to go back to $(\sum_H A)^*$. To do this, consider the term
\[
    (\Sigma A)^* \xrightarrow{\checkterm_H} \myunderbrace{(\Sigma A)^*}{obtained from $H$} \hspace*{2mm} +  \hspace*{2mm} \myunderbrace{(\Sigma A)^*}{the rest} \xrightarrow{\either{\unprefterm}{(1\comp[])}} (\Sigma A)^* \drop{(\inv \rho)^*} \left( \sum_H A \right)^*,
\]
where $\inv \rho$ is a (nonnatural) drop one-sided inverse of $\rho$ and $\checkterm_H$ is a term which denotationally checks whether the input list has been obtained by prefix multiplication from a list of elements in $H$, and puts it in the left variant if so, the right variant otherwise. We leave it as an exercise to the reader to show that $\checkterm_H$ exists, and that it is aperiodic. 

Our goal now is to show that this procedure works as expected, i.e.~that it creates a loop in $(\sum_H A)^*$ which is provably equivalent to $\idterm$. First, we prove that the term $\checkterm_H$ can be replaced by $\leftterm$.

\begin{lemma}\label{lem:checkterm-leftterm}
    The following diagram commutes provably:
    \[
    \begin{tikzcd}
        (\sum_H A)^*
        \ar[r, \dropar, "\rho^*"]
        \ar[d, \dropar, "\rho^*"'{xshift=-0.5ex}]
        &(\Sigma A)^*
        \ar[r, "\prefterm"]
        &(\Sigma A)^*
        \ar[d, "\checkterm_H"]
        \\ 
        (\Sigma A)^*
        \ar[r, "\prefterm"'{xshift=-0.5ex}]
        &(\Sigma A)^*
        \ar[r, \dropar, "\leftterm"]
        &(\Sigma A)^* + (\Sigma A)^*
    \end{tikzcd}
    \]
\end{lemma}
\begin{proof}
    Instead of applying $\checkterm_H$ after the list is already multiplied, we can first check whether the input list $(\Sigma A)^*$ consists of elements from $H$, based on this split it into two cases, and then apply the prefix multiplication to both of them. Concretely, consider the term \[
        (\Sigma A)^* \xrightarrow{\precheckterm_H} (\Sigma A)^* + (\Sigma A)^*,
    \]
    which denotationally checks whether the input list consists only of elements from $H$ and puts it in the left variant if so, the right variant otherwise. The term $\precheckterm_H$ is clearly aperiodic. Moreover, we obtain the following diagram:
    \[
    \begin{tikzcd}[column sep=huge]
        (\Sigma A)^* 
        \ar[r, "\prefterm"]
        \ar[d, "\precheckterm_H"'{xshift=-0.5ex}]
        &(\Sigma A)^*
        \ar[d, "\checkterm_H"]
        \\ 
        (\Sigma A)^* + (\Sigma A)^*
        \ar[r, "\prefterm + \prefterm"']
        &(\Sigma A)^* + (\Sigma A)^*
    \end{tikzcd}
    \]
    The diagram clearly commutes denotationally, and by the isomorphism trick, is equivalent to the equality of \kl{aperiodic terms} $\unprefterm \comp \precheckterm_H$ and $\checkterm_H \comp (\unprefterm + \unprefterm)$. Therefore, it is provably commutative by \cref{cor:aperiodic-completeness}.

    With this in hand, the top part of the diagram in the statement of the lemma is equivalent to 
    $\rho^* \comp \precheckterm_H \comp (\prefterm + \prefterm)$. Now we can replace $\precheckterm_H$ by $\leftterm$ provably: the terms $\rho^* \comp \precheckterm_H$ and $\rho^* \comp \leftterm$ are denotationally equal, and hence provably equivalent by \cref{cor:aperiodic-completeness}. The last piece is the equivalence of $\leftterm \comp (\prefterm + \prefterm)$ and $\prefterm \comp \leftterm$, which follows from naturality of $\leftterm$ (\cref{ax:naturality}).
\end{proof}

With \cref{lem:checkterm-leftterm} in hand, we can now prove that the entire loop of partial prefix multiplication is provably equivalent to $\idterm$.

\begin{lemma}\label{lem:partial-prefix-multiplication}
    \[
        \rho^* \comp \prefterm \comp \checkterm_H \comp \either{\unprefterm}{(1\comp[])} \comp (\inv \rho)^* \equiv \idterm.
    \]
\end{lemma}
\begin{proof}
    By \cref{lem:checkterm-leftterm}, the term $\checkterm_H$ in the left-hand side of the postulated equivalence can be provably replaced by $\leftterm$. The rest is computation:
    \begin{align*}
        \rho^* \comp \prefterm \comp \leftterm \comp \either{\unprefterm}{(1\comp[])} \comp (\inv \rho)^* \equalbecause{\cref{lem:case-equivalence}} \\
        \rho^* \comp \prefterm \comp \unprefterm \comp (\inv \rho)^* \equalbecause{\cref{ax:isomorphisms}} \\ 
        \rho^* \comp (\inv \rho)^* \equalbecause{the choice of $\inv \rho$} \\ 
        \idterm.
    \end{align*}
\end{proof}

\subsection{Group-Lifts}

We will prove completeness of our system in the presence of $\prefterm$ analogously to the proof of completeness in the aperiodic case. The first step is to adjust the definition of \kl{lift} and \kl{drop terms}, as currently $\prefterm$ is not derivable as a composition of a lift followed by a drop, which invalidates \cref{lem:lift-then-drop} in the general setting. 

We will not change the definition of drop terms, but lifts need to incorporate new primitive operations. What we want is to equip lifts with the ability to perform the (partial) \kl{group prefix multiplication}. This will unfortunately invalidate \cref{lem:lift-provable-inverse}, because lifts will no longer be \kl{surjective}. However, we still want them to be \kl{injective}, and have drop terms as their one-sided inverses -- this property is crucial for the proof of completeness.

To achieve all of the above, we define a new kind of lift, called \emph{Simon lift} -- indirectly inspired by the Factorisation Forest Theorem of Simon~\cite{Simon1990} -- which is capable of partial group prefix multiplication (in some relaxed sense) and has a drop one-sided inverse.

\begin{lemma}[Simon lift]\label{lem:simon-lift}
    Suppose that $H \subseteq G$ is a subset of elements of the group $G$ and $A$ is a \kl{type}. There exists a term, called the \emph{Simon lift} associated to $H$ and $A$, which satisfies the following properties:
    \begin{enumerate}
        \item it is a composition of a \kl{drop}, $\prefterm$, and then a \kl{lift} $\lambda$,
        \item it is \kl{injective}, and its one-sided inverse is a \kl{drop term} $\gamma$:
        \[
        \begin{tikzcd}[column sep=huge]
            \left(\sum_H A \right)^* 
            \ar[r, "\text{Simon lift}"]
            \ar[dr, "\idterm"'{xshift=-0.5ex}]
            &S
            \ar[d, \dropar, "\gamma"]
            \\
            &\left(\sum_H A \right)^*
        \end{tikzcd}
        \]
        \item there exists a \kl{drop term} $\delta$ such that the following diagram commutes provably:
        \[
        \begin{tikzcd}
            (\sum_H A)^* 
            \ar[r, \dropar, "\rho^*"]
            &(\Sigma A)^*
            \ar[d, "\prefterm"]
            \\ 
            S
            \ar[r, \dropar, "\delta"']
            \ar[u, \dropar, "\gamma"{xshift=-0.5ex}]
            &(\Sigma A)^*
        \end{tikzcd}
        \]
    \end{enumerate}
\end{lemma}
\begin{proof}
    Consider the term $\checkterm_H$ from the previous section. By the \kl{lift-drop factorisation} for \kl{aperiodic terms} (\cref{lem:lift-then-drop}) and the fact that lifts are \kl{isomorphisms} with drop inverses (\cref{lem:lift-provable-inverse}), there is a lift $\lambda$ and drop terms $f, g$ such that the following diagram commutes provably:
    \[
    \begin{tikzcd}[column sep=huge, row sep=huge]
        (\Sigma A)^* 
        \ar[r, "\checkterm_H"]
        \ar[dr, \liftar, "\lambda"'{xshift=-0.5ex}]
        &(\Sigma A)^* + (\Sigma A)^*
        \ar[r, "\unprefterm + \unprefterm"]
        &(\Sigma A)^* + (\Sigma A)^*
        \\ 
        &S
        \ar[u, \dropar, "f"{xshift=-0.5ex}]
        \ar[ur, \dropar, "g"']
    \end{tikzcd}
    \]
    With this in hand, we define the \kl{Simon lift} and drops $\gamma, \delta$ as follows: 
    \begin{align*}
        \text{Simon lift} &\eqdef \rho^* \comp \prefterm \comp \lambda \\
        \gamma            &\eqdef g \comp \either{\left( \inv \rho \right)^*}{(1\comp[])} \\
        \delta            &\eqdef f \comp \either{\idterm}{(1\comp[])}.
    \end{align*}
    We now verify the properties (2) and (3). The property (2) follows from \cref{lem:partial-prefix-multiplication} with some straightforward computation:
    \begin{align*}
        \text{Simon lift} \comp \gamma &= \why{by definition} \\ 
        \rho^* \comp \prefterm \comp \lambda \comp g \comp \either{\left( \inv \rho \right)^*}{(1\comp[])} \equalbecause{the choice of $\lambda$ and $g$} \\ 
        \rho^* \comp \prefterm \comp \checkterm_H \comp \unprefterm + \unprefterm \comp \either{\left( \inv \rho \right)^*}{(1\comp[])} \equalbecause{\cref{lem:functoriality-of-either-left}} \\ 
        \rho^* \comp \prefterm \comp \checkterm_H \comp \either{\left( \unprefterm \comp \left( \inv \rho \right)^* \right)}{\left( \unprefterm \comp (1\comp []) \right)} \equalbecause{\cref{cor:aperiodic-completeness}} \\ 
        \rho^* \comp \prefterm \comp \checkterm_H \comp \either{\left( \unprefterm \comp \left( \inv \rho \right)^* \right)}{\left( (1\comp []) \comp (\inv \rho)^* \right)} \equalbecause{\cref{lem:functoriality-of-either}} \\ 
        \rho^* \comp \prefterm \comp \checkterm_H \comp \either{\unprefterm}{(1\comp [])} \comp (\inv \rho)^* \equalbecause{\cref{lem:partial-prefix-multiplication}} \\ 
        \idterm.
    \end{align*}
    For property (3), after multiplying on the left by $\lambda$ and on the right by $\unprefterm$ (both are isomorphisms), we obtain an equivalent equality to prove:
    \[
        \lambda \comp \gamma \comp \rho^* \equiv \lambda \comp \delta \comp \unprefterm.
    \]
    Both sides are already aperiodic, so their provable equivalence would follow from denotational equality by \cref{cor:aperiodic-completeness}. To see that they are denotationally equal, we rewrite both sides. For the left-hand side:
    \begin{align*}
        \lambda \comp \gamma \comp \rho^* &= \why{by definition of $\gamma$} \\ 
        \lambda \comp g \comp \either{\left( \inv \rho \right)^*}{(1\comp[])} \comp \rho^* \equalbecause{the choice of $\lambda$ and $g$} \\
        \checkterm_H \comp \unprefterm + \unprefterm \comp \either{\left( \inv \rho \right)^*}{(1\comp[])} \comp \rho^* \equalbecause{Lemmas \ref{lem:functoriality-of-either-left} and \ref{lem:functoriality-of-either}} \\
        \checkterm_H \comp \either{\left( \unprefterm \comp (\inv \rho)^* \comp \rho^* \right)}{\left( \unprefterm \comp (1 \comp []) \comp \rho^* \right)} \equalbecause{\cref{cor:aperiodic-completeness}} \\ 
        \checkterm_H \comp \either{\left( \unprefterm \comp (\inv \rho)^* \comp \rho^* \right)}{(1 \comp [])}.
    \end{align*}
    Similarly, the right-hand side can be rewritten as follows:
    \begin{align*}
        \lambda \comp \delta \comp \unprefterm &= \why{by definition of $\delta$} \\
        \lambda \comp f \comp \either{\idterm}{(1\comp[])} \comp \unprefterm \equalbecause{the choice of $\lambda$ and $f$} \\
        \checkterm_H \comp \either{\idterm}{(1\comp[])} \comp \unprefterm \equalbecause{\cref{lem:functoriality-of-either} and \cref{cor:aperiodic-completeness}} \\ 
        \checkterm_H \comp \either{\unprefterm}{(1 \comp [])}.
    \end{align*}
    Therefore, we are left with proving that after applying $\checkterm_H$ and then $\unprefterm$ to the obtained left variant, we obtain a list on which $(\inv \rho)^* \comp \rho^*$ is the identity. This is exactly the intended purpose of $\checkterm_H$, which sends to the left variant the lists consisting of only elements in $H$ -- and these are exactly the lists that after unprefixing are fixed points of $(\inv \rho)^* \comp \rho^*$ (i.e.~they are in the image of $\rho^*$).

    Therefore, both sides are denotationally equal, and since they are also aperiodic, their provable equivalence follows from \cref{cor:aperiodic-completeness}, and so follows property (3).
\end{proof}

Apart from the Simon lift, we also define \emph{group injections}, which lift the domain of a drop term with a group codomain into a sum annotated by a subset of the group.

\begin{lemma}[Group injection]\label{lem:group-injection}
    Consider a \kl{drop term} $f : A \dropmap \Sigma B$. There exists a term, called the \emph{group injection} associated to $f$ with codomain $\sum_H C$ for some \kl{type} $C$ and $H \subseteq G$, which satisfies the following properties:
    \begin{enumerate}
        \item it is aperiodic,
        \item it is \kl{injective} and its one-sided inverse is a \kl{drop term} $\sigma$:
        \[
        \begin{tikzcd}[column sep=2.5cm]
            A
            \ar[r, "\text{group injection}"]
            \ar[dr, "\idterm"'{xshift=-0.5ex}]
            &\sum_H C
            \ar[d, \dropar, "\sigma"]
            \\
            &A
        \end{tikzcd}
        \]
        \item there exists a collection of \kl{drop terms} $\vec f$ indexed by $H$, such that the following diagram commutes provably ($\rho_h$ is the coprojection of $B$ into the $h$-th variant of $\Sigma B$):
        \[
        \begin{tikzcd}[column sep=2.5cm]
            A
            \ar[r, \dropar, "f"]
            &\Sigma B
            \\
            \sum_H C
            \ar[u, \dropar, "\sigma"{xshift=-0.5ex}]
            \ar[ur, \dropar, "\eithermany{h}{(f_h \comp \rho_h)}"']
        \end{tikzcd}
        \]
    \end{enumerate}
\end{lemma}
\begin{proof}
    By a repeated application of the normalisation of drop terms with sum codomains (\cref{lem:drops-with-sum-co-domain}), we obtain a lift $\alpha$ and collections of drop terms $\vec g, \vec \rho$ such that $f \equiv \alpha \comp \eithermany{i}{(g_i \comp \rho_i)}$, where every $g_i$ is of the type $D_i \dropmap B$ for some $D_i$, and $\rho_i : B \dropmap \Sigma B$ is a coprojection into some variant of the group coproduct. By grouping summands, we can assume that every $\rho_i$ sends its domain into a different group element. Let $H \subseteq G$ be the subset that $\vec \rho$ sends elements to, and label terms $\vec g, \vec \rho$ with corresponding elements of $H$. 

    Now consider the following injective drop term, which sends every $D_h$ first into $\sum_H D_h$, and then labels it with $h$ again:
    \[
        \sum_{h \in H} D_h \drop{\varepsilon} \sum_H \left( \sum_{h \in H} D_h \right).
    \]  
    Define the \kl{group injection} associated to $f$ to be $\alpha \comp \varepsilon$. Note that (1) is satisfied. We now want to choose an inverse $\inv \varepsilon$ of $\varepsilon$, which composed with the unique inverse $\inv \alpha$ of $\alpha$ will give $\sigma$. 
    
    For every $h \in H$, we choose a \kl{value} $d_h \in D_h$. Then $\inv \varepsilon$ is defined as $\sum_{h \in H} \eithermany{k \in H}{(\inv \varepsilon_{h k})}$, where $\inv \varepsilon_{h k}$ is equal to $\idterm$ for $h = k$, and $(1\comp d_h)$ otherwise. It is clear that $\varepsilon \comp \inv \varepsilon \equiv \idterm$, and as a consequence $\sigma$ is a one-sided inverse of the group injection, hence (2) is satisfied.

    For (3), we have to be careful with the choice of $\vec f$, so that it is compatible with $\sigma$. We define $f_h : \sum_{k \in H} D_k \dropmap B$ as $\eithermany{k}{(f_{h k})}$, where $f_{h k}$ is equal to $(\inv \varepsilon_{h k} \comp g_h)$. It remains to verify the condition (3):
    \begin{align*}
        \sigma \comp f &= \why{the choice of $\sigma$} \\ 
        \inv \varepsilon \comp \inv \alpha \comp f \equalbecause{the choice of $\alpha$ by \cref{lem:drops-with-sum-co-domain}} \\ 
        \inv \varepsilon \comp \eithermany{h}{(g_h \comp \rho_h)} &= \why{the definition of $\inv \varepsilon$} \\
        \left( \sum_{h \in H} \eithermany{k \in H}{\left( \inv \varepsilon_{h k} \right)} \right) \comp \eithermany{h}{(g_h \comp \rho_h)} \equalbecause{\cref{lem:functoriality-of-either-left}} \\ 
        \eithermany{h \in H}{\left( \eithermany{k \in H}{(\inv \varepsilon_{h k})} \comp g_h \comp \rho_h \right)} \equalbecause{\cref{lem:functoriality-of-either}} \\ 
        \eithermany{h \in H}{\left( \eithermany{k \in H}{(\inv \varepsilon_{h k} \comp g_h)} \comp \rho_h \right)} &= \why{the choice of $f_{h k}$} \\ 
        \eithermany{h \in H}{\left( \eithermany{k \in H}{(f_{h k})} \comp \rho_h \right)} &=  \why{the choice of $f_h$} \\ 
        \eithermany{h}{\left( f_h \comp \rho_h \right)}.
    \end{align*}
\end{proof}

We are now in position to define the corresponding notion of lift terms in the presence of groups, which we call \intro{group-lifts}.

\begin{definition}[Group-lifts]\label{def:group-lifts}
    \kl{Group-lifts} is the class of terms defined identically to \kl{lifts} (\cref{def:lift-drop}), except that we add the \kl{Simon lift} (\cref{lem:simon-lift}) and \kl{group injections} (\cref{lem:group-injection}) to the list of \kl{atomic lift terms}.
\end{definition}

\subsection{Proof of Completeness}

It is sufficient to reduce the completeness proof to the case of equality of two \kl{drop terms}, similarly as in the aperiodic case in \cref{cor:aperiodic-completeness}, because introducing groups did not affect the definition of drop terms. To achieve this, we need an analogue of the \kl{lift-drop factorisation} (\cref{lem:lift-then-drop}), suitable for \kl{group-lifts}. However, we have to be careful: a verbatim ``\kl{group-lift--drop factorisation}'' statement would not suffice to prove completeness, since group-lifts are not \kl{surjective}, and hence cannot be cancelled on the left-hand side in equivalence proofs. Instead, we prove a strictly stronger version of a factorisation, which asserts that the diagram commutes after inverting the group-lift. 

\begin{lemma}[Group-lift--drop factorisation]\label{lem:grouplift-then-drop}
    Every term $f : A \to B$ decomposes as a \kl{group-lift} followed by a \kl{drop}, in the sense that the following diagram commutes provably:
    \[ 
    \begin{tikzcd}[column sep=huge]
        A
        \ar[r, "f"]
        &B
        \\ 
        C 
        \ar[u, \dropar, "\text{a group-lift inverse}"{xshift=-0.5ex}]
        \ar[ur, \dropar, "\text{a drop}"']
    \end{tikzcd}
    \]
\end{lemma}

Note that the version of factorisation with the inverted group-lift is strictly stronger than the naive factorisation, precisely because group-lifts are not surjective. In particular, if we restrict \cref{lem:grouplift-then-drop} to \kl{aperiodic terms} $f$ and \kl{lifts}, it becomes equivalent to \cref{lem:lift-then-drop}.

\cref{lem:grouplift-then-drop} is sufficient to reduce completeness of the equational axiomatisation for all terms to the case of drop terms. Similarly to the proof of \cref{cor:aperiodic-completeness}, given two denotationally equal terms, we find their joint group-lift--drop factorisation, and then we cancel the group-lift inverse on the left-hand side of the equivalence by \cref{lem:pre-postcompose-with-injective-surjective}, which can be done because the group-lift inverses are surjective.

Therefore, we are left with proving \cref{lem:grouplift-then-drop}. Similarly to the proof of the aperiodic version (\cref{lem:lift-then-drop}), we proceed by structural induction on the term $f$. The only new \kl{atomic term} to consider is $\prefterm$, whose decomposition follows from property (3) of the definition of the \kl{Simon lift} (\cref{lem:simon-lift}) applied with $\rho = \idterm$. Again, the cases of product, coproduct and list combinators are purely functorial and the only interesting case is the composition, which requires an analogue of the exchange lemma (\cref{lem:exchange}) for group-lifts (with inverted group-lift arrows), which we now state and prove in the next section.

\begin{lemma}[Group exchange]\label{lem:group-exchange}
    For every composition of a \kl{drop term} $f$ with a \kl{group-lift} $\alpha$, there is a \kl{group-lift} and a \kl{drop} such that the following diagram commutes provably:
    \[
    \begin{tikzcd}[column sep=huge]
        A
        \ar[r, \dropar, "f"]
        &B
        \\
        C
        \ar[u, \dropar, "\text{a group-lift inverse}"{xshift=-0.5ex}]
        \ar[r, \dropar, "\text{a drop}"']
        &D
        \ar[u, \dropar, "{\inv \alpha}"'{xshift=0.5ex}]
    \end{tikzcd}
    \]
\end{lemma}

\subsection{Proof of the Group Exchange Lemma}

To prove \cref{lem:group-exchange}, it is enough to show that every \kl{drop} followed by an atomic \kl{group-lift} has a \kl{group-lift--drop factorisation}, and the rest follows functorially similarly to the aperiodic case (\cref{lem:exchange}). We only need to do this for the new atomic group-lifts, since the original \kl{atomic lifts} are already handled in \cref{lem:exchange}. 

The exchange for \kl{group injections} is straightforward, because they are aperiodic. Note that every drop term $f : A \dropmap B$ composed with a group injection gives an \kl{aperiodic term}, which can be lift-drop factorised by \cref{lem:lift-then-drop}. Therefore, the construction is as on the following diagram:
\[
\begin{tikzcd}[row sep=1cm, column sep=3cm]
    A 
    \ar[r, \dropar, "f"]
    \ar[dr, "{f \, \comp \, \text{group injection}}" description]
    \ar[d, \liftar, shift right=1ex]
    &B
    \\ 
    D
    \ar[u, shift right=1ex, \dropar]
    \ar[r, \dropar]
    &\sum_H C 
    \ar[u, \dropar, "\sigma"'{xshift=0.5ex}]
\end{tikzcd}
\]

Therefore, we are left with the case of a drop followed by a \kl{Simon lift}. Before we tackle the case of general drop terms, we begin with a special case, where the drop is of the type $(\sum_H A)^* \dropmap (\sum_H B)^*$ and applies a collection of drops $A \dropmap B$ component-wise.

\begin{lemma}\label{lem:simon-exchange-component-wise}
    Consider a collection $\vec f$ of \kl{drop terms} $A \dropmap B$ indexed by a set $H \subseteq G$. There is a \kl{group-lift} and a \kl{drop} such that the following diagram commutes provably:
    \[
    \begin{tikzcd}[column sep=1.5cm]
        \left( \sum_H A \right)^* 
        \ar[r, \dropar, "{\left( \sum_{h \in H} f_h \right)^*}"{yshift=0.5ex}]
        &\left( \sum_H B \right)^*
        \ar[d, shift right=0.5ex, "\text{Simon lift}"'{xshift=-0.5ex}]
        \\ 
        C 
        \ar[u, \dropar, "\text{a group-lift inverse}"{xshift=-0.5ex}]
        \ar[r, \dropar, "\text{a drop}"']
        &S
        \ar[u, \dropar, shift right=0.5ex, "\gamma"'{xshift=0.5ex}]
    \end{tikzcd}
    \]
\end{lemma}
\begin{proof}
    The first step of the construction is to apply a Simon lift to the input \kl{type} $(\sum_H A)^*$. Now, suppose that we found an aperiodic term $g$ such that the following diagram would commute provably:
    \[
    \begin{tikzcd}
        \left( \sum_H A \right)^* 
        \ar[r, \dropar, "{\left( \sum_{h \in H} f_h \right)^*}"{yshift=0.5ex}]
        &\left( \sum_H B \right)^*
        \\ 
        (\Sigma A)^* + (\Sigma A)^*
        \ar[u, \dropar, "{\either{\left( \inv \rho \right)^*}{(1 \comp [])}}"{xshift=-0.5ex}]
        &(\Sigma B)^* + (\Sigma B)^*
        \ar[u, \dropar, "{\either{\left( \inv \rho \right)^*}{(1 \comp [])}}"'{xshift=0.5ex}]
        \\ 
        (\Sigma A)^* + (\Sigma A)^*
        \ar[u, \dropar, "\unprefterm + \unprefterm"{xshift=-0.5ex}]
        &(\Sigma B)^* + (\Sigma B)^*
        \ar[u, \dropar, "\unprefterm + \unprefterm"'{xshift=0.5ex}]
        \\ 
        (\Sigma A)^* 
        \ar[u, "\checkterm_H"{xshift=-0.5ex}]
        \ar[r, "g"]
        &(\Sigma B)^* 
        \ar[u, "\checkterm_H"'{xshift=0.5ex}]
        \\ 
        T 
        \ar[u, \dropar, "\inv{\lambda_T}"{xshift=-0.5ex}]
        \ar[r, "{\inv{\lambda_T} \comp g \comp \lambda_S}"']
        &S 
        \ar[u, \dropar, "\inv{\lambda_S}"'{xshift=0.5ex}]
    \end{tikzcd}
    \]
    Then we would be done, because the left and right upward paths starting from $T$ and $S$ are inverses of the Simon lifts by the proof of \cref{lem:simon-lift}, and the bottom arrow is an aperiodic term that can be further lifted by \cref{lem:lift-then-drop}, which would complete the construction.

    The bottom rectangle commutes trivially, independently of the choice of $g$. Therefore, we are left with the task of finding an aperiodic term $g$ such that the upper face commutes. It suffices to commute denotationally, because the diagram consists only of aperiodic terms, and hence it will commute provably by \cref{cor:aperiodic-completeness}.

    The goal of $g$ is to apply some term $f_h$ to every element of the input list. For the diagram to commute, every element must be mapped by $f_h$, where $h \in H$ is the group element that the input element will be labelled with after applying $\unprefterm$. Similarly as in the case of $\checkterm_H$, inspecting which label it should be is an aperiodic operation. There is only one caveat: the input list may contain elements that after unprefixing are labelled by an element outside of $H$. In this case, we can apply any $f_h$ to it, because such an input list will be identified as ``invalid'' by $\checkterm_H$ and then discarded in the $(1\comp[])$ branch either way.
\end{proof}

We now proceed to the exchange of Simon lift for general drop terms. Since a Simon lift is a composition of a drop, $\prefterm$, and then a \kl{lift}, in light of the aperiodic exchange lemma (\cref{lem:exchange}), it is enough to show that a drop followed by $\prefterm$ has a group-lift--drop factorisation. This is exactly the content of the following lemma, which is the last piece of \cref{thm:main}.

\begin{lemma}[Prefix exchange]
    For every \kl{drop} $f : A \dropmap (\Sigma B)^*$, there is a \kl{group-lift} and a \kl{drop} such that the following diagram commutes provably:
    \[
    \begin{tikzcd}[column sep=huge]
        A
        \ar[r, \dropar, "f"]
        &(\Sigma B)^*
        \\
        C
        \ar[u, \dropar, "\text{a group-lift inverse}"{xshift=-0.5ex}]
        \ar[r, \dropar, "\text{a drop}"']
        &(\Sigma B)^*
        \ar[u, "\unprefterm"'{xshift=0.5ex}]
    \end{tikzcd}
    \]
\end{lemma}

\begin{proof}
    By \cref{lem:drops-with-list-co-domain}, it suffices to prove the statement for drops $f$ in \kl[normalised drop term]{normal form} $F(\vec f) \comp \concmap$. We proceed by induction on the structure of $F$.
    \begin{description}
        \item[Leaf node, $F = X_i$.] In this case, $\left(F(\vec f) \comp \concmap\right) = f_i$. We omit the index $i$ for simplicity. We consider three subcases depending on the shape of $f$.
        \begin{description}
        \item[Value, $f = (1 \comp v)$.] There is a \kl{value} $w$ such that $v \equiv w \comp \unprefterm$ by \cref{lem:value-completeness}. We get that $\idterm \comp (1 \comp v) \equiv (1 \comp w) \comp \unprefterm$.

        \item[Singleton, $f = {[g]}$.] Here $g : A \dropmap \Sigma B$ is a drop term, and the exchange is $\idterm \comp [g] \equiv [g] \comp \unprefterm$ by \cref{cor:aperiodic-completeness}.

        \item[List, $f = g^*$.] Here $g : A \dropmap \Sigma B$ is a drop term. The construction uses both a group injection and a Simon lift, as illustrated on the following diagram:
        \[
        \begin{tikzcd}[column sep=huge]
            A^*
            \ar[r, \dropar, "{g^*}"]
            \ar[d, shift right=1ex, "\text{group injection}^*"'{xshift=-0.5ex}]
            &(\Sigma B)^*
            \\ 
            \left(\sum_H C\right)^* 
            \ar[u, shift right=1ex, \dropar, "{\sigma^*}"'{xshift=0.5ex}]
            \ar[r, \dropar, "{\left(\sum_{h \in H} g_h\right)^*}"{yshift=0.5ex}]
            &\left(\sum_H B\right)^*
            \ar[u, \dropar, "\rho^*"'{xshift=0.5ex}]
            \ar[d, shift right=1ex, "\text{Simon lift}"'{xshift=-0.5ex}]
            &(\Sigma B)^*
            \ar[ul, "\unprefterm"']
            \\
            T 
            \ar[r, \dropar]
            \ar[u, \dropar]
            &S 
            \ar[u, shift right=1ex, \dropar, "\gamma"'{xshift=0.5ex}]
            \ar[ur, \dropar, "\delta"'{xshift=0.5ex}]
        \end{tikzcd}
        \]
        The diagram commutes: the top-left rectangle commutes by the definition of the group injection, the bottom rectangle commutes by \cref{lem:simon-exchange-component-wise}, and the right face commutes by property (3) of the Simon lift (\cref{lem:simon-lift}). The group-lift inverse we need is the left upward path, and the drop is the bottom path.
        \end{description}

        \item[Sum node, $F = F_1 + F_2$.] In this case, we have $\left( F(\vec f) \comp \concmap \right) = \either{\left( F_1(\vec f) \comp \concmap \right)}{\left( F_2(\vec f) \comp \concmap \right)}$. The construction is straightforward: we apply the induction hypothesis to both $F_1$ and $F_2$, and then use the functoriality of $\mathtt{either}$ (Lemmas \ref{lem:functoriality-of-either-left} and \ref{lem:functoriality-of-either}) to obtain a group-lift--drop factorisation for the sum.
        
        \item[Product node, $F = F_1 \times F_2$.] In this case, we have $\left( F(\vec f) \comp \concmap \right) = \left( F_1(\vec f) \comp \concmap \right) \odot \left( F_2(\vec f) \comp \concmap \right)$. We first demonstrate the construction, which uses the induction hypothesis on both $F_1$ and $F_2$ together with \cref{lem:lift-then-drop}, and then explain why it works:
        \[
        \begin{tikzcd}[column sep=2.5cm]
            F_1(\vec A) \times F_2(\vec A) 
            \ar[r, \dropar, "{\left( F_1(\vec f) \comp \concmap \right) \times \left( F_2(\vec f) \comp \concmap \right)}"]
            &(\Sigma B)^* \times (\Sigma B)^*
            \ar[r, \dropar, "\binconcatterm"]
            &(\Sigma B)^*
            \\ 
            C_1 \times C_2
            \ar[u, \dropar]
            \ar[d, \liftar, shift right=1ex]
            \ar[r, \dropar]
            &(\Sigma B)^* \times (\Sigma B)^*
            \ar[u, "{\unprefterm^2}"{xshift=-0.5ex}]
            \ar[d, "\text{aperiodic term}"'{xshift=-0.5ex}]
            &(\Sigma B)^*
            \ar[u, "\unprefterm"]
            \\ 
            D 
            \ar[u, \dropar, shift right=1ex]
            \ar[r, \dropar]
            &\Sigma \left( (\Sigma B)^* \times (\Sigma B)^* \right)
            \ar[ur, \dropar]
        \end{tikzcd}
        \]
        The top rectangle commutes by induction hypothesis on $F_1$ and $F_2$. The aperiodic term denotationally works as follows: it checks the last element of the first list and annotates the whole input with its corresponding element in the group, or the group identity if the list is empty. Then the diagonal drop term consolidates the two lists by multiplying every element in the second list by the group element obtained from the first list, and then concatenates the two lists. Clearly, the right face commutes denotationally, and hence provably by \cref{cor:aperiodic-completeness}. Lastly, the bottom rectangle is obtained by the \kl{lift-drop factorisation} (\cref{lem:lift-then-drop}) of the aperiodic term precomposed with the drop $C_1 \times C_2 \dropmap (\Sigma B)^* \times (\Sigma B)^*$.

        Therefore, the group-lift--drop factorisation for $F$ is obtained by taking the leftmost and bottom paths on the diagram.

        \item[List node, $F = G^*$.] In this case, we have $\left( F(\vec f) \comp \concmap \right) = \hom \left( G(\vec f) \comp \concmap \right)$. This is the most involved case, which combines techniques used in the previous cases. We first demonstrate the construction, and then explain why it works:
        \[
        \begin{tikzcd}[column sep=2cm]
            G(\vec A)^* 
            \ar[r, \dropar, "{\left( G(\vec f) \comp \concmap \right)^*}"]
            &(\Sigma B)^{**}
            \ar[r, \dropar, "\concatterm"]
            &(\Sigma B)^*
            \\ 
            C^* 
            \ar[u, \dropar]
            \ar[r, \dropar, "{g^*}"]
            \ar[d, \liftar, shift right=1ex]
            &(\Sigma B)^{**}
            \ar[u, "\unprefterm^*"{xshift=-0.5ex}]
            \ar[d, "\text{aperiodic term}^*"'{xshift=-0.5ex}]
            &(\Sigma B)^*
            \ar[u, "\unprefterm"'{xshift=0.5ex}]
            \\ 
            D^*
            \ar[u, \dropar, shift right=1ex]
            \ar[r, \dropar, "k^*"]
            \ar[d, shift right=1ex, "\text{group injection}^*"'{xshift=-0.5ex}]
            &\left( \Sigma \left( \Sigma B \right)^* \right)^*
            \ar[dr, "\prefterm"]
            \\ 
            \left( \sum_H E \right)^*
            \ar[u, shift right=1ex, \dropar, "\sigma^*"'{xshift=0.5ex}]
            \ar[r, \dropar, "{\left( \sum_{h \in H} k_h \right)^*}"{yshift=0.5ex}]
            &\left( \sum_H (\Sigma B)^* \right)^*
            \ar[u, \dropar, "\rho^*"'{xshift=0.5ex}]
            \ar[d, shift right=1ex, "\text{Simon lift}"'{xshift=-0.5ex}]
            &\left( \Sigma \left( \Sigma B \right)^* \right)^*
            \ar[uu, "\text{aperiodic term}"'{xshift=0.5ex}]
            \\
            T 
            \ar[r, \dropar]
            \ar[u, \dropar]
            &S
            \ar[u, shift right=1ex, \dropar, "\gamma"'{xshift=0.5ex}]
            \ar[ur, \dropar, "\delta"'{xshift=0.5ex}]
        \end{tikzcd}
        \]
        Regarding the top-right face, we will explain in a moment how to choose the aperiodic terms so that it commutes. The rest of the diagram is constructed similarly to the product and list cases: the top-left rectangle is obtained by induction on $G$, the second-to-top rectangle is obtained by the lift-drop factorisation (\cref{lem:lift-then-drop}) of $g$ composed with the aperiodic term, the second-to-bottom rectangle commutes by the definition of the group injection (\cref{lem:group-injection}), the bottom rectangle is obtained by \cref{lem:simon-exchange-component-wise}, and the bottom-right face commutes by property (3) of the Simon lift (\cref{lem:simon-lift}).

        This almost yields a group-lift--drop factorisation for $F$ by taking paths starting from $T$, but the bottom path is not exactly a drop followed by $\unprefterm$. However, this can be fixed by factoring the first three arrows of the bottom path as lift-then-drop (\cref{lem:lift-then-drop}) -- this lifts $T$ one more time and gives the desired group-lift--drop factorisation. This step is omitted from the diagram for clarity.

        It now remains to choose the aperiodic terms of the diagram in such a way that the top-right face commutes provably. The intention is straightforward: moving along the downwards path starting from $(\Sigma B)^{**}$, we want the first aperiodic term to annotate every sublist (i.e.~element of type $(\Sigma B)^*$) with the label of its last element, or the group identity if the list is empty. Then, $\prefterm$ is applied, which performs prefix multiplication on the outer labels that we have just added. After that, for every sublist, we want to multiply all of its elements by the product of labels of sublists strictly before it. The outer label contains the product of labels of all sublists up to this point inclusively, so it has to be first multiplied by the inverse of the last element of the sublist. After all elements of every sublist are multiplied, we concatenate $(\Sigma (\Sigma B)^*)^* \concmap (\Sigma B)^*$, and this is the second aperiodic term.

        We can see that denotationally this procedure consolidates a list of prefix-multiplied lists into one concatenated prefix-multiplied list -- which is exactly the same as the upwards path (after equivalently reversing the $\unprefterm$ arrow): $\unprefterm^* \comp \concatterm \comp \prefterm^*$, and hence the diagram commutes denotationally. However, it is unclear how to directly prove it in the equational system. Instead, we will reduce it to an aperiodic equation by reversing some of the arrows in the top-right face, which will then be proved by \cref{cor:aperiodic-completeness}. Unfortunately, this will require us to reverse a nonsurjective arrow, and therefore we have to make a nonnatural choice: from now on we fix a value $b : \onetype \to B$.
        
        First, consider a term $\repairterm$ of the type:
        \[
            \Sigma \left( \Sigma B \right)^* \xrightarrow{\repairterm} \Sigma \left( \Sigma B \right)^*.
        \]
        Denotationally, the job of $\repairterm$ is to modify the input list, so that its last element has a label matching the outer label. Suppose that the outer label corresponds to $g \in G$. Then $\repairterm$ inspects the list $(\Sigma B)^*$: (1) if it is empty and $g$ is the group identity, then it does nothing, (2) if it is empty and $g$ is not the identity, then it replaces the list with a one-element list $[b\comp\rho_g]$, and (3) otherwise it changes the label of the last element to $g$. In particular, if the last element's label is already $g$, then it does nothing. 

        Next, we define the term $\postrepairterm$, of the same type as $\repairterm^*$, whose job is to apply the same ``fix'' to the last element of every sublist, but after the outer labels have already been prefix-multiplied with $\prefterm$. In other words, we want to choose $\postrepairterm$ such that the following diagram commutes denotationally:
        \[
        \begin{tikzcd}
            \left( \Sigma \left( \Sigma B \right)^* \right)^*
            \ar[r, "\prefterm"]
            &\left( \Sigma \left( \Sigma B \right)^* \right)^*
            \\ 
            \left( \Sigma \left( \Sigma B \right)^* \right)^*
            \ar[u, "{\repairterm^*}"]
            \ar[r, "\prefterm"]
            &\left( \Sigma \left( \Sigma B \right)^* \right)^*
            \ar[u, "\postrepairterm"']
        \end{tikzcd}
        \]
        At this point, it should be clear how $\postrepairterm$ should be constructed: the outer labels from before applying $\prefterm$ can be recovered by inspecting windows of width two, similarly as $\unprefterm$ does, and based on that the last elements of the sublists can be adjusted using the same value $b$ as in the $\repairterm$ procedure. Since both $\repairterm$ and $\postrepairterm$ are aperiodic, the diagram above is equivalent to the aperiodic rectangle obtained by inverting both $\prefterm$ arrows, and so is provably commutative by \cref{cor:aperiodic-completeness}.

        Finally, we arrive at the following diagram:
        \[
        \begin{tikzcd}[column sep=1.5cm]
            \left( \Sigma B \right)^{**}
            \ar[r, \dropar, "\concatterm"]
            &
            \left( \Sigma B \right)^*
            \\ 
            \left( \Sigma B \right)^{**}
            \ar[u, "{\unprefterm^*}"]
            &\left( \Sigma B \right)^*
            \ar[u, "\unprefterm"']
            \\ 
            \left( \Sigma \left( \Sigma B \right)^* \right)^*
            \ar[u, "\text{forget outer label}^*"]
            &\left( \Sigma \left( \Sigma B \right)^* \right)^*
            \ar[u, "\text{multiply and concat}"']
            \\ 
            \left( \Sigma \left( \Sigma B \right)^* \right)^*
            \ar[u, "{\repairterm^*}"]
            \ar[r, "\prefterm"]
            &\left( \Sigma \left( \Sigma B \right)^* \right)^*
            \ar[u, "\postrepairterm"']
        \end{tikzcd}
        \]
        where the ``multiply and concat'' term is the one described above, which multiplies every element of every sublist by the product of sublists strictly before it, and then concatenates everything into $(\Sigma B)^*$. With everything we established, we can see that the diagram commutes denotationally, and thus also provably by \cref{cor:aperiodic-completeness}, as it can be turned into an equivalent aperiodic diagram by reversing the bottom $\prefterm$ arrow. 
        
        Moreover, this diagram implies that the top-right face of the main diagram commutes if we choose the first aperiodic term to be as described above, of which $(\repairterm \comp \text{forget outer label})$ is a one-sided inverse (which again follows from \cref{cor:aperiodic-completeness}), and the second aperiodic term to be $(\postrepairterm \comp \text{multiply and concat})$.
    \end{description}
\end{proof}

\section{Restricted Prefix Multiplication}
\label{sec:restricted-prefix-multiplication}

\newcommand{\rprefterm}{\mathsf{rpref}}
We finish this paper with an observation that is not directly used in the rest of the paper, but is interesting in its own right. 
As defined in \cite{BojanczykDaviaudKrishna2018} and used in this paper, the \kl{group prefix multiplication} has type 
\begin{align*}
(A \times G)^* \xrightarrow{\prefterm} (A  \times G)^*.
\end{align*}
This means that there is a \kl{type} $A$ that is not really involved in the operation, and only copied verbatim from input to output.  What would happen if we removed this type? In other words, what if we had only a restricted prefix multiplication of type 
\begin{align*}
G^* \xrightarrow{\rprefterm} G^*,
\end{align*}
which would correspond to the original one in the special case of $A = \onetype$, up to isomorphism.

In this section, we show that the restricted prefix multiplication is expressive enough to define the extended prefix multiplication, and hence the system would remain expressively complete even if we only had the restricted prefix multiplication. Nevertheless, as we explain at the end of this section,  we do not know how to modify our equational axiomatisation to use only the restricted prefix multiplication, and hence we keep the extended version in our system.

The expressive completeness of the restricted operation follows from the following lemma.

\begin{lemma}\label{lem:reduction-to-non-extended-prefix-multiplication}
    For every finite group $G$ and \kl{type} $A$ there is a finite group $H$ and \kl{aperiodic terms}  encode and decode such that the following diagram commutes:
    \[
    \begin{tikzcd}[column sep=2cm]
    (A \times G)^* 
    \ar[rr, "\text{prefix multiplication}"]
    \ar[d, "\text{encode}"']
    &  & (A \times G)^* \\
    H^*
    \ar[rr,"\text{prefix multiplication}"'] &  &
    H^*
    \ar[u, "\text{decode}"']
    \end{tikzcd}
    \]
\end{lemma}
\begin{proof}
    We begin by reducing  to  the special case when $A$ is a type of the form $N^*$ for some \kl{finite alphabet}. To see this reduction, consider an \kl{injective} string representation of type $A \to N^*$ for some finite alphabet $N$, as in \cref{lem:provably-injective-string-representation}. By naturality, we can decompose the prefix multiplication as 
    \begin{align*}
    (A \times G)^* \xrightarrow{(\text{encode}, \idterm)^*}
        (N^* \times G)^* \xrightarrow{\prefterm}
    (N^* \times G)^* \xrightarrow{(\text{decode}, \idterm)^*}
(A \times G)^*
    \end{align*}
    Therefore, it remains to show that the middle part can be expressed in terms of the restricted prefix multiplication. 
Let $F$ be some finite group that is strictly larger than the finite alphabet $N$, and consider some injective term $f : N \to F$ that does not use the group identity as an output. Using the product group $H = F \times G$, we decompose  prefix multiplication as follows: 
    \begin{enumerate}
        \item Suppose that the input list is 
        \begin{align*}
        [(w_1,g_1),\ldots,(w_n,g_n)] \in (N^* \times G)^*.
        \end{align*}
        In the first step, we replace each pair in the input list by a list as follows
        \begin{align*}
        ([b_1,\ldots,b_m],g) \quad \mapsto \quad [(f(b_1),1_G),\ldots,(f(b_m),1_G),(1_F,g)] \in H^*,
        \end{align*}
        where $1_F$ and $1_G$ are the identities of the groups $F$ and $G$, respectively. Then we concatenate all these lists to obtain a single list in $H^*$. 
        \item In the second step, we apply the restricted  prefix multiplication operation 
        \begin{align*}
        H^* \xrightarrow \rprefterm H^*
        \end{align*}
         to the result of the previous step. 
        \item In the third and final step, we need to reverse the encoding. We apply the aperiodic inverse of $\rprefterm$. After doing this, the  group elements from $G$ correspond to list elements where $1_F$ is used in the first coordinate, and the letters from $F$ correspond to the remaining coordinates. Using this, we can reconstruct the original strings $w_1,\ldots,w_n$ hence we obtain the desired output.
    \end{enumerate}
    The terms in the first and third steps above are easily seen to be aperiodic, as required in the lemma.        
    \end{proof}

    Thanks to the above lemma, the system would remain expressively complete if we had only the restricted prefix multiplication, and not the extended one. Nevertheless, our equational axiomatisation is still based on the (nonrestricted) version of prefix multiplication. This is because we have not yet managed to formalise the above reasoning in an equational axiomatisation. The issue is that the  proof of the above lemma crucially relies on naturality of the (nonrestricted) prefix multiplication, and hence such naturality would need to be either phrased as an axiom (thus requiring the operation) or proved from simpler axioms (which we do not know how to do).

\end{document}